\documentclass[hyper,letterpaper]{JHEP3}

\usepackage{epsfig}
\usepackage{amsbsy}
\usepackage{varioref} 
\usepackage{pifont}
\usepackage{amsmath} 
\usepackage{graphicx}
\usepackage{axodraw}

\def\slashii#1{\setbox0=\hbox{$#1$}             
   \dimen0=\wd0                                 
   \setbox1=\hbox{\sl/} \dimen1=\wd1            
   \ifdim\dimen0>\dimen1                        
      \rlap{\hbox to \dimen0{\hfil\sl/\hfil}}   
      #1                                        
   \else                                        
      \rlap{\hbox to \dimen1{\hfil$#1$\hfil}}   
      \hbox{\sl/}                               
   \fi}                                         %
\def\slashiii#1{\setbox0=\hbox{$#1$}#1\hskip-\wd0\hbox to\wd0{\hss\sl/\/\hss}}
\title{One-Loop Corrections to the $S$ and $T$ Parameters \\
 in a Three Site Higgsless Model}

\author{
Shinya Matsuzaki\\
Department of Physics, Nagoya University\\
Nagoya 464-8602, Japan\\
	E-mail: synya@eken.phys.nagoya-u.ac.jp}

\author{R. Sekhar Chivukula and Elizabeth H. Simmons\\
Department of Physics and Astronomy, Michigan State University\\
East Lansing, MI 48824, USA\\
E-mail: sekhar@msu.edu, esimmons@msu.edu}

\author{
Masaharu Tanabashi\\
Department of Physics, Tohoku University\\
Sendai 980-8578, Japan\\
	E-mail: tanabash@tuhep.phys.tohoku.ac.jp}

\abstract{
In this paper we compute  the one-loop chiral logarithmic corrections to the $S$ and $T$ parameters in a highly deconstructed Higgsless model with only three sites. In addition to the electroweak gauge bosons, this model contains a single extra triplet of vector states (which we denote ${\rho}^{\pm}$ and $\rho^0$), rather than an infinite tower of ``KK" modes.  We compute the corrections to $S$ and $T$  in 't~Hooft-Feynman gauge, including the ghost, unphysical Goldstone-boson, and appropriate ``pinch" contributions required to obtain gauge-invariant results for the one-loop self-energy functions. We demonstrate that the chiral-logarithmic corrections naturally separate into two parts, a model-independent part arising from scaling below the $\rho$ mass, which has the same form as the large Higgs-mass dependence of the $S$ or $T$ parameter in the standard model, and a second model-dependent contribution arising from scaling between the $\rho$ mass and the cutoff of the model. The form of the universal part of the one-loop result allows us to correctly interpret the phenomenologically derived limits on the $S$ and $T$ parameters (which depend on a ``reference" Higgs-boson mass) in this three-site Higgsless model. Higgsless models may be viewed as dual to models of dynamical symmetry breaking akin to ``walking technicolor", and in these terms our calculation is the first to compute the subleading  $1/N$ corrections to the $S$ and $T$ parameters. We also discuss the reduction of the model
to the ``two-site" model, which is the usual electroweak chiral lagrangian, noting
the ``non-decoupling" contributions present in the limit $M_\rho \to \infty$.\\ \\ 
\centerline{February 27, 2007}}


\preprint{DPNU-06-04\\
MSUHEP-060717\\
TU-783}

\begin{document}

\section{Introduction}

Higgsless models \cite{Csaki:2003dt}  accommodate electroweak symmetry breaking without the introduction of a fundamental scalar  Higgs boson \cite{Higgs:1964ia}. In these models, the unitarity of longitudinally-polarized electroweak gauge-boson scattering is achieved through the exchange of extra vector bosons \cite{SekharChivukula:2001hz,Chivukula:2002ej,Chivukula:2003kq,He:2004zr}, rather than scalars. Based on TeV-scale \cite{Antoniadis:1990ew} compactified five-dimensional gauge theories with appropriate boundary conditions \cite{Agashe:2003zs,Csaki:2003zu,Burdman:2003ya,Cacciapaglia:2004jz}, these models provide effectively unitary descriptions of the electroweak sector beyond the TeV energy scale. They are not, however, renormalizable, and must be viewed as effective theories valid below a cutoff energy scale inversely proportional to the five-dimensional gauge-coupling squared. Above this energy scale, some new ``high-energy" completion, which is valid to higher energies, must be present.

Deconstruction \cite{Arkani-Hamed:2001ca,Hill:2000mu}  is a technique to build four-dimensional gauge theories, with appropriate gauge symmetry breaking patterns. which approximate -- at least over some energy range -- the properties of a five-dimensional theory.  Deconstructed Higgsless models \cite{Foadi:2003xa,Hirn:2004ze,Casalbuoni:2004id,Chivukula:2004pk,Perelstein:2004sc,Georgi:2004iy,SekharChivukula:2004mu} have been used as tools to compute the general properties of Higgsless theories, and to illustrate the phenomological properties of this class of models. 

In the simplest realization of Higgsless models, the ordinary fermions are localized (on ``branes") in the extra dimension. Such models necessarily \cite{SekharChivukula:2004mu} give rise to large tree-level corrections to the electroweak $S$ parameter, and are not phenomenologically viable. It has been shown, however, that by relaxing the fermion locality constraint  \cite{Cacciapaglia:2004rb,Cacciapaglia:2005pa,Foadi:2004ps,Foadi:2005hz,Chivukula:2005bn} -- more correctly, by allowing fermions to propagate in the compactified fifth dimension and identifying the ordinary fermions with the lowest KK fermion states -- it is always  \cite{SekharChivukula:2005xm} possible to choose the fermion wavefunction in the fifth dimension so that all four-fermion electroweak quantities at tree-level have their standard model forms.\footnote{It should be emphasized, however, that there is no explanation
in any of these models (which are only low-energy effective theories)
for the amount of delocalization. In particular, there is no 
dynamical reason why the fermion delocalization present
{\it must} be such as to make the value of $\alpha S$
small.}

Recently, a detailed investigation of a highly deconstructed three site Higgsless model \cite{SekharChivukula:2006cg}  -- in which the only vector states are the ordinary electroweak gauge bosons and a single triplet of $\rho^\pm$ and $\rho^0$ vector states -- has been completed.\footnote{Note that the $\rho^\pm$ and $\rho^0$ here correspond to the ${W'}^\pm$ and $Z'$ in that paper.} Although relatively simple in form, the model was shown to be sufficiently rich to incorporate the interesting physics issues related to fermion masses and electroweak observables. Calculations were presented addressing the size of corrections\footnote{In the original version of \protect\cite{SekharChivukula:2006cg}, we used the notation $\Delta\rho$ rather than $\alpha T$.  To the order we are working, they are identical: $Y \propto (\Delta\rho - \alpha T)$ vanishes in an ideally delocalized model \cite{SekharChivukula:2005xm}.} to $\alpha T$, $b \to s \gamma$, and $Z\to b \bar{b}$.

In this paper we compute  the one-loop chiral logarithmic corrections to the $S$ and $T$ parameters \cite{Peskin:1992sw,Altarelli:1990zd,Altarelli:1991fk} in the three site Higgsless model, in the limit $M_W \ll M_{\rho} \ll \Lambda$, where $\Lambda$ is the cutoff of the effective theory. We compute these corrections in 't~Hooft-Feynman gauge, including the ghost, unphysical Goldstone-boson, and appropriate  ``pinch" contributions \cite{Degrassi:1992ue,Degrassi:1992ff} required to obtain gauge-invariant results for the one-loop self-energy functions. 

For the $S$-paramter, we find the result
\begin{eqnarray}
\alpha S_{3-site} & = & \left[\frac{4 s^2 M^2_W}{M^2_\rho}
\left(1- \frac{x_1 M^2_\rho}{2 M^2_W}\right)\right]_{\mu=\Lambda} + \frac{\alpha}{12 \pi} \log{\frac{M^2_\rho}{M^2_{Href}}}\nonumber \\
& - & \frac{41 \alpha}{24 \pi} \log{\frac{\Lambda^2}{M^2_{\rho}}}
+ \frac{3\alpha}{8\pi} \left( \frac{x_1M_\rho^2}{2M_W^2} \right) 
\log\frac{\Lambda^2}{M_{\rho}^2} 
\nonumber \\ 
&& -8 \pi \alpha (c_1(\Lambda) + c_2(\Lambda))~.
\label{result-sparam}
\end{eqnarray}
where the parameter $x_1$ measures the amount of fermion delocalization, $M_{Href}$
is the reference Higgs boson mass used in the definition of the $S$-parameter, and 
$c_{1,2}$ are higher order counter-terms \cite{Perelstein:2004sc}. The parameters 
$M^2_\rho$, $M^2_W$, and $x_1$ are renormalized at one-loop\footnote{In a forthcoming paper
\cite{newpaper} we will report the results of a full renormalization-group analysis of the ${\cal O}(p^4)$
terms in the three-site Higgsless model effective theory, allowing us to independently
confirm the results in Eqns.
(\protect\ref{result-sparam}) and (\protect\ref{result-t}), and to express the values of $\alpha S$
and $\alpha T$ in terms of low-energy parameters.}   and, to this order,
in the first term of Eqn. (\ref{result-sparam})
 they should be understood to be evaluated at the scale $\Lambda$.
Note that the chiral-logarithmic corrections naturally separate into two parts, a model-independent part arising from scaling below the $\rho$ mass, which has the same form as the large Higgs-mass dependence of the $S$-parameter in the standard model, and a second model-dependent contribution arising from scaling between the $\rho$ mass and the cutoff of the model. The form of the model-independent part of the one-loop result allows us to correctly interpret the phenomenologically derived limits on the $S$ parameter (which depend on a ``reference" Higgs-boson mass \cite{Peskin:1992sw}) in this three-site Higgsless model. 

Similarly, we obtain for $T$ 
 \begin{equation} 
      \alpha T_{3-site} = - \frac{3 \alpha}{16 \pi c^2} \log \frac{M_\rho^2}{M_{Href}^2} 
              - \frac{3 \alpha}{32 \pi c^2} \log \frac{\Lambda^2}{M_\rho^2}
              + \frac{4\pi \alpha\,c_0(\Lambda)}{c^2}
\label{result-t}
\,,  
 \end{equation}
where $M_{Href}$ is the reference Higgs-boson mass, $c$ is approximately cosine of the standard weak mixing angle (see Eqn. (\ref{eq:scdef})) , and $c_0(\Lambda)$
is the relevant ${\cal O}(p^4)$ custodial isospin-violating counterterm renormalized at
scale $\Lambda$.   Again, note the separation into model-independent and model-dependent pieces 
and the standard-model-like dependence on the ``reference'' Higgs-boson mass.

The next few sections of the paper introduce the model and the form of the Lagrangian in terms of the gauge eigenstates and mass eigenstates.  We then present the results of our computations of the one-loop corrections to the self-energy functions of the $W$ and $Z$ bosons.  Subsequently, we compute the one-loop corrections to the $S$ and $T$ parameters arising from the gauge sector and arrive at the results summarized above.  We then turn to the relationship between
the $M_\rho \to \infty$ limit of the three-site model and the usual electroweak chiral lagrangian \cite{Appelquist:1980ae,Appelquist:1980vg}, discussing the importance of the ``non-decoupling" contributions \cite{Appelquist:1974tg} which
arise in this limit. 

We conclude the paper by discussing the relationship of our results to the general expectations for the form of these corrections in models with a strongly-interacting symmetry breaking sector. 
Higgsless models may be viewed as dual \cite{Maldacena:1998re,Gubser:1998bc,Witten:1998qj,Aharony:1999ti} to models of dynamical symmetry breaking \cite{Weinberg:1979bn,Susskind:1979ms} akin to ``walking technicolor" \cite{Holdom:1981rm,Holdom:1985sk,Yamawaki:1986zg,Appelquist:1986an,Appelquist:1987tr,Appelquist:1987fc}, and in these terms our calculation is the first to compute the subleading  $1/N$ corrections to the $S$ and $T$ parameters.  The model we discuss is in the same class as models of extended  electroweak gauge symmetries \cite{Casalbuoni:1985kq,Casalbuoni:1996qt}  motivated by hidden local symmetry models \cite{Bando:1985ej,Bando:1985rf,Bando:1988ym,Bando:1988br,Harada:2003jx} of chiral dynamics in QCD.  We specifically compare our findings to the corresponding results in the ``vector limit" \cite{Georgi:1989xy} of  hidden local symmetry models.


\section{The Three-Site Model}

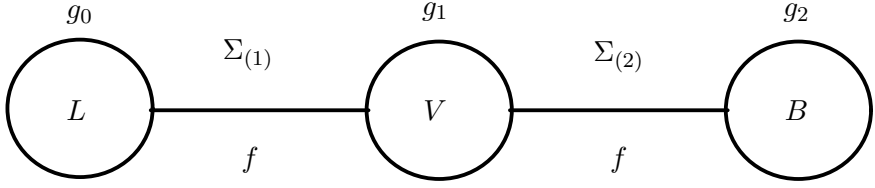
\begin{figure}
\begin{center}
\vspace{10pt}
\unitlength 0.1in
\begin{picture}( 45.2000, 13.0500)(  9.9000,-16.0000)
%
\special{pn 20}%
\special{ar 1370 1070 380 360  0.0000000 6.2831853}%
%
\special{pn 20}%
\special{pa 1740 1080}%
\special{pa 2880 1080}%
\special{fp}%
%
\special{pn 20}%
\special{pa 3620 1080}%
\special{pa 4760 1080}%
\special{fp}%
%
\special{pn 20}%
\special{ar 3250 1080 380 360  0.0000000 6.2831853}%
%
\special{pn 20}%
\special{ar 5130 1080 380 360  0.0000000 6.2831853}%
\put(13.5000,-10.8000){\makebox(0,0){$L$}}%
\put(32.3000,-10.8000){\makebox(0,0){$V$}}%
\put(51.2000,-10.8000){\makebox(0,0){$B$}}%
\put(22.5000,-7.9000){\makebox(0,0){$\Sigma_{(1)}$}}%
\put(41.9000,-8.0000){\makebox(0,0){$\Sigma_{(2)}$}}%
\put(22.6000,-13.4000){\makebox(0,0){$f$}}%
\put(41.9000,-13.4000){\makebox(0,0){$f$}}%
%
%
%
\put(13.7000,-5.8000){\makebox(0,0){$g_0$}}%
\put(32.3000,-5.6000){\makebox(0,0){$g_1$}}%
\put(51.2000,-5.7000){\makebox(0,0){$g_2$}}%
\end{picture}%
\vspace{10pt}
\caption{The three-site Higgsless model analyzed in this paper is illustrated using ``moose notation'' \cite{Georgi:1985hf}.  The model  incorporates an
$SU(2)_L \times SU(2)_V \times U(1)_B$ gauge group with couplings $g_0$, $g_1$, and $g_2$
respectively, and two 
nonlinear $(SU(2)\times SU(2))/SU(2)$ sigma models in which the global symmetry groups 
in adjacent sigma models are identified with the corresponding factors of the gauge group.
}
\label{threesite}

\end{center}
\end{figure}

The three-site Higgsless model analyzed in this paper is illustrated in Fig. \ref{threesite} using ``moose notation'' \cite{Georgi:1985hf}.  The model  incorporates an
$SU(2)_L \times SU(2)_V \times U(1)_B$ gauge group with couplings $g_0$, $g_1$, and $g_2$
respectively, and $2$ 
nonlinear $(SU(2)\times SU(2))/SU(2)$ sigma models in which the global symmetry groups 
in adjacent sigma models are identified with the corresponding factors of the gauge group.
The symmetry breaking between the middle $SU(2)$ and the $U(1)$ follows an $SU(2)_L \times SU(2)_R/SU(2)_V$ symmetry breaking pattern with the $U(1)$ embedded as the $T_3$-generator of $SU(2)_R$.    The  leading order lagrangian in the model is given by 
\begin{eqnarray} 
\mathcal{L}_{(2)} 
&=&  \frac{f^2}{4} \sum_{i=1}^2 {\rm tr}[D_{\mu} {\Sigma}_{(i)}^{\dagger} D^{\mu} {\Sigma}_{(i)}]  
\nonumber \\
&& 
- \frac{1}{2g_0^2} {\rm tr}[L_{\mu\nu}]^2 - \frac{1}{2g_1^2} {\rm tr}[V_{\mu\nu}]^2 
- \frac{1}{2g_2^2} {\rm tr}[R_{\mu\nu}]^2
\,, \label{p2 lag}
\end{eqnarray}
where $L_{\mu\nu}$, $V_{\mu\nu}$, and $R_{\mu\nu}$ are the matrix field-strengths of the three gauge groups,  $R_\mu= B_{\mu} \frac{\sigma_3}{2}$, and the covariant derivatives acting on $\Sigma_{(i)}$ are defined as  
\begin{eqnarray} 
D_{\mu} \Sigma_{(1)} &=&  
\partial_{\mu} \Sigma_{(1)} - i L_{\mu} \Sigma_{(1)} + i 
\Sigma_{(1)} V_{\mu} 
\,, \\
D_{\mu} \Sigma_{(2)} &=&  
\partial_{\mu} \Sigma_{(2)} - i V_{\mu} \Sigma_{(2)} + i 
\Sigma_{(2)} R_{\mu} 
\, .
\end{eqnarray}
The $2 \times 2$ unitary matrix fields $\Sigma_{(1)}$ and $\Sigma_{(2)}$ may be parametrized by the Nambu-Goldstone (GB) boson fields
$\pi_{(1)}$ and $\pi_{(2)}$:
\begin{eqnarray} 
\Sigma_{(i)} = e^{2 i {\pi}_{(i)}/f} \,, 
\qquad 
{\rm for}
\quad 
i=1,2 
\,, 
\end{eqnarray}
with the decay constant\footnote{For simplicity, here we take the same decay constant $f$ for both links.}  $f$. 

This model (see \cite{SekharChivukula:2006cg} for details) approximates the standard model
in the limit
\begin{equation}
x={g_0/g_1} \ll 1~,\ \ \ \ \ y={g_2/g_1}\ll 1~,
\end{equation}
in which case we expect a massless photon, light $W$ and $Z$ bosons, and
a heavy set of bosons $\rho^\pm$ and $\rho^0$ with $M_W \ll M_{\rho}$ .  
Numerically, then, $g_{0,2}$ are approximately
equal to the standard model $SU(2)_W$ and $U(1)_Y$ couplings, and we therefore
define an angle $\theta$ such that  $s=\sin\theta$, $c=\cos\theta$, and
\begin{equation}
g^2_0 \approx \frac{4\pi\alpha}{s^2}={e^2\over s^2}~,\ \ \ \
g^2_2 \approx \frac{4 \pi \alpha}{c^2}={e^2\over c^2}~,\ \ \ \ 
 \frac{s}{c} = \frac{g_2}{g_0} 
\label{eq:scdef}
\end{equation}
where $\alpha$ is the fine-structure constant and $e$ the charge of the electron.

\subsection{Fermion Couplings and $\alpha S$ at Tree-Level}

In general, the standard model fermions may be delocalized along the three-site moose
in the sense that their weak couplings arise from both sites  0 and 1 \cite{Anichini:1994xx,Chivukula:2005bn}
\begin{equation}
{\cal L}_f = \vec{J}^\mu_L \cdot \left((1-x_1) L_\mu + x_1 V_\mu\right) + J^\mu_Y B_\mu~,
\label{fermioni}
\end{equation}
where $\vec{J}^\mu_L$ and $J^\mu_Y$ are the fermionic weak and hypercharge currents,
respectively, and $0\le x_1 \ll 1$ is a measure of the amount of fermion delocalization.
This expression is not separately gauge invariant under
$SU(2)_0$ and $SU(2)_1$. Rather, the fermions should be viewed as being charged under
$SU(2)_0$, and the terms proportional to $x_1$
should be interpreted as arising from the operator of the form
\begin{equation}
{\cal L'}_f = - x_1\cdot \bar{\psi}_L (i\slashiii{D} \Sigma_{(1)} \Sigma^\dagger_{(1)} ) \psi_L~,
\label{eq:delocop}
\end{equation}
in ÔÔunitaryÕÕ gauge. We will be interested only in the light fermions ({\it i.e.} all standard
model fermions except for the top-quark), and will therefore ignore the couplings giving
rise to fermon masses (these are discussed in detail in \cite{SekharChivukula:2006cg}).

Some degree of fermion delocalization is desirable for phenomenological reasons.
Diagonalizing the gauge-boson mass matrix and computing the relevant tree-level four-fermion
processes, one may compute the value of the $S$-parameter at tree-level, with the
result \cite{Anichini:1994xx,Chivukula:2005bn}
\begin{equation}
\alpha S^{tree} = \frac{4 s^2 M^2_W}{M^2_\rho}
\left(1- \frac{x_1 M^2_\rho}{2 M^2_W}\right)~.
\label{eq:stree}
\end{equation}
Current experimental bounds on $\alpha S$ are ${\cal O}(10^{-3})$ \cite{Barbieri:2004qk}.
Since exchange of the $\rho$ meson is necessary to maintain the unitarity of longitudinally
polarized $W$-boson scattering, we must require that $M_\rho \le {\cal O}(1\,{\rm TeV})$ --
leading, for localized fermions with $x_1=0$,  to a value of $\alpha S^{tree}$ which is too large.
For the three-site model to be viable, therefore, the value of the fermion delocalization parameter
must be chosen to reduce the value of $\alpha S^{tree}$  \cite{Cacciapaglia:2004rb,Cacciapaglia:2005pa,Foadi:2004ps,Foadi:2005hz,Chivukula:2005bn,SekharChivukula:2005xm}.\footnote{See footnote 1.}

\subsection{Duality and The Size of Radiative Electroweak Corrections}

By duality \cite{Maldacena:1998re}, tree-level computations in the 5-dimensional 
theory represent the leading terms in a large-$N$ expansion \cite{'tHooft:1973jz}
of the strongly-coupled dual gauge theory akin to ``walking technicolor" \cite{Holdom:1981rm,Holdom:1985sk,Yamawaki:1986zg,Appelquist:1986an,Appelquist:1987tr,Appelquist:1987fc}.  The mass of the $W$ boson, $M_W$, is proportional to a weak gauge-coupling, $g_{ew}$,
(which is fixed in the large-$N$ approximation) times the $f$-constant for the electroweak 
chiral symmetry breaking of the strongly-coupled theory. Therefore, we expect $M^2_W/M^2_\rho$
to scale as \cite{'tHooft:1973jz,Chivukula:1992gi}
\begin{equation}
\frac{M^2_W}{M^2_\rho} = {\cal O}\left(\frac{g_{ew}^2 N}{(4\pi)^2}\right)~,
\end{equation}
which is the expected behavior of the $S$-parameter in the large-$N$
limit \cite{Peskin:1992sw,Burdman:2003ya}.

We now specify the limit in which we will perform our analysis.
As shown below (see Eqns. (\ref{mwwpm}, \ref{mrhopm})), in the small $x$ limit
\begin{equation}
\frac{M^2_W}{M^2_\rho} \approx \frac{x^2}{4}~,
\end{equation}
so that the tree-level value of $\alpha S$ vanishes if $x_1=x^2/2$.
In what follows, therefore, we will assume that $x_1 ={\cal O}(x^2)$.
Overall, then, we work in the limit
\begin{equation}
1 \gg |\alpha S^{tree}|
={\cal O}\left(\frac{g^2_{ew} N}{(4\pi)^2}\right)
\gg  |\alpha S^{one-loop}|={\cal O}\left(\frac{g^2_{ew}}{(4\pi)^2}\right) >  0~,
\end{equation}
which is manifestly consistent with the large-$N$ approximation.
Once we choose the value of $x_1$ to make the size of
$\alpha S^{tree}$ consistent with the phenomenological bound
of ${\cal O}(10^{-3})$,
one-loop electroweak corrections $\alpha S^{one-loop}$
become potentially relevant.  

Note also that 
$\alpha T^{tree} \approx 0$ in these models, independent of the degree of fermion delocalization \cite{SekharChivukula:2004mu,Chivukula:2005bn}.  
The one-loop corrections to $\alpha T$ are therefore of interest.  Those arising from the extended fermion sector have been shown$^2$ to place strong lower bounds on the masses of the KK fermions \cite{SekharChivukula:2006cg}.  Those arising from the gauge sector are considered below.

In practice, in calculating corrections to the gauge-boson self-energy functions,
we will work in the  leading-log approximation, and to order $\alpha$; we will
neglect corrections ${\cal O}(\alpha x^2 M^2_W)$ or ${\cal O}(\alpha x^2 p^2)$, but keep those of order ${\cal O}(\alpha x^2 M^2_\rho)$.


\section{Gauge Sector Lagrangian}

\label{gaugelagrangian}

In order to obtain the relevant interaction terms to compute the one-loop
electroweak corrections,  
we expand the link variables $\Sigma_1$ and $\Sigma_2$ as follows
\begin{eqnarray} 
\Sigma_{(i)} = 1 + 2 i \frac{\pi_{(i)}}{f} - \frac{2 \pi_{(i)}^2}{f^2} + \mathcal{O}(\pi^3) 
\,, \qquad \textrm{for $i=1, 2$}
\,. 
\end{eqnarray} 
Furthermore, it is convenient to change the normalization of the gauge-boson fields
so that the gauge-boson kinetic energy terms in Eqn. (\ref{p2 lag}) are canonically
normalized, and to introduce the following vectors in ``link" and ``site" space, respectively
\begin{eqnarray} 
\vec{\pi^a} 
= 
\left( 
\begin{array}{c} 
\pi_{(1)}^a \\ 
\pi_{(2)}^a 

\end{array}
\right)
\,, \qquad 
\vec{A}^a_{\mu} 
= 
\left( 
\begin{array}{c} 
L_{\mu}^a \\ 
V_{\mu}^a \\
R_{\mu}^a
\end{array}
\right)
\,, 
\end{eqnarray}
with $R_\mu^1=R_\mu^2=0$, and $R_\mu^3=B_\mu$.  
 In terms of these quantities, the lagrangian (\ref{p2 lag}) decomposes into the following pieces: 
\begin{eqnarray} 
    \mathcal{L}_{(2)} 
    &=& \mathcal{L}_{\pi\pi}^{AA} + \mathcal{L}_{\pi AA} + \mathcal{L}_{\pi\pi A} + \mathcal{L}_{\pi\pi AA} 
    \nonumber \\ 
    && + \mathcal{L}_{AAA} + \mathcal{L}_{AAAA} + \cdots 
    \,, \label{lagfull}
\end{eqnarray}
where we have ignored the interaction terms including more than three GB fields 
since these terms do not generate vertices relevant to the one-loop processes of interest.

\subsection{ $\mathcal{L}_{\pi\pi}^{AA}$: Kinetic Energy and  Gauge-Fixing Terms}

Terms in the lagrangian $\mathcal{L}_{\pi\pi}^{AA}$ are quadratic in the 
GB  fields $\vec{\pi}^a$ or  gauge fields $\vec{A}^a_\mu$
\begin{eqnarray}
\mathcal{L}_{\pi\pi}^{AA} 
    = 
\mathcal{L}^{\rm kin}_{\rm gauge} 
+ \frac{1}{2} \left( \partial_{\mu} \vec{\pi}^a - f \left(D \cdot G \cdot \vec{A}^a_\mu \right) \right)^{T} 
\cdot 
\left( \partial^{\mu} \vec{\pi}^a - f \left( D \cdot G \cdot \vec{A}^{\mu a}\right) \right)
\,, \label{p2 lag quad}
\end{eqnarray}
where the kinetic terms of the gauge fields $\vec{A}_{\mu}^a$ are included in 
 $\mathcal{L}^{\rm kin}_{\rm gauge}$, 
$D$ is a $2 \times 3$ difference matrix in the link/site space  defined as 
\begin{eqnarray} 
D =  \left(  
\begin{array}{ccc} 
1 & -1 & 0 \\ 
0 & 1 & -1 
\end{array}
\right)
\, ,
\end{eqnarray}
and $G$ is the gauge coupling-constant matrix with the diagonal elements $(g_0, g_1, g_2)$.  

       It is convenient to introduce charge eigenstate fields 
\begin{eqnarray} 
&& \vec{A}^{\pm}_\mu=(L_\mu^\pm, V_\mu^\pm)^T  \, , \qquad 
   \vec{A}^0_\mu= (L_\mu^3, V_\mu^3, B_\mu)^T
\,, \\ 
&& \vec{\pi}^{\pm}=(\pi_{(1)}^\pm, \pi_{(2)}^\pm)^T 
\,, \qquad 
   \vec{\pi}^0=(\pi_{(1)}^3, \pi_{(2)}^3)^T
\, , 
\end{eqnarray} 
where
\begin{eqnarray} 
           L_{\mu}^{\pm} = \frac{L_{\mu}^1 \mp i L_{\mu}^2}{\sqrt{2}}  
           \,, \qquad 
           V_{\mu}^{\pm} = \frac{V_{\mu}^1 \mp i V_{\mu}^2}{\sqrt{2}} 
           \,,   
           \end{eqnarray} 
and with $\pi_{(i)}^\pm$ and $\pi_{(i)}^0$ ($i=1,2$) defined analogously. 
Using the charge eigenstate fields, the mass terms of the gauge fields are expressed as 
\begin{eqnarray} 
     \vec{A}_\mu^{+T}  M^2_{\rm CC} \vec{A}^{\mu -} 
         + \frac{1}{2} \vec{A}_\mu^{0 T} M^2_{\rm NC} \vec{A}^{\mu 0} 
\,, 
\end{eqnarray}
where $M_{CC}^2$ and $M_{NC}^2$ are the mass matrices for the charged 
and neutral gauge bosons 
\begin{eqnarray} 
M^2_{\rm CC} 
            &=& \frac{f^2}{4} 
\left( 
        \begin{array}{cc} 
            g_0^2  & - g_0 g_1 \\
         - g_0 g_1 & 2 g_1^2  
        \end{array}
\right) \,,\\
M^2_{\rm NC} 
            &=& \frac{f^2}{4} 
\left( 
        \begin{array}{ccc} 
                g_0^2 & - g_0g_1 & 0 \\
             - g_0g_1 & 2 g_1^2  &- g_1g_2 \\ 
                    0 & - g_1 g_2 & g_2^2 
        \end{array}
\right) \,.
\end{eqnarray}  

The lagrangian $\mathcal{L}_{\pi\pi}^{AA}$ includes quadratic mixing terms 
between the GB fields $\vec{\pi}^a$ and the gauge fields $\vec{A}_\mu^a$. 
These terms are eliminated by adding the following $R_\xi$ gauge fixing term\footnote{   
                   We take the same gauge parameter for all the gauge groups. 
                  }  \cite{SekharChivukula:2004mu}:
\begin{eqnarray} 
\mathcal{L}_{\rm GF} 
= - \frac{1}{2 \xi} \left( \vec{\mathcal{G}}^T \right)^a \cdot \left( \vec{\mathcal{G}}\right)^a 
\,, \label{gf}
\end{eqnarray}
where 
\begin{eqnarray} 
\vec{\mathcal{G}}^a =  
\left[ \left( \partial_{\mu} \vec{A}^{\mu a} \right) 
+ \frac{\xi f}{2} \left( G \cdot D^T \cdot \vec{\pi}^a \right) \right]
\, .
\end{eqnarray} 
After fixing the gauge, the unphysical Goldstone boson fields 
acquire the gauge-dependent masses $M_{\pi^\pm}$ and $M_{\pi^0}$
\begin{eqnarray}
  M_{\pi^\pm}^2 
 &=& \frac{\xi f^2}{4} 
\left( 
        \begin{array}{ccc} 
              g_0^2 + g_1^2 &\ &-g_1^2 \\
                     -g_1^2 &\ & g_1^2  
        \end{array}
\right)
\,, \label{mpipm}\\
M_{\pi^0}^2 
 &=& \frac{\xi f^2}{4} 
\left( 
        \begin{array}{ccc} 
              g_0^2 + g_1^2 &\ & -g_1^2 \\
                     -g_1^2 &\ & g_1^2 + g_2^2 
        \end{array}
\right)
\, .\label{mpi0}
\end{eqnarray} 
The lagrangian $\mathcal{L}_{\pi\pi}^{AA}$ combined with 
    the gauge-fixing term in $\mathcal{L}_{\rm GF}$ then become
\begin{eqnarray} 
\mathcal{L}_{\pi\pi}^{AA} + \mathcal{L}_{\rm GF}  
             &=& 
    - \vec{A}^{+ T}_\mu [D_{A^\pm}^{\mu\nu}] \vec{A}_\nu^- 
    - \frac{1}{2} \vec{A}^{0 T}_\mu [D_{A^0}^{\mu\nu}] \vec{A}_\nu^0 
    \nonumber \\ 
              && 
    - \vec{\pi}^{+T}  [D_{\pi^\pm}]  \vec{\pi}^- 
    - \frac{1}{2} \vec{\pi}^{0T} [D_{\pi^0}] \vec{\pi}^0 
\,, \label{lag quad} 
\end{eqnarray}
where 
\begin{eqnarray} 
 \left[ D_{A^\pm, A^0}^{\mu\nu} \right]   
           &=&  
         \left( - {\bf 1}\,\partial^2  - M_{CC,NC}^2 \right) g^{\mu\nu} 
  + \left(1 -  \frac{1}{\xi} \right) {\bf 1}\,\partial^\mu \partial^\nu 
 \,, \\ 
 \left[ D_{\pi^{\pm,0}} \right] 
           &=& 
         \left( - {\bf 1}\,\partial^2 - M_{\pi^{\pm,0}}^2 \right)
\,. 
\end{eqnarray} 

\subsection{The Fadeev-Popov  Ghost Lagrangian $\mathcal{L}_{\rm FP}$}

Next we introduce the ghost terms corresponding to the gauge fixing terms in Eqn. (\ref{gf})
\begin{eqnarray} 
 \mathcal{L}_{\rm FP} = - \bar{C}^a_I \cdot (\Gamma^{ab})_{IJ} \cdot C^b_J 
 \,, 
\end{eqnarray}
where $C_I^a$ and $\bar{C}_J^a$ $(I,J= 0,1,2)$ are respectively the Fadeev-Popov (FP) ghost and the anti-ghost fields corresponding to 
the gauge groups on the $I$th-and $J$th-site, and
\begin{eqnarray} 
  (\Gamma^{ab})_{IJ} = g_J \cdot \frac{ \delta \mathcal{G}_I^a }{\delta \Theta_J^b} 
  \,, 
\end{eqnarray}
with $\Theta^a_I$ being the infinitesmal generator of the gauge transformations. 
The infinitesimal transformation laws for the gauge fixing functions $\mathcal{G}^a_I$ are 
immediately derived from those for the gauge fields $\vec{A}^a_\mu$ and the GB fields $\vec{\pi}^a$
            \footnote{  Here we omit terms including more than two GB fields, 
                        since these interactions are irrelevant to the processes we are concerned with. 
                     }
\begin{eqnarray}
\delta L^a_\mu &=&  D_\mu \Theta_0^a 
        = \left( \partial_\mu \delta^{ac}+ g_0 \epsilon^{abc} L^b_\mu \right) \Theta^c_0 
\,, \\
\delta V^a_\mu &=&  D_\mu \Theta_1^a 
        = \left( \partial_\mu \delta^{ac}+ g_1 \epsilon^{abc} V^b_\mu \right) \Theta^c_1 
\,, \\
\delta B_\mu &=&  \partial_\mu \Theta_2 
\,, \\ 
\delta \pi_{(1)}^a  &=& \frac{f}{2} ( g_0 \Theta_0^a -  g_1 \Theta_1^a) + \mathcal{O}(\pi^2) 
\,, \\
\delta \pi_{(2)}^a  &=& \frac{f}{2} ( g_1 \Theta_1^a - g_2 \Theta_2^a) + \mathcal{O}(\pi^2) 
\,. 
\end{eqnarray}
Defining the charge eigenstates for the FP ghost fields, we find
\begin{eqnarray} 
  \mathcal{L}_{\rm FP} &=& 
                          \mathcal{L}_{\rm FP}^{\rm kin.} + \mathcal{L}_{\rm FP}^{\rm int.}  
                \,, \label{fp term} 
\end{eqnarray}
where 
\begin{eqnarray} 
   \mathcal{L}_{\rm FP}^{\rm kin.} 
   &=& - \bar{C}_i^+ \left( \partial^2 \delta_{ij} - \xi [M_{CC}^2]_{ij} \right) C_j^- + {\rm h.c.} 
   \nonumber \\
   &&  
       - \bar{C}_I^0 \left( \partial^2 \delta_{IJ} - \xi [M_{NC}^2]_{IJ} \right) C_J^0
   \,, \label{fpmass} \\ 
   \mathcal{L}_{\rm FP}^{\rm int.} 
   &=&  i g_i  \left( A_{\mu i}^+ \left\{ \partial^\mu \bar{C}^-_i C^3_i - \partial^\mu \bar{C}^3_i C^-_i \right\}
                     + A_{\mu i}^3 \partial^\mu \bar{C}^+_i C^-_i \right)  + {\rm h.c.}
   \,,   
   \label{re fpint}
\end{eqnarray}
with $A_{\mu i}^+ = (L_\mu^+, V_\mu^+)^T$ and $A_{\mu i}^3 = (L_\mu^3, V_\mu^3)^T$, and
where we sum  over the repeated indices ($i,j=0,1$ and $I,J=0,1,2$).

\subsection{The Non-Abelian Interactions $\mathcal{L}_{AAA}$ and $\mathcal{L}_{AAAA}$}

The Non-Abelian interaction terms among the gauge fields $L_\mu^a$ and $V_\mu^a$ are 
\begin{eqnarray}
\mathcal{L}_{AAA} 
           &=& 
i g_0 \Bigg[ \left( \partial_\mu L_\nu^+ - \partial_\nu L_\mu^+ \right) L^{\mu -} L^{\nu 3} 
             + \partial_\mu L_\nu^3 L^{\mu +} L^{\nu -}
      \Bigg]  
      \nonumber \\  
           && 
+ i g_1 \Bigg[ \left( \partial_\mu V_\nu^+ - \partial_\nu V_\mu^+ \right) V^{\mu -} V^{\nu 3} 
             + \partial_\mu V_\nu^3 V^{\mu +} V^{\nu -}
      \Bigg] + {\rm h.c.}
\,,  \label{re aaa}\\ 
\mathcal{L}_{AAAA} 
           & = & 
\frac{g_0^2}{2} \Bigg[ L_\mu^+ L_\nu^- \left( L^{\mu +} L^{\nu -}  +  L^{\mu -} L^{\nu +} \right)  
                         - 2 L_\mu^+ L_\nu^+  L^{\mu -} L^{\nu -}   \Bigg]  
           \nonumber \\ 
            && 
+ g_0^2 \Bigg[ L_\mu^+ L_\nu^- L^{\mu 3} L^{\nu 3}  - L_\mu^+ L^{\mu -} L_\nu^3 L^{\nu 3} \Bigg] 
           \nonumber \\ 
            && 
+ \frac{g_1^2}{2} \Bigg[ V_\mu^+ V_\nu^- \left( V^{\mu +} V^{\nu -}  +  V^{\mu -} V^{\nu +} \right)  
                         - 2 V_\mu^+ V_\nu^+  V^{\mu -} V^{\nu -}   \Bigg]  
           \nonumber \\ 
            && 
+ g_1^2 \Bigg[ V_\mu^+ V_\nu^- V^{\mu 3} V^{\nu 3}  - V_\mu^+ V^{\mu -} V_\nu^3 V^{\nu 3} \Bigg]            
\,. 
\label{re aaaa}
\end{eqnarray}

\subsection{The Goldstone Boson Interactions $\mathcal{L}_{\pi AA}, \mathcal{L}_{\pi\pi A}$ and $\mathcal{L}_{\pi\pi AA}$ }

The remaining Goldstone Boson interactions necessary for our computations are expressed as follows:
\begin{eqnarray} 
\mathcal{L}_{\pi A A}
&=& 
- \frac{g_1 f}{2} \epsilon^{abc} \left( g_0 L_{\mu}^a V^{\mu b} \pi_{(1)}^c + g_2 V_{\mu}^a R^{\mu b} \pi_{(2)}^c \right) 
\,, \label{int1} \\
\mathcal{L}_{\pi \pi A} 
&=& 
- \frac{1}{2} \epsilon^{abc} \partial_{\mu} \pi_{(1)}^a \pi_{(1)}^b \left(g_0  L^{\mu c} + g_1 V^{\mu c} \right) 
- \frac{1}{2} \epsilon^{abc} \partial_{\mu} \pi_{(2)}^a \pi_{(2)}^b \left( g_1 V^{\mu c} + g_2 R^{\mu c} \right) 
\,, \label{int2} \\
\mathcal{L}_{\pi \pi A A } 
&=& 
\frac{1}{2} \epsilon^{eab} \epsilon^{ecd} \left( g_0 g_1 L_{\mu}^a \pi_{(1)}^b V^{\mu c} \pi_{(1)}^d + 
g_1 g_2 V_{\mu}^a \pi_{(2)}^b R^{\mu c} \pi_{(2)}^d 
\right)
\,. \label{int3}
\end{eqnarray}
These interaction terms may be rewritten in terms of the charge eigenstate fields as\footnote{Here we
define $ (A \stackrel{\leftrightarrow}{\partial}_\mu B) \equiv 
     (\partial_\mu A ) B  - A (\partial_\mu B) $~.}
\begin{eqnarray} 
\mathcal{L}_{\pi AA} 
&=&  i\frac{g_1f}{2} \Bigg[g_0 \left( L_\mu^3 V^{\mu +} - V_\mu^3 L^{\mu +} \right) \pi_{(1)}^- 
               - g_2 B_\mu V^{\mu +} \pi_{(2)}^-  
               +  g_0 L_\mu^+ V^{\mu -} \pi_{(1)}^3  
               \Bigg]  + {\rm h.c.} 
               \,, \label{re pi aa} \\ 
\mathcal{L}_{\pi \pi A} 
&=& \frac{i}{2}  \Bigg[ g_0 \partial_\mu \pi_{(1)}^+ \pi_{(1)}^- L^{\mu 3}  
                       + g_1 \left( \partial_\mu \pi_{(1)}^+ \pi_{(1)}^- 
                                + \partial_\mu \pi_{(2)}^+ \pi_{(2)}^- \right) V^{\mu 3}  
                      + g_2 \partial_\mu \pi_{(2)}^+ \pi_{(2)}^- B^\mu  
                      \nonumber \\ 
 && 
 -\left( \pi_{(1)}^3  \stackrel{\leftrightarrow}{\partial}_\mu \pi_{(1)}^- \right) \left(g_0  L^{\mu +} + g_1 V^{\mu +} \right)
 - g_1 \left( \pi_{(2)}^3 \stackrel{\leftrightarrow}{\partial}_\mu  \pi_{(2)}^-\right)  V^{\mu +} 
\Bigg] +  {\rm h.c.} 
\,, \label{re pipi a}\\ 
\mathcal{L}_{\pi \pi AA} 
&=& 
  \frac{g_1}{2} \Bigg[ g_0 \pi_{(1)}^+ \pi_{(1)}^- \left( L_\mu^+ V^{\mu -} + L_\mu^3 V^{\mu 3} \right) 
- g_0 \pi_{(1)}^+ \pi_{(1)}^3 \left( L_\mu^- V^{\mu 3} + L_\mu^3 V^{\mu -} \right) 
\nonumber \\ 
&& 
+ g_0 \pi_{(1)}^3 \pi_{(1)}^3 L_\mu^+ V^{\mu -} - g_0 \pi_{(1)}^+ \pi_{(1)}^+ L_\mu^- V^{\mu -}
- g_2 \pi_{(2)}^- \pi_{(2)}^3 B_\mu V^{\mu +} + g_2 \pi_{(2)}^+ \pi_{(2)}^- B_\mu V^{\mu 3} \Bigg] 
\nonumber \\ 
&& 
+  {\rm h.c.} 
\,.
\label{re pipi aa}
\end{eqnarray}  
%


\section{Mass Eigenstate Fields}

To facilitate our computation of the one-loop corrections to $\alpha S$ and $\alpha T$, we express
the interactions derived above in terms of mass eigenstate fields. As we are interested
in the limit $x = {g_0/g_1} \ll 1$, we will diagonalize the mass matrices perturbatively in $x$.

The charged gauge boson mass matrix $M^2_{\rm CC}$ has the eigenvalues 
\begin{eqnarray} 
M_W^2 &=& 
          \frac{g^2_1 f^2}{4}\Bigg[ \frac{x^2}{2} - \frac{x^4}{8} + \mathcal{O}(x^6) \Bigg] 
\,, \label{mwwpm}\\ 
M_{\rho^\pm}^2 &=& 
          \frac{g^2_1 f^2}{4}\Bigg[ 2 + \frac{x^2}{2} - \frac{x^4}{8} + \mathcal{O}(x^6) \Bigg]  
\,. 
\label{mrhopm}
\end{eqnarray}         
Expanding the gauge-eigenstate fields in terms of the
mass eigenstates, we find
\begin{eqnarray} 
  L_\mu^\pm &=& v_{W^\pm}^L W_\mu^\pm + v_{\rho^\pm}^L \rho_\mu^\pm 
  \,, \label{lpm}\\ 
  V_\mu^\pm &=& v_{W^\pm}^V W_\mu^\pm + v_{\rho^\pm}^V \rho_\mu^\pm 
\,,  \label{vpm}
\end{eqnarray}
where 
\begin{equation} 
\begin{array}{cc} 
   v_{W^\pm}^L = 1- \frac{x^2}{8} + \cdots   \,, 

 & v_{\rho^\pm}^L =  -\frac{x}{2} \left( 1 + \frac{x^2}{8} + \cdots \right) \,,
\\
    v_{W^\pm}^V = \frac{x}{2} \left( 1 +  \frac{x^2}{8} + \cdots \right)  \,, 

 & v_{\rho^\pm}^V =   1 - \frac{x^2}{8} + \cdots  \, .  

\end{array}  
\label{wavefunc:Wrhopm} 
\end{equation} 

The neutral gauge boson mass matrix $M^2_{NC}$ has 
one zero eigenvalue, corresponding to the photon,
and the two non-zero eigenvalues
\begin{eqnarray} 
M_Z^2 &=& 
          \frac{g^2_1 f^2}{4} \Bigg[ \frac{x^2}{2c^2} - \frac{(1-t^2)^2 x^4}{8c^4} + \mathcal{O}(x^6) \Bigg] 
\,, \label{mzee}\\ 
M_{\rho^0}^2 &=& 
          \frac{g^2_1 f^2}{4} \Bigg[ 2 + \frac{x^2}{2c^2} + \frac{(1-t^2)^2x^4}{8} + \mathcal{O}(x^6) \Bigg]  
\label{mrho0}\,,
\end{eqnarray}         
where the angles $s=\sin\theta$ and $c=\cos\theta$ are defined in Eqn. (\ref{eq:scdef}), 
and $t=\tan\theta = s/c$. 
Expanding the neutral gauge-eigenstate fields in terms of mass eigenstates,
we find
\begin{eqnarray} 
 L_\mu^3 &=& v_A^L A_\mu + v_Z^L Z_\mu + v_{\rho^0}^L \rho_\mu^0 
 \,, \label{l3}\\ 
 V_\mu^3 &=& v_A^V A_\mu + v_Z^V Z_\mu + v_{\rho^0}^V \rho_\mu^0  
 \,, \label{v3}\\ 
 B_\mu   &=& v_A^B A_\mu + v_Z^B Z_\mu + v_{\rho^0}^B \rho_\mu^0
\,,  \label{b} 
\end{eqnarray}
where 
\begin{equation} 
\begin{array}{ccc} 
   v_A^L = s \left( 1 - \frac{1}{2} s^2 x^2  + \cdots \right) \,, &  

   v_Z^L = c \left( 1 - \frac{c^2x^2(1+2t^2-3t^4)}{8} + \cdots \right) \,, &   

   v_{\rho^0}^L = - \frac{x}{2} \left( 1 + \frac{x^2 (1-3t^2)}{8} \right) \,, \\ 

   v_A^V = s x \left( 1 - \frac{1}{2} s^2 x^2 + \cdots \right) \,, &  

   v_Z^V = \frac{cx(1-t^2)}{2} \left( 1 + \frac{c^2 x^2 (1-t^2)^2}{8}  + \cdots \right) \,, &

v_{\rho^0}^V = 1 - \frac{x^2(1+t^2)}{8} + \cdots  \,, \\ 

   v_A^B = c \left( 1 - \frac{1}{2} s^2 x^2 + \cdots \right) \,, &  

   v_Z^B = - s \left( 1 + \frac{c^2x^2 (3-2t^2-t^4)}{8} + \cdots \right) \,, &  
v_{\rho^0}^B = - \frac{xt}{2} \left( 1 - \frac{x^2 (3-t^2)}{8}  + \cdots \right) 
\,. 
\end{array} 
\label{wavefunc:AZrho0}
\end{equation} 
  In obtaining the photon wavefunctions $v_A^{L,V,B}$, 
we have expanded the electromagnetic coupling $e$ 
in powers of $x$ as 
\begin{eqnarray} 
  \frac{1}{e^2} &=& \frac{1}{g_0^2} + \frac{1}{g_1^2} + \frac{1}{g_2^2} 
  \nonumber \\  
                &=& \frac{1}{g_0^2 s^2} \left( 1 + s^2 x^2 + \cdots \right) 
  \,. \label{e-g0}
\end{eqnarray}

Since the mass matrices for the ghost fields are (see Eqn. (\ref{fpmass})) equal to those of
the vector bosons, up to an overall factor of $\xi$, the corresponding relationships
between the gauge-eigenstate and mass-eigenstate ghost fields are 
\begin{eqnarray} 
   C^\pm_{(0)} &=& v_{W^\pm}^L C_{W^\pm} + v_{\rho^\pm}^L C_{\rho^\pm} 
   \label{c0}
\,, \\ 
   C^\pm_{(1)} &=& v_{W^\pm}^V C_{W^\pm} + v_{\rho^\pm}^V C_{\rho^\pm} 
   \label{c1}
\, ,
\end{eqnarray}
and 
\begin{eqnarray} 
C_{(0)}^3 &=& v_A^L \ C_A + v_Z^L \ C_Z + v_{\rho^0}^L C_{\rho^0} 
\,, \\ 
C_{(1)}^3 &=& v_A^V \ C_A + v_Z^V \ C_Z + v_{\rho^0}^V \ C_{\rho^0} 
\,, \\ 
C_{(2)}^3 &=& v_A^B \ C_A  + v_Z^V \ C_Z  + v_{\rho^0}^B  C_{\rho^0}  
\, .
\end{eqnarray}

Similarly, the charged GB matrix $M_{\pi^\pm}^2$ has the eigenvalues $\xi M_W^2$ and 
$\xi M_{\rho^\pm}^2$, and
the neutral GB matrix $M_{\pi^0}^2$  has the eigenvalues $ \xi M_Z^2$ and $ \xi M_{\rho^0}^2$.
The mass matrices for the Goldstone bosons are given in Eqns. (\ref{mpi0}) and
(\ref{mpipm}).
Expanding the eigenvectors in powers of $x$ we find that the GB fields  
are expressed in terms of the mass eigenstate fields $\pi_{W^\pm,Z}$ and $\pi_{\rho^\pm,\rho^0}$
as
\begin{eqnarray} 
  \pi_{(1)}^{\pm,3} &=& v_{\pi_{W^\pm,Z}}^{(1)} \pi_{W^\pm,Z} 
                          + v_{\pi_{\rho^{\pm,0}}}^{(1)} \pi_{\rho^{\pm,0}} 
  \,, \label{gbeigenstatea}\\
  \pi_{(2)}^{\pm,3} &=& v_{\pi_{W^\pm,Z}}^{(2)} \pi_{W^\pm,Z} 
                          + v_{\pi_{\rho^{\pm,0}}}^{(2)} \pi_{\rho^{\pm,0}} 
  \,, \label{gbeigenstate}
\end{eqnarray}
where 
\begin{equation} 
\begin{array}{cc} 
   v_{\pi_Z}^{(1)} = \frac{1}{\sqrt{2}} \left( 1 - \frac{(1-t^2) x^2}{4} + \cdots \right)  \,, 
 & v_{\pi_{\rho^0}}^{(1)} = \frac{1}{\sqrt{2}} \left( - 1 - \frac{(1-t^2) x^2}{4} + \cdots \right)  \,, 
\\ 
    v_{\pi_Z}^{(2)} = \frac{1}{\sqrt{2}} \left( 1 + \frac{(1-t^2) x^2}{4} + \cdots \right)  \,, 
 & v_{\pi_{\rho^0}}^{(2)} = \frac{1}{\sqrt{2}} \left( 1 - \frac{(1-t^2) x^2}{4} + \cdots \right)  \,, 
\end{array}  
\label{wavefunc:Pi}  
\end{equation}  
with $t=0$ for the wavefunctions of $\pi_{W^\pm,\rho^\pm}$.


\section{One-Loop Corrections to the Gauge Boson Self-Energies}

In order to compute the one-loop corrections to the $S$ and $T$ parameters, we
must evaluate the relevant contributions to the gauge-boson self-energies
\cite{Peskin:1992sw,Altarelli:1990zd}. Using the results of the previous section,
the gauge-sector interactions may be written in terms 
of the mass-eigenstate fields, yielding (order by order in $x$) the interactions
necessary for our calculations. The gauge-sector interactions, written
in the mass-eigenstate basis, are summarized in Appendix
\ref{interactions}, and the relevant diagrams are shown in figs. \ref{selfenergy} and \ref{selfenergyW}.  

We define the self-energy amplitudes for the SM gauge bosons as
\begin{equation}  
\int d^4x  \, e^{- ipx} \langle 0| {\rm T} \mathcal{A}_i^\mu(x) \mathcal{A}_j^\nu (0) | 0 \rangle 
= i g^{\mu\nu} \Pi_{\mathcal{A}_i\mathcal{A}_j}(p) + ( p^\mu p^\nu \, {\rm term} ) 
\,, 
\end{equation}
where $i$ and $j$ denote the species of the SM gauge bosons. 
In the present calculation, we choose the 't~Hooft-Feynman gauge $\xi=1$. 
The amplitudes are evaluated by using the Feynman integral formulae given in
Appendix~\ref{formula}; as described there, the formulae are derived using dimensional 
regularization and renormalized at the cutoff scale $\Lambda$ of the effective theory.

\subsection{Neutral Gauge Boson Self-Energies}
\label{sec:oneloopneutral}

\begin{figure}
\begin{center}
\input{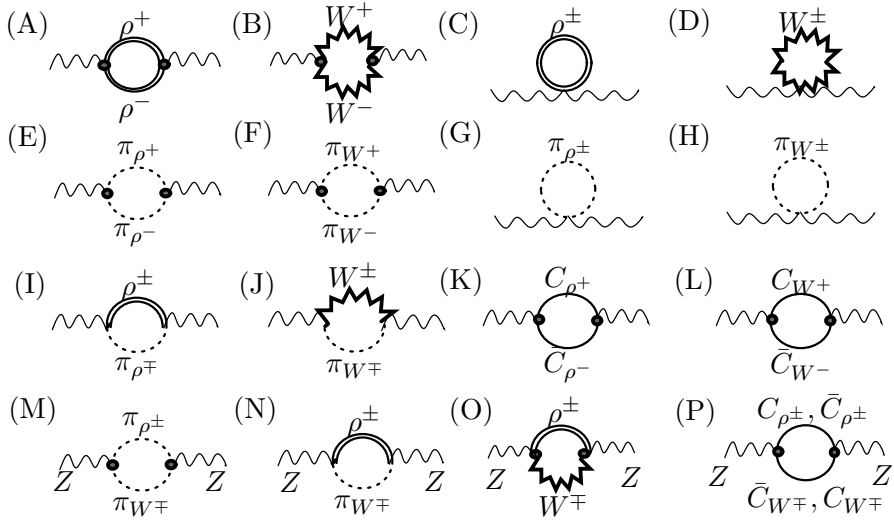}
\vspace{15pt}
\caption{One-loop diagrams for the $Z$ boson and photon self-energies $\Pi_{ZZ,ZA,AA}$ in the three-site model. Each external line in diagrams (A) -- (L) can be
either a photon or a $Z$ boson;  (M), (N), (O) and (P)
         apply only to the $Z$ boson. Expressions for the relevant vertices are given
         in Appendix \protect\ref{interactions}. Dots on vertices denote derivative couplings. As described in the text, the calculation is done in 't~Hooft-Feynman gauge. The $\pi_{W/Z,\rho}$ and $C_{W/Z,\rho}$ fields, respectively, denote the 't~Hooft-Feynman gauge unphysical Goldstone bosons and the ghost fields corresponding to  the electroweak and $\rho$ bosons.
}
\label{selfenergy}
\end{center}
\end{figure}

The values of the individual diagrams in Fig. \ref{selfenergy} are shown
in Appendix \ref{feynman-nc}. Putting these contributions together, we obtain the photon self-energy
\begin{equation} 
\Pi_{AA}(p^2) = \frac{e^2}{(4\pi)^2} \cdot p^2 
\Bigg[\left( 3 \log \frac{\Lambda^2}{M_W^2}  
+ 3 \log \frac{\Lambda^2}{M_{\rho^\pm}^2} \right) 
\Bigg] 
     \,,
\end{equation}
the $ZA$ mixing self-energy
\begin{eqnarray} 
     \Pi_{ZA}(p^2) &=& \frac{e^2}{(4\pi)^2 sc}\, 
 \Bigg[ 
  2 M_W^2 \log \frac{\Lambda^2}{M_W^2}
  +  (2c^2-1) M_Z^2 \log \frac{\Lambda^2}{M_{\rho^\pm}^2}  
 \nonumber \\ 
 && + p^2\Bigg( \frac{18c^2+1}{6} \log \frac{\Lambda^2}{M_W^2}    
     + \frac{3(2c^2-1)}{2} \log \frac{\Lambda^2}{M_{\rho^\pm}^2}
                \Bigg)  
 \Bigg] \,,
\end{eqnarray}
and the $Z$-boson self-energy
\begin{eqnarray} 
     \Pi_{ZZ}(p^2) &=& \frac{e^2}{(4\pi)^2 s^2c^2} 
\Bigg[ 
\left\{ 4c^2 - \frac{3}{2} \right\} M_W^2 \log\frac{\Lambda^2}{M_W^2} 
+  \left\{ \frac{56c^2-47}{8} M_W^2 + \frac{11}{8} M_Z^2 
- \frac{3}{8} M_{\rho^\pm}^2 \right\} 
\log\frac{\Lambda^2}{M_{\rho^\pm}^2} 
\nonumber \\  
&&  \hspace{60pt} 
+  p^2 \Bigg(
\left\{ 3c^4 + \frac{1}{3} c^2 - \frac{1}{12} \right\} 
\log \frac{\Lambda^2}{M_W^2}   
+ \left\{ 3c^4 - 3c^2 + \frac{17}{24} \right\} 
\log \frac{\Lambda^2}{M_{\rho^\pm}^2} 
\Bigg)
\Bigg] 
\,. \nonumber \\
\end{eqnarray}  
These expressions are correct to leading-log approximation, and to order $\alpha$; we neglect corrections ${\cal O}(\alpha x^2 M^2_W)$ or ${\cal O}(\alpha x^2 p^2)$, but keep those of order ${\cal O}(\alpha x^2 M^2_\rho)$ -- and must therefore also account for the difference\footnote{$M^2_{\rho^0}
= M^2_{\rho^+} + s^2 M^2_W/c^2 + \ldots$} between $M^2_{\rho^+}$
and $M^2_{\rho^0}$ in contributions proportional to $\alpha M^2_\rho$.

We note that these results for $\Pi_{ZA}$ and $\Pi_{ZZ}$ are not transverse. While in the case of the $Z$-boson, one expects a scalar contribution renormalizing the $Z$-boson mass, the $ZA$ mixing self-energy, properly defined, {\it must} be transverse by electromagnetic gauge-invariance.  
Therefore the calculation is not yet complete.  As is well-known, a gauge-invariant result is obtained only after inclusion of the appropriate pieces (the so-called ``pinch contributions") of one-loop vertex corrections and box diagrams \cite{Degrassi:1992ue,Degrassi:1992ff}.  In 't~Hooft-Feynman
gauge the only such contributions arise from  diagrams
containing triple-vector-boson vertices, as illustrated for the electroweak and $\rho$ gauge bosons
in Fig. \ref{pinchsm}.

\subsection{Charged Gauge Boson Self-Energies}
\label{sec:oneloopcharged}

\begin{figure}
\begin{center}    
\input{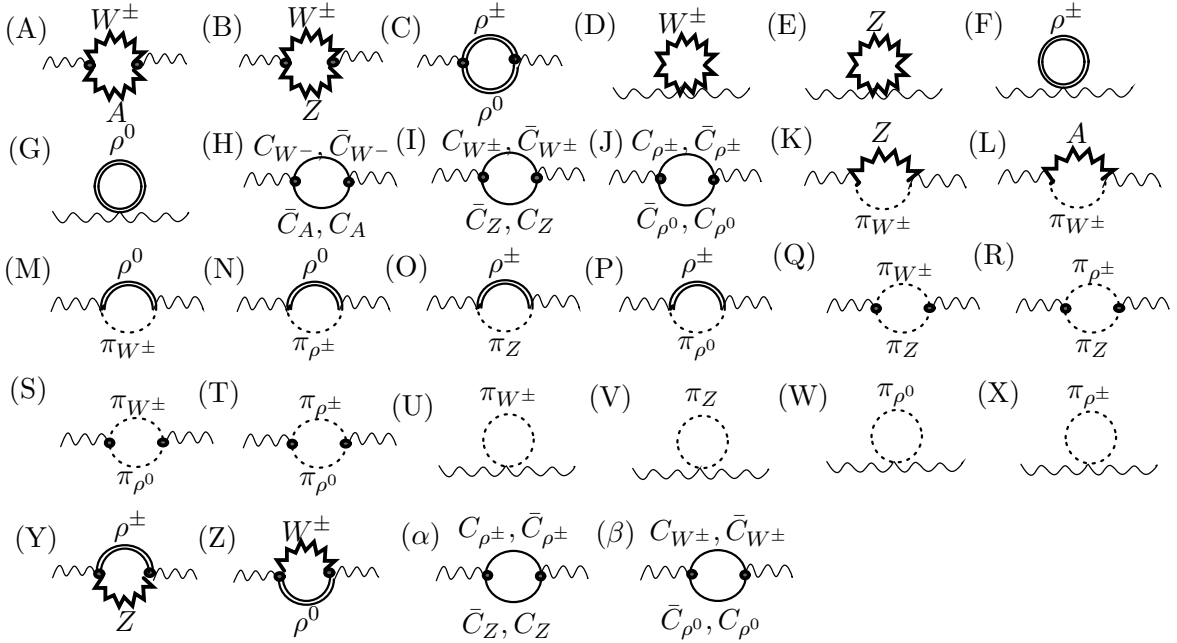}
\vspace{15pt}
\caption{ One-loop diagrams for the $W$ boson self-energy $\Pi_{WW}$ in the three-site model. The relevant vertices are listed given
         in Appendix \protect\ref{interactions}. Dots on vertices denote derivative couplings. As described in the text,
 the calculation is done in 't~Hooft-Feynman
          gauge. The $\pi_{W/Z,\rho}$ and $C_{W/Z,\rho}$ fields denote the 't~Hooft-Feynman
          gauge unphysical
          Goldstone bosons and ghost fields corresponding to  the electroweak and $\rho$
          bosons.
           }
\label{selfenergyW}
\end{center}

\end{figure}

The values of the individual diagrams in Fig. \ref{selfenergyW} are shown
in Appendix \ref{feynman-cc}. 
Putting these contributions together, and using the relation
$M^2_{\rho^0} \approx M^2_{\rho^\pm} + {s^2 \over c^2} M^2_W$, we obtain
\begin{eqnarray} 
     \Pi_{WW}(p^2) &=& \frac{e^2}{(4\pi)^2 s^2} 
\Bigg[ 
\left\{ \frac{13}{4}M_W^2 - \frac{3}{4} M_Z^2 \right\} 
\log\frac{\Lambda^2}{M_W^2} 
+  \left\{ \frac{12c^2+5}{8} M_W^2 + \frac{3}{8} M_Z^2 
- \frac{3}{8} M_{\rho^\pm}^2 \right\} 
\log\frac{\Lambda^2}{M_{\rho^\pm}^2} 
\nonumber \\  
&&  \hspace{60pt} 
+  p^2 \Bigg(
\frac{13}{4} \log \frac{\Lambda^2}{M_W^2}   
+ \frac{17}{24} \log \frac{\Lambda^2}{M_{\rho^\pm}^2} 
\Bigg)
\Bigg] 
\,. \nonumber \\
\label{eq:piwwnaive}
\end{eqnarray}  
Again, the complete result will include pinch contributions.

\section{Pinch Contributions}

\subsection{Pinch Contributions in the Standard Model}

\begin{figure}

\begin{center}
\unitlength 0.1in
\begin{picture}( 34.5600, 18.7200)( 24.6000,-32.4700)
%
\special{pn 8}%
\special{pa 2574 2452}%
\special{pa 3870 2452}%
\special{fp}%
%
\special{pn 20}%
\special{pa 3040 2466}%
\special{pa 2982 2554}%
\special{pa 3066 2554}%
\special{pa 3066 2644}%
\special{pa 3136 2612}%
\special{pa 3180 2710}%
\special{pa 3220 2630}%
\special{pa 3286 2702}%
\special{pa 3310 2602}%
\special{pa 3380 2650}%
\special{pa 3374 2546}%
\special{pa 3456 2546}%
\special{pa 3376 2474}%
\special{pa 3376 2474}%
\special{pa 3376 2474}%
\special{fp}%
%
\special{pn 20}%
\special{pa 3226 2650}%
\special{pa 3260 2674}%
\special{pa 3288 2700}%
\special{pa 3302 2724}%
\special{pa 3300 2748}%
\special{pa 3278 2772}%
\special{pa 3254 2798}%
\special{pa 3240 2826}%
\special{pa 3250 2854}%
\special{pa 3270 2884}%
\special{pa 3288 2912}%
\special{pa 3288 2942}%
\special{pa 3268 2968}%
\special{pa 3244 2994}%
\special{pa 3226 3020}%
\special{pa 3232 3046}%
\special{pa 3256 3072}%
\special{pa 3280 3098}%
\special{pa 3292 3122}%
\special{pa 3282 3148}%
\special{pa 3256 3174}%
\special{pa 3228 3196}%
\special{sp}%
\put(27.9100,-26.1300){\makebox(0,0){$\rho^+$}}%
\put(36.0500,-26.2000){\makebox(0,0){$\rho^-$}}%
\put(34.5800,-29.8200){\makebox(0,0){$Z,A$}}%
%
\special{pn 20}%
\special{pa 5242 2812}%
\special{pa 5278 2836}%
\special{pa 5304 2860}%
\special{pa 5312 2882}%
\special{pa 5294 2906}%
\special{pa 5268 2930}%
\special{pa 5256 2956}%
\special{pa 5270 2984}%
\special{pa 5292 3014}%
\special{pa 5298 3042}%
\special{pa 5278 3068}%
\special{pa 5252 3092}%
\special{pa 5242 3118}%
\special{pa 5262 3142}%
\special{pa 5288 3166}%
\special{pa 5300 3192}%
\special{pa 5286 3216}%
\special{pa 5256 3240}%
\special{pa 5244 3248}%
\special{sp}%
\put(48.6500,-26.3500){\makebox(0,0){$\rho^+$}}%
\put(55.7900,-26.4300){\makebox(0,0){$\rho^-$}}%
\put(54.5100,-30.4800){\makebox(0,0){$Z,A$}}%
%
\special{pn 20}%
\special{pa 5110 2640}%
\special{pa 5072 2572}%
\special{pa 5128 2572}%
\special{pa 5128 2502}%
\special{pa 5176 2526}%
\special{pa 5204 2452}%
\special{pa 5230 2514}%
\special{pa 5276 2458}%
\special{pa 5290 2536}%
\special{pa 5340 2498}%
\special{pa 5334 2578}%
\special{pa 5390 2578}%
\special{pa 5336 2634}%
\special{pa 5336 2634}%
\special{pa 5336 2634}%
\special{fp}%
%
\special{pn 20}%
\special{pa 5102 2640}%
\special{pa 5064 2710}%
\special{pa 5120 2710}%
\special{pa 5120 2780}%
\special{pa 5168 2756}%
\special{pa 5196 2830}%
\special{pa 5222 2768}%
\special{pa 5268 2824}%
\special{pa 5282 2746}%
\special{pa 5332 2784}%
\special{pa 5328 2704}%
\special{pa 5382 2704}%
\special{pa 5328 2646}%
\special{pa 5328 2646}%
\special{pa 5328 2646}%
\special{fp}%
%
\special{pn 8}%
\special{pa 4622 2452}%
\special{pa 5916 2452}%
\special{fp}%
%
\special{pn 20}%
\special{pa 3916 2790}%
\special{pa 4528 2790}%
\special{fp}%
\special{sh 1}%
\special{pa 4528 2790}%
\special{pa 4462 2770}%
\special{pa 4476 2790}%
\special{pa 4462 2810}%
\special{pa 4528 2790}%
\special{fp}%
\put(42.3300,-26.6500){\makebox(0,0){Pinch Part}}%
%
\special{pn 8}%
\special{pa 2558 1376}%
\special{pa 3854 1376}%
\special{fp}%
%
\special{pn 20}%
\special{pa 3024 1390}%
\special{pa 2968 1478}%
\special{pa 3050 1478}%
\special{pa 3050 1568}%
\special{pa 3120 1536}%
\special{pa 3164 1634}%
\special{pa 3204 1554}%
\special{pa 3270 1626}%
\special{pa 3294 1526}%
\special{pa 3366 1574}%
\special{pa 3358 1470}%
\special{pa 3440 1470}%
\special{pa 3362 1398}%
\special{pa 3362 1398}%
\special{pa 3362 1398}%
\special{fp}%
%
\special{pn 20}%
\special{pa 3210 1574}%
\special{pa 3244 1600}%
\special{pa 3272 1624}%
\special{pa 3288 1648}%
\special{pa 3284 1672}%
\special{pa 3262 1698}%
\special{pa 3236 1722}%
\special{pa 3224 1750}%
\special{pa 3234 1778}%
\special{pa 3254 1808}%
\special{pa 3272 1838}%
\special{pa 3274 1866}%
\special{pa 3254 1892}%
\special{pa 3228 1918}%
\special{pa 3212 1944}%
\special{pa 3218 1970}%
\special{pa 3242 1996}%
\special{pa 3266 2022}%
\special{pa 3276 2046}%
\special{pa 3266 2072}%
\special{pa 3240 2098}%
\special{pa 3214 2120}%
\special{sp}%
\put(27.7500,-15.3700){\makebox(0,0){$W^+$}}%
\put(35.9000,-15.4400){\makebox(0,0){$W^-$}}%
\put(35.0400,-19.0500){\makebox(0,0){$Z,A$}}%
%
\special{pn 20}%
\special{pa 5226 1736}%
\special{pa 5262 1760}%
\special{pa 5288 1784}%
\special{pa 5296 1806}%
\special{pa 5278 1830}%
\special{pa 5250 1854}%
\special{pa 5240 1880}%
\special{pa 5254 1908}%
\special{pa 5278 1938}%
\special{pa 5284 1966}%
\special{pa 5264 1992}%
\special{pa 5236 2016}%
\special{pa 5228 2042}%
\special{pa 5246 2066}%
\special{pa 5274 2090}%
\special{pa 5286 2116}%
\special{pa 5270 2140}%
\special{pa 5240 2164}%
\special{pa 5230 2172}%
\special{sp}%
\put(48.5000,-15.5900){\makebox(0,0){$W^+$}}%
\put(55.6300,-15.6600){\makebox(0,0){$W^-$}}%
\put(54.8900,-19.7200){\makebox(0,0){$Z,A$}}%
%
\special{pn 20}%
\special{pa 5094 1564}%
\special{pa 5056 1496}%
\special{pa 5112 1496}%
\special{pa 5112 1426}%
\special{pa 5160 1450}%
\special{pa 5188 1376}%
\special{pa 5214 1438}%
\special{pa 5260 1382}%
\special{pa 5276 1460}%
\special{pa 5324 1420}%
\special{pa 5320 1502}%
\special{pa 5376 1502}%
\special{pa 5320 1558}%
\special{pa 5320 1558}%
\special{pa 5320 1558}%
\special{fp}%
%
\special{pn 20}%
\special{pa 5086 1564}%
\special{pa 5048 1634}%
\special{pa 5104 1634}%
\special{pa 5104 1704}%
\special{pa 5152 1680}%
\special{pa 5180 1754}%
\special{pa 5206 1692}%
\special{pa 5252 1748}%
\special{pa 5268 1670}%
\special{pa 5316 1708}%
\special{pa 5312 1628}%
\special{pa 5368 1628}%
\special{pa 5314 1570}%
\special{pa 5314 1570}%
\special{pa 5314 1570}%
\special{fp}%
%
\special{pn 8}%
\special{pa 4606 1376}%
\special{pa 5900 1376}%
\special{fp}%
%
\special{pn 20}%
\special{pa 3900 1714}%
\special{pa 4512 1714}%
\special{fp}%
\special{sh 1}%
\special{pa 4512 1714}%
\special{pa 4446 1694}%
\special{pa 4460 1714}%
\special{pa 4446 1734}%
\special{pa 4512 1714}%
\special{fp}%
\put(42.1800,-15.8800){\makebox(0,0){Pinch Part}}%
\end{picture}%
\vspace{15pt}
\caption{Vertex correction diagrams constructed from the SM and $\rho$ gauge boson loops which 
         contribute to the pinch part of the self-energies \cite{Degrassi:1992ue,Degrassi:1992ff} for the
          $Z$ and $A$ bosons. The external fermion lines are arbitrary, and may represent
           fermions charged under any combination of $SU(2)_0$ or $SU(2)_1$.}
\label{pinchsm}
\end{center}

\end{figure}
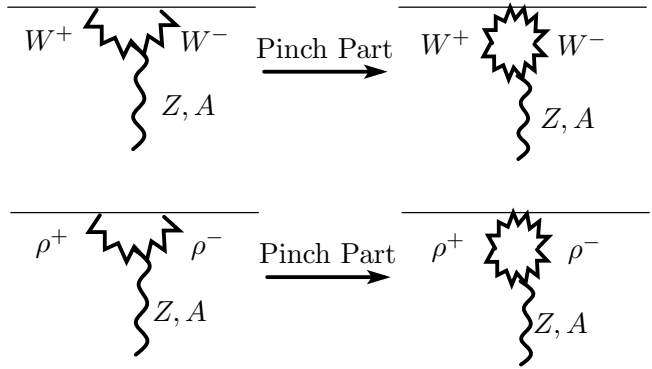

We begin by reviewing the results of the pinch contributions to the $Z$ and
$A$ self-energy functions in the standard model: see the first row of diagrams in 
Fig. \ref{pinchsm}. As discussed in detail in refs. \cite{Degrassi:1992ue,Degrassi:1992ff},
the pinch contributions arise (in 't-Hooft-Feynman gauge) from the momentum
dependence of the triple gauge-boson vertex, and yield effects
proportional to the following commutator of generators times coupling constants
\begin{equation}
[{g\over \sqrt{2}} T^+, {g\over \sqrt{2}} T^-] =  g^2 T_3~,
\label{eq:smcommutator}
\end{equation}
where the two terms in the commutator arise from the contraction of the momentum
with the two charged-$W$ vertices, and $g=e/s$ is the weak coupling constant. The
factors $g T^\pm/\sqrt{2}$ arise from the $W^\pm$ couplings to the external fermion line
currents $J^\mu_W = g J^\mu_\pm/\sqrt{2}$.

The pinch parts of the vertex corrections are proportional
to the value of $ g^2 J^\mu_3$ on the relevant external fermion line. 
They are therefore universal, {\it i.e.} they depend only on the charges of the external
fermion lines. As we will see, this property allows their effects to be incorporated
into the gauge-boson self-energy functions. To do so, we will
need to re-express this current in terms of the (tree-level) currents to which
the photon and $Z$ bosons couple. In the standard model, the relationship between
the neutral mass-eigenstate and gauge-eigenstate fields is given by
\begin{equation}
\begin{pmatrix}
Z^\mu \\
A^\mu
\end{pmatrix}
= 
\begin{pmatrix}
c&-s \\
s & c 
\end{pmatrix}
\begin{pmatrix}
W_3^\mu \\
B^\mu
\end{pmatrix}~,
\end{equation}
and, therefore, the relationship between the currents to which the mass eigenstates
couple to the symmetry currents of the theory is given by
\begin{equation}
\begin{pmatrix}
J_Z^\mu \\
J_A^\mu
\end{pmatrix}
= 
\begin{pmatrix}
c&-s \\
s & c 
\end{pmatrix}
\begin{pmatrix}
{e\over s}J_3^\mu \\
{e\over c}J_Y^\mu
\end{pmatrix}~.
\label{eq:smchargedcurrents}
\end{equation}
Note that, in this notation, the $J_A$ and $J_Z$ currents {\it include} 
the relevant tree-level couplings.
Inverting this relationship, we may solve for $ g^2 J^\mu_3$ in terms
of $J_{A,Z}$, and we find
\begin{equation}
 {e^2\over s^2} J^\mu_3 =  \left(e {c\over s} J^\mu_Z + e J^\mu_A \right)~.
\label{eq:smdecomp}
\end{equation}

Let us first consider the pinch contribution to $\Pi_{AA}$, where
\begin{equation}
\Pi^{\mu\nu}_{AA} = i\, g^{\mu\nu} \Pi_{AA} + \ldots
\label{eq:pidef}
\end{equation}
and we neglect the terms proportional to $p^\mu p^\nu$, which will not contribute to
the universal corrections. 
The pinch parts of photon vertex corrections to the 
four-fermion scattering amplitudes, then, are of the form
\begin{equation}
{\cal M}|^\gamma_{one-loop} \propto
{\cal A} \cdot {1\over p^2}\cdot  e\,G(p^2,M^2_W) \cdot (e{c\over s} {\cal Z}' + e {\cal A}')
+ (e{c\over s} {\cal Z}+ e {\cal A})\cdot e\,G(p^2, M^2_W)\cdot {1\over p^2}\cdot {\cal A}'~.
\label{eq:mgammasm}
\end{equation}
The factors in the first term, read from left to right, represent the coupling of the photon
to one external fermion line, the photon propagator, the $\gamma W^+ W^-$ coupling proportional
to $e$, the relevant one-loop pinch vertex correction  function $G(p^2,M^2_W)$ to the second 
photon-fermion vertex, 
and lastly the coupling of $ g^2 J^\mu_3$ to the other
external fermion line (with ``primed" charges). 
The second term arises from applying the pinch vertex correction to the first
fermion-vertex instead.
A contribution to the self-energy function $\Delta \Pi_{AA}$, on the other hand, generally gives
rise to a correction of the form
\begin{equation}
{\cal M}|^{AA}_{one-loop} \propto
{\cal A} \cdot {1\over p^2}\cdot \Delta \Pi_{AA}(p^2) \cdot {1\over p^2} \cdot {\cal A}'~.
\label{eq:maa}
\end{equation}
Thus Eqn. (\ref{eq:mgammasm}) may be viewed as yielding a contribution to $\Pi_{AA}$
\begin{equation}
\Delta \Pi_{AA}(p^2) = 2 e^2 p^2 G(p^2,M^2_W)~.
\end{equation}

Comparing this result for the standard model 
pinch contribution to that of \cite{Degrassi:1992ue,Degrassi:1992ff}, we find
\begin{equation}
\Delta \Pi^{SM}_{AA}(p^2) = 4 e^2 p^2 F_2(M_W,M_W;p^2)~,
\end{equation}
we see that $G(p^2,M^2_W) \equiv 2 F_2(M_W,M_W;p^2)$, where $F_2$
is defined in Appendix \ref{formula}. Note that the
function $G(p^2,M^2_W)$ will be the same in every standard model pinch contribution to the
self-energies of the $Z$ and $A$ bosons. In addition, since the loop-functions depends only
on the (universal) form of the triple gauge-boson vertex and the masses of
the gauge-bosons, by substituting the appropriate
gauge-bosons in the more general expression $F_2(M_A, M_B;p^2)$, 
we may use this result immediately to compute
the relevant loop-function in any pinch contribution in either the standard model or
the three-site model.

The one-loop contribution in Eqn. (\ref{eq:mgammasm}) also gives rise to contributions
proportional to the product of the photon and $Z$ charges of the external fermions, and
hence also corrects $\Pi_{ZA}(p^2)$. In general, a correction to $\Pi_{ZA}$ would
give rise to a contribution to the four-fermion amplitude of the form
\begin{equation}
{\cal M}\vert^{ZA}_{one-loop} \propto
{\cal A} \cdot {1\over p^2} \cdot \Delta \Pi_{ZA}(p^2) \cdot {1\over p^2-M^2_Z}\cdot
{\cal Z}' + 
{\cal Z} \cdot {1\over p^2-M^2_Z} \cdot \Delta \Pi_{ZA}(p^2) \cdot {1\over p^2}\cdot
{\cal A}'~.
\label{eq:mza}
\end{equation}
Hence, from Eqn. (\ref{eq:mgammasm}), we find a contribution
\begin{equation}
\Delta \Pi^\gamma_{AZ} = 2e^2 {c\over s}(p^2-M^2_Z) F_2(M_W,M_W;p^2)~.
\end{equation}

The pinch parts of the $Z$ vertex corrections to the four-fermion scattering
amplitudes are of the form
\begin{equation}
{\cal M}\vert^Z_{one-loop} \propto
{\cal Z} \cdot {1\over p^2-M^2_Z} \cdot e{c\over s} 2 F_2 \cdot (e{c\over s}{\cal Z}'+e{\cal A'})
+ (e{c\over s}{\cal Z} + e{\cal A}) \cdot e{c\over s} 2 F_2 \cdot {1\over p^2-M^2_Z} \cdot {\cal Z}'~,
\label{eq:mzsm}
\end{equation}
where we have abbreviated $F_2 = F_2(M_W,M_W;p^2)$ and the $e{c\over s}$ factor arises
from the $Z W^+ W^-$ vertex. We see that this gives rise to a correction to $\Pi_{ZA}$
\begin{equation}
\Delta \Pi^Z_{ZA} = 2 e^2 {c\over s} p^2 F_2(M_W,M_W;p^2)~,
\end{equation}
and, hence, the total pinch contribution to $\Delta \Pi_{AZ}$ is
\begin{equation}
\Delta \Pi^{SM}_{AZ}=\Delta\Pi^\gamma_{ZA} + \Delta\Pi^Z_{ZA} = 2e^2 {c\over s}(2p^2-M^2_Z) F_2(M_W,M_W;p^2)~,
\end{equation}
in agreement with \cite{Degrassi:1992ue,Degrassi:1992ff}.

Eqn. (\ref{eq:mzsm}) also makes a contribution proportional to the
product of the $Z$ charges of the external fermions. If we write
corrections to $\Pi_{ZZ}$ in the general form,
\begin{equation}
{\cal M}\vert^{ZZ}_{one-loop} \propto
{\cal Z} \cdot {1\over p^2-M^2_Z} \cdot \Delta \Pi_{ZZ}(p^2) \cdot {1\over p^2-M^2_Z} \cdot {\cal Z}'~,
\label{eq:mzz}
\end{equation}
then, Eqn. (\ref{eq:mzsm}) yields the pinch contribution
\begin{equation}
\Delta \Pi^{SM}_{ZZ} = 4 e^2 {c^2 \over s^2} (p^2-M^2_Z) F_2(M_W,M_W;p^2)~,
\end{equation}
in agreement with \cite{Degrassi:1992ue,Degrassi:1992ff}.

An analogous calculation, arising from $W$-boson vertex corrections 
and corresponding to the commutator of one charged
and one neutral current,  yields the corresponding pinch correction for the
$W$ boson propagator \cite{Degrassi:1992ue,Degrassi:1992ff}
\begin{equation}
\Delta \Pi^{SM}_{WW} = \frac{4e^2}{s^2} (p^2-M_W^2) 
\left[ c^2 F_2(M_Z,M_W;p^2) + s^2 F_2(0,M_W;p^2) \right]~,
\end{equation}
in which the two terms represent the contributions from vertex corrections
with  internal $ZW$ and $\gamma W$ states, respectively.

\subsection{Additional Pinch Contributions in the Three Site Model}
\label{subsec:pinch}

We next consider the pinch contributions in the three site model \cite{SekharChivukula:2006cg},
as illustrated in the second row of diagrams in Fig. \ref{pinchsm}.
In this model, up to corrections of order ${\cal O}(x^2)$, the mass-eigenstate charged gauge bosons
are related to the gauge-eigenstates through ({\it cf}. Eqn. (\ref{wavefunc:Wrhopm}))
\begin{equation}
\begin{pmatrix}
W_\mu^+ \\
\rho_\mu^+
\end{pmatrix} 
=
\begin{pmatrix}
1 & {x\over 2} \\
-{x\over 2} & 1
\end{pmatrix}
\begin{pmatrix}
L_\mu^+ \\
V_\mu^+
\end{pmatrix}~.
\end{equation}
In order to understand the form of the pinch contributions, we need to understand
the currents to which these gauge-bosons couple
\begin{equation}
\begin{pmatrix}
J^\mu_W \\
J^\mu_{\rho\pm}
\end{pmatrix}
=
\begin{pmatrix}
1 & {x\over 2} \\
-{x\over 2} & 1
\end{pmatrix}
\begin{pmatrix}
(1-x_1) {e\over \sqrt{2} s} J^\mu_\pm \\
{e\over \sqrt{2} sx} ({J^\mu_\pm}' + x_1 J^\mu_\pm)
\end{pmatrix} ~,
\end{equation}
where $J^\mu$ represents the current associated with fermions
primarily charged under $SU(2)_0$ delocalized by an amount $x_1$
-- the ordinary fermions -- and ${J^\mu}'$ represents any matter charged primarily
$SU(2)_1$ (see Eqn. (\ref{eq:delocop})). Here we approximate $g_0\approx e/s$
and $g_1 = e/sx$. From this, we find
\begin{equation}
J^\mu_{\rho\pm} = {e \over \sqrt{2} s x} {J^\mu_\pm}' - {xe \over 2\sqrt{2}s}(1-{2x_1\over x^2})J^\mu_\pm~,
\end{equation}
where we have neglected terms of order ${\cal O}(x_1 x^2) = {\cal O}(x^3)$.
Note that, as required by ideal delocalization, the ordinary fermions do not couple
to the charged-$\rho$ bosons when $x_1 = x^2/2$.

Comparing to Eqn. (\ref{eq:smcommutator}), we see that the pinch contributions 
arising from $\rho$-boson vertex corrections
in the three site model will be proportional to  
\begin{equation}
\left[{e\over\sqrt{2}s x} {T^+}' - {xe\over 2\sqrt{2}s}\left(1-{2x_1\over x^2}\right) T^+,
{e\over\sqrt{2}s x} {T^-}' - 
{xe\over 2\sqrt{2}s}\left(1-{2x_1\over x^2}\right) T^-\right]
= {e^2\over s^2} {T_3}' + {x^2 e^2\over 4 s^2} \left(1-{2x_1\over x^2}\right)^2\, T_3~.
\end{equation}
The second term above is proportional to $x^2$, and is therefore irrelevant
in what follows. 

The pinch contributions are therefore proportional to the value
of $e^2{J^\mu_3}'/s^2x^2$ on the relevant external fermion line. 
As in Eqn. (\ref{eq:smdecomp}), the key to understanding the pinch contributions 
is to determine the relationship between ${J^\mu_3}'$ and the currents
to which the neutral  mass-eigenstates couple. Diagonalizing the mass-squared
matrix, we find that the relationship between the neutral boson gauge- and mass-eigenstates 
is given by ({\it cf}. Eqn. (\ref{wavefunc:AZrho0}))
\begin{equation}
\begin{pmatrix}
Z^\mu \\
A^\mu \\
\rho^\mu
\end{pmatrix}
=
\begin{pmatrix}
c & -s & {{c^2-s^2}\over 2c} x \\
s & c & s x \\
-{x\over 2} & -{sx\over 2c} & 1 
\end{pmatrix}
\begin{pmatrix}
L^\mu_3 \\
B^\mu \\
V^\mu_3
\end{pmatrix}~,
\end{equation}
where the rotation matrix is orthogonal up to corrections of ${\cal O}(x^2)$.
The relationship between the currents to which the neutral mass-eigenstates
couple and the symmetry currents is given, therefore, by
\begin{equation}
\begin{pmatrix}
J^\mu_Z \\
J^\mu_A \\
J^\mu_\rho
\end{pmatrix}
=
\begin{pmatrix}
c & -s & {{c^2-s^2}\over 2c} x \\
s & c & s x \\
-{x\over 2} & -{sx\over 2c} & 1 
\end{pmatrix}
\begin{pmatrix}
(1-x_1){e\over s} J^\mu_3 \\
{e\over c} J^\mu_Y \\
{e\over sx}({J^\mu_3}'+x_1 J^\mu_3)
\end{pmatrix}~,
\label{eq:fermion-couplings}
\end{equation}
allowing for fermion delocalization.
Inverting the matrix, we find the relations
\begin{equation}
{e \over sx} ({J^\mu_3}' + x_1 J^\mu_3) = 
J^\mu_\rho + sx J^\mu_A + {{c^2-s^2}\over 2c} x J^\mu_Z~,
\label{eq:j3prime}
\end{equation}
and
\begin{equation}
{e \over s} (1-x_1) J^\mu_3 = c J^\mu_Z + s J^\mu_A -{x\over 2}J^\mu_\rho~.
\label{eq:j3}
\end{equation}
Noting that $x_1={\cal O}(x^2)$, and therefore neglecting it on the left hand side
of Eqn. (\ref{eq:j3}), we may rearrange these equations to find
\begin{equation}
{e^2 \over s^2 x^2} {J^\mu_3}' =
{e\over sx} \left(1-{x_1\over 2}\right)J^\mu_\rho + e\left(1-{x_1\over x^2}\right)J^\mu_A
+e {{c^2-s^2}\over 2 c s}\left(1-{2c^2 \over c^2-s^2}{x_1\over x^2}\right)J^\mu_Z~.
\label{eq:rhocouplings}
\end{equation}
This equation will allow us to extract the pinch contributions in the three site model -- note
that the operator $e^2 {J^\mu_3}'/s^2x^2$ has, counter-intuitively, relevant weak-size couplings
to the ordinary fermion currents $J^\mu_A$ and $J^\mu_Z$!

In analogy with our calculations in the standard model, we may immediately read off the
form of the $\rho$-boson vertex correction contributions to photon exchange
\begin{equation}
{\cal M}\vert^\gamma_{one-loop}
= {\cal A} \cdot {1\over p^2} \cdot 2 e \tilde{F_2} \cdot  \left[
e\left(1-{x_1\over x^2}\right){\cal A}' + e {{c^2-s^2}\over 2 c s}\left(1-{2c^2 \over c^2-s^2}{x_1\over x^2}\right)
{\cal Z}'\right]+({\cal A, Z} \leftrightarrow {\cal A}', {\cal Z}')~,
\label{eq:mrhoa}
\end{equation}
where the $2 e \tilde{F_2}$ includes both the $\gamma \rho^+ \rho^-$
coupling $e$, and the loop-function $2 \tilde{F_2} = 2 F_2(M_\rho,M_\rho;p^2)$. 
Similarly, the $\rho$-boson vertex corrections to $Z$ exchange may be written
\begin{align}
{\cal M}\vert^Z_{one-loop} = &
{\cal Z} \cdot {1\over p^2-M^2_Z} \cdot 2{e(c^2-s^2)\over 2cs}\tilde{F_2} \cdot 
\left[
e\left(1-{x_1\over x^2}\right){\cal A}' +e  {{c^2-s^2}\over 2 c s}\left(1-{2c^2 \over c^2-s^2}{x_1\over x^2}\right)
{\cal Z}'\right] \cr
& +({\cal A, Z} \leftrightarrow {\cal A}', {\cal Z}')~,
\label{eq:mrhoz}
\end{align}
where the $Z\rho^+ \rho^-$ coupling is proportional to $e(c^2-s^2)/2cs$ to
leading order (see Table \ref{3-point-gauge-couplings}  of Appendix \ref{sec:3pt}).

We may then compute the corresponding pinch contributions, by comparing Eqns. (\ref{eq:mrhoa})
and (\ref{eq:mrhoz}) to Eqns. (\ref{eq:maa}), (\ref{eq:mza}), and (\ref{eq:mzz}).
From this we obtain
\begin{align}
\Delta \Pi^{3-site}_{AA}(p^2) & = 4 e^2 \left(1-{x_1\over x^2}\right) p^2  F_2(M_\rho,M_\rho;p^2)~,
\label{eq:pinchi}\\
\Delta \Pi^{\gamma,3-site}_{ZA}(p^2) & = {e^2(c^2-s^2)\over c s} \left(1-{2c^2 \over c^2-s^2}{x_1\over x^2}\right)(p^2-M^2_Z)
F_2(M_\rho,M_\rho;p^2)~,
\label{eq:pinchii}\\
\Delta \Pi^{Z,3-site}_{ZA}(p^2) & = {e^2 (c^2-s^2)\over cs} \left(1-{x_1\over x^2}\right) p^2  F_2(M_\rho,M_\rho;p^2)~,
\label{eq:pinchiii}\\
\Delta \Pi^{3-site}_{ZZ}(p^2) & = {e^2 (c^2-s^2)^2 \over s^2 c^2} \left(1-{2c^2 \over c^2-s^2}{x_1\over x^2}\right)
(p^2-M^2_Z) F_2(M_\rho,M_\rho;p^2)~.
\label{eq:pinchiv}
\end{align}
Precisely analogous computations, arising from $W$-boson vertex corrections and
corresponding to the commutator $[J^\mu_{\rho\pm}, J^\nu_{\rho}]$, yields the pinch
correction for the $W$ boson propagator
\begin{equation}
\Delta \Pi^{3-site}_{WW}(p^2) = {e^2 \over s^2}\left(1-{2x_1 \over x^2}\right) (p^2-M^2_W) F_2(M_\rho,M_\rho;p^2)~.
\label{eq:pinchv}
\end{equation}
Note that, consistent with isospin symmetry, $\Delta \Pi_{ZZ}$ in Eqn. 
(\ref{eq:pinchiv}) reduces to $\Delta\Pi_{WW}$ in Eqn. (\ref{eq:pinchv})
in the limit $c\to 1$ with $e/s$ held fixed.

Finally to the order in which we work, expanding $F_2(M_\rho,M_\rho;p^2)$ for
$p^2\simeq M^2_{W,Z} \ll M^2_\rho$, we see that we may approximate
$F_2(M_\rho,M_\rho;p^2) \approx F_2(M_\rho,M_\rho;0)$, and these pinch contributions
only affect the values of the self-energies at zero momentum.

As described here, the ``pinch" contributions are determined entirely by the gauge symmetry
and fermion coupling structure(s) of the theory. Examining Eqn. (\ref{eq:fermion-couplings})
however, one notes that the couplings of the fermions (with current $J^\mu$)
to the $\rho$ are suppressed by $x$. The vertex diagrams illustrated in the lower
row of Fig. \ref{pinchsm},
therefore, do not contribute in the case of ordinary fermions on the external lines! Diagrammatically,
as shown explicitly in Appendix \ref{appendix:mixing},  for ordinary fermions the 
contributions in Eqns. (\ref{eq:pinchi}) -- (\ref{eq:pinchv}) may be shown to arise instead  
from $\gamma-\rho$, $Z-\rho$, and $W-\rho$ mixing corrections to the four-fermion scattering
amplitudes.

\section{Fermion Delocalization Contributions}

\begin{figure}

\begin{center} 
\input{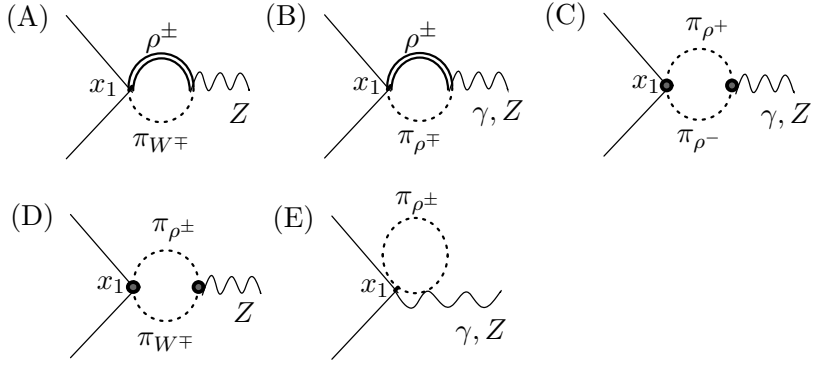}

\end{center}
\caption{One-loop vertex contributions to neutral-current
processes arising  from 
the delocalization operator $\mathcal{L}_f'$. 
As described in the appendix, since they are universal -- {\it i.e.} proportional
to the charges of the external fermions --  these
contributions can be incorporated into the neutral gauge boson self-energy 
functions $\Pi_{ZA,ZZ}$.
Eots on vertices denote derivative couplings.
The delocalization operator is proportional to 
$x_1={\cal O}(x^2)$, and therefore only the contributions proportional
to $M^2_\rho$, which are illustrated here,  contribute to this order. There
are analogous contributions to charged-current processes, which result
in corrections to $\Pi_{WW}$.}
\label{x1-reno-graphs} 
\end{figure}

Consider the fermion delocalization operator of Eqn. (\ref{eq:delocop})
\begin{equation}
{\cal L'}_f = - x_1\cdot \bar{\psi}_L (i\slashiii{D} \Sigma_{(1)} \Sigma^\dagger_{(1)} ) \psi_L~,
\nonumber
\end{equation}
where $D_\mu \Sigma_{(1)} = \partial_\mu \Sigma_{(1)}
- i g_0 L_\mu \Sigma_{(1)} + i g_1 \Sigma_{(1)} V_\mu$. 
The link variable $\Sigma_{(1)}$ is expanded as 
\begin{equation} 
\Sigma_{(1)} = e^{i 2\pi_{(1)}/f} 
= 1 + 2i \frac{\pi_{(1)}}{f} - \frac{2\pi_{(1)}^2}{f^2} + \cdots 
\,. 
\end{equation}
  From this, we see that the delocalization operator of Eqn. (\ref{eq:delocop}) 
is expressed as 
\begin{eqnarray} 
\mathcal{L}_f' &=& 
- x_1 g_1 \frac{2}{f} \epsilon^{abc} \pi_{(1)}^a V_\mu^b J_L^{\mu c} 
- x_1 \frac{2}{f^2}  \epsilon^{abc} \pi_{(1)}^a \partial_\mu \pi_{(1)}^b J_L^{\mu c} 
\nonumber \\ 
&& 
+ x_1 g_1 \frac{2}{f^2} \epsilon^{eab} \epsilon^{ecd} V_\mu^a \pi_{(1)}^b \pi_{(1)}^c J_L^{\mu d} 
+ \cdots 
\,, \label{delocal-op:iso}
\end{eqnarray}
where $J^{\mu a}_L = \bar{\psi}_L \gamma^\mu T^a \psi_L$. 
In terms of the mass- and charged-eigenstate fields, to leading order in $x$, we find
\begin{eqnarray} 
  \mathcal{L}_{f}'|_{NC} &=& 
- i \frac{e^2}{s^2M_{\rho^\pm}} 
\left( \frac{x_1}{x^2} \right) J^{\mu 3}_L \Bigg[ 
\rho_\mu^+ \pi_{W^-} - \rho_\mu^+ \pi_{\rho^-} 
\Bigg] + {\rm h.c.} 
\nonumber \\ 
&& 
- i \frac{e^2}{2s^2 M_{\rho^\pm}^2} 
\left( \frac{x_1}{x^2} \right) J^{\mu 3}_L \Bigg[ 
\left( \pi_{W^+} \stackrel{\leftrightarrow}{\partial}_\mu \pi_{W^-} \right) 
+ \left( \pi_{\rho^+} \stackrel{\leftrightarrow}{\partial}_\mu \pi_{\rho^-} \right) 
\nonumber \\ 
&& \hspace{100pt} 
- \left( \pi_{W^+} \stackrel{\leftrightarrow}{\partial}_\mu \pi_{\rho^-} \right) 
- \left( \pi_{\rho^-} \stackrel{\leftrightarrow}{\partial}_\mu \pi_{W^+} \right) 
\Bigg] 
\nonumber \\ 
&& 
- \frac{e^3}{s^2 M_{\rho^\pm}^2} 
\left( \frac{x_1}{x^2} \right) J^{\mu 3}_L 
\left( A_\mu + \frac{c^2-s^2}{2sc} Z_\mu \right) 
\left(  \pi_{W^+}\pi_{W^-} + \pi_{\rho^+}\pi_{\rho^-} \right) 
+ \cdots 
\,, \label{delocal-op:me}
\end{eqnarray}
where we have used $f \approx \sqrt{2} M_{\rho^\pm}/g_1$ and $g_0 \approx e/s$ 
which are valid only to leading order in $x$. 

As illustrated in Fig. \ref{x1-reno-graphs}, these couplings in the delocalization
operator give rise to vertex-corrections to neutral- and charged-current four-fermion
processes.\footnote{As described in \protect\cite{newpaper}, these contributions
correspond to renormalizations of the parameter $x_1$. As noted previously (see
footnote 1), there is no explanation for the amount of delocalization in this
low-energy effective theory.}  These diagrams are evaluated in Appendix \ref{appendix:delocal},
and their effects are summarized by the effective interactions in Eqn. (\ref{eff-op}). 
Using  the relation ({\it cf}. Eqn. (\ref{eq:smchargedcurrents}))
\begin{equation} 
   \frac{e}{s} J^\mu_{3 \, L} = (c J_Z^\mu + s J_A^\mu) 
   \,, \label{J3-JA-JZ}
\end{equation} 
we may rewrite Eqn. (\ref{eff-op}) in terms of the currents $J_A^\mu$ and 
$J_Z^\mu$, 
\begin{equation} 
  \mathcal{L}_{\rm eff} = 
  \left\{ G_1 J_Z^\mu + \frac{s}{c} G_1 J_A^\mu \right\}Z_\mu 
+ \left\{ sc G_2 J_Z^\mu + s^2G_2 J_A^\mu \right\}A_\mu
\,, 
\end{equation}
where we have abbreviated $G_{1,2}=G_{1,2}(M_{\rho^\pm}^2; x_1)$, and
these functions are defined in Eqns. (\ref{G1}) and (\ref{G2}).
Note that these vertex corrections are {\it universal} -- {\it i.e.} proportional
to the charges of the external fermions. As in the case of the ``pinch" contributions,
their effects may be incorporated into corrections of the gauge-boson self-energy
functions.

The contribution of these corrections to four-fermion
amplitudes mediated by the photon ($\mathcal{M}_\gamma^{one-loop}$) 
and $Z$ boson ($\mathcal{M}_Z^{one-loop}$) exchange may then
be written
\begin{eqnarray} 
\mathcal{M}_\gamma^{one-loop} 
&=& \left\{ sc G_2 \mathcal{Z} + (1+s^2G_2) \mathcal{A} \right\} 
\frac{1}{p^2} \left\{ sc G_2 \mathcal{Z}' + (1+s^2G_2) \mathcal{A}' \right\} 
\,, \label{M-gamma} \\ 
\mathcal{M}_Z^{one-loop} 
&=& \left\{ (1+G_1)\mathcal{Z} + \frac{s}{c} G_1 \mathcal{A} \right\} 
\frac{1}{p^2-M_Z^2} \left\{ (1+G_1)\mathcal{Z}' + \frac{s}{c} G_1 \mathcal{A}' \right\} 
\, \label{M-Z},
\end{eqnarray}
where, as before,  $\mathcal{A}(')$ and $\mathcal{Z}(')$ are photon- and $Z$-charges
carried by the external fermions. 
From the forms of Eqns.(\ref{M-gamma}) and (\ref{M-Z}), 
we can read off the corrections to the self-energy functions, 
\begin{eqnarray} 
  \Delta \Pi_{AA}(p^2) 
&=& 2s^2 G_2 \cdot p^2 
\,, \label{AA-pinchpart:x1}\\ 
\Delta \Pi_{ZA}^\gamma(p^2) 
&=& sc G_2 \cdot (p^2-M_Z^2)  
\,, \\ 
\Delta \Pi_{ZA}^Z(p^2) 
&=& \frac{s}{c} G_1 \cdot p^2  
\,, \\ 
\Delta \Pi_{ZZ}(p^2) 
&=& 2 G_1 \cdot (p^2-M_Z^2)  
\,.  \label{ZZ-pinchpart:x1}
\end{eqnarray}
By an analogous computation, or by noting the isospin symmetry relation, 
$\Delta\Pi_{WW}=\Delta\Pi_{ZZ}|_{c\to 1}$ with 
$(e/s)$ fixed, we can also easily read off the 
corresponding correction to 
the $W$ boson self-energy function.

Substituting Eqns.(\ref{G1}) and (\ref{G2}) 
into Eqns.(\ref{AA-pinchpart:x1})-(\ref{ZZ-pinchpart:x1}) 
we find
\begin{eqnarray} 
  \Delta \Pi^{delocal}_{AA}(p^2) 
&=& \frac{4e^2}{(4\pi)^2}  
\left( \frac{x_1}{x^2} \right) p^2 \log\frac{\Lambda^2}{M_{\rho^\pm}^2} 
\,, \\ 
\Delta \Pi^{delocal}_{ZA}(p^2) 
&=& \frac{e^2}{(4\pi)^2sc} 
\left( \frac{x_1}{x^2} \right) 
\Bigg[ \left(4c^2-\frac{1}{4}\right)p^2 -2M_W^2 \Bigg]
\log\frac{\Lambda^2}{M_{\rho^\pm}^2} 
\,, \\ 
\Delta \Pi^{delocal}_{ZZ}(p^2) 
&=& 
\frac{e^2}{(4\pi)^2s^2} \left( 4c^2 - \frac{1}{2} \right) 
(p^2-M_Z^2) 
\left( \frac{x_1}{x^2} \right) \log\frac{\Lambda^2}{M_{\rho^\pm}^2} 
\,, \\
\Delta \Pi^{delocal}_{WW}(p^2) 
&=& 
\frac{7e^2}{2(4\pi)^2s^2} 
(p^2-M_W^2) 
\left( \frac{x_1}{x^2} \right) \log\frac{\Lambda^2}{M_{\rho^\pm}^2} 
\,.
\end{eqnarray} 

\section{Total Gauge Boson Self-Energies}

\subsection{Neutral Gauge Bosons}

Including the standard model and three-site ``pinch corrections", and the corrections
arising from the fermion delocalization operator, we find the neutral gauge
boson self-energies
\begin{eqnarray} 
\overline{\Pi}_{AA}(p^2) &=& \Pi_{AA}(p^2) + 4 e^2 p^2 
\left\{ F_2(M_W,M_W; p^2) + F_2(M_{\rho^\pm},M_{\rho^\pm}; p^2) \right\} 
                        \,, \\ 
\overline{\Pi}_{ZA}(p^2) &=& \Pi_{ZA}(p^2) + \frac{2 e^2c}{s} ( 2 p^2 - M_Z^2 ) F_2(M_W,M_W; p^2)  
\nonumber \\ 
&& + \frac{e^2(c^2-s^2)}{sc} 
\Bigg( p^2 \left\{ 2 + \frac{3}{4(c^2-s^2)} \left(\frac{x_1}{x^2}\right) 
\right\} - M_Z^2 \Bigg) F_2(M_{\rho^\pm},M_{\rho^\pm}; p^2) 
                        \,, \\
\overline{\Pi}_{ZZ}(p) &=& \Pi_{ZZ}(p^2) + \frac{4 e^2 c^2}{s^2} (p^2 - M_Z^2)
F_2(M_W,M_W; p^2)
\nonumber \\ 
&& + \frac{e^2(c^2-s^2)^2}{s^2c^2} \left\{ 1 + \frac{3c^2}{2(c^2-s^2)^2} 
\left(\frac{x_1}{x^2}\right) \right\} (p^2-M_Z^2)
F_2(M_{\rho^\pm},M_{\rho^\pm}; p^2) 
                        \,, 
\end{eqnarray}
where the $\Pi_{AA, ZA, ZZ}$ were computed in Section \ref{sec:oneloopneutral},
and the  function $F_2$ is defined in Appendix~\ref{formula}. 
Evaluating and simplifying, we have 
\begin{eqnarray} 
\overline{\Pi}_{AA}(p^2) 
&=& \frac{e^2}{(4\pi)^2} p^2 \Bigg[ 7 \log\frac{M_{\rho^\pm}^2}{M_W^2}   
+ 14 \log \frac{\Lambda^2}{M_{\rho^\pm}^2}\Bigg] 
\,,\label{inv-AA} \\ 
\overline{\Pi}_{ZA}(p^2) 
&=& \frac{e^2}{(4\pi)^2 s c} p^2 \Bigg[ 
\left\{ 7c^2+\frac{1}{6} \right\} \log\frac{M_{\rho^\pm}^2}{M_W^2} 
+ \left\{ 14c^2-\frac{10}{3} + \frac{3}{4} \left( \frac{x_1}{x^2} \right) 
\right\} \log \frac{\Lambda^2}{M_{\rho^\pm}^2} 
\Bigg]
\,, \label{inv-ZA}\\ 
\overline{\Pi}_{ZZ} (p^2)
&=& \frac{e^2}{(4\pi)^2 s^2c^2} \Bigg[ 
- \frac{3}{2} M_W^2 \log\frac{M_{\rho^\pm}^2}{M_W^2} 
\nonumber \\ 
&& \hspace{60pt} 
+ \Bigg( \left\{ 3c^2 - \frac{27}{8} 
- \frac{3}{2} \left( \frac{x_1}{x^2} \right) \right\} M_W^2 
+ \frac{3}{8} M_Z^2 - \frac{3}{8} M_{\rho^\pm}^2 \Bigg) 
\log \frac{\Lambda^2}{M_{\rho^\pm}^2} 
\nonumber \\ 
&& \hspace{60pt}
+ p^2 \Bigg( \left\{ 7c^4 + \frac{1}{3} c^2 - \frac{1}{12} \right\} 
\log\frac{M_{\rho^\pm}^2}{M_W^2} 
\nonumber \\ 
&& \hspace{60pt} 
+ \left\{ 14 c^4 - \frac{20}{3} c^2 + \frac{39}{24} 
+ \frac{3}{2}c^2 \left( \frac{x_1}{x^2} \right) \right\} 
\log \frac{\Lambda^2}{M_{\rho^\pm}^2} 
\Bigg) \Bigg] 
\,. \label{inv-ZZ} 
\end{eqnarray}
up to $\mathcal{O}(\alpha x^2 p^2 )$ or $\mathcal{O}(\alpha x^2 M_W^2 )$. 
Note that the modified $AA$ and  $ZA$ self-energies are purely transverse. The $ZZ$
self-energy, $\overline{\Pi}_{ZZ}(p^2)$, represents, in part,  a renormalization of the
electroweak symmetry breaking scale\footnote{That is, a renormalization of 
 the electroweak $F$-constant, 
equal to the vacuum expectation value of the Higgs in the standard model}
similar to the corresponding one-loop renormalization proportional to the Higgs boson mass-squared
in the standard model \cite{Appelquist:1980ae}.

\subsection{Charged Gauge Boson}

Including the standard model and three-site ``pinch corrections", and the corrections
arising from the fermion delocalization operator, we find the charged gauge
boson self-energy
\begin{eqnarray}
\overline{\Pi}_{WW}(p^2) 
 &=& \Pi_{WW}(p^2) + \frac{4e^2}{s^2} (p^2-M_W^2) 
\left[ c^2 F_2(M_Z,M_W;p^2) + s^2 F_2(0,M_W;p^2) \right]  
\nonumber \\                     
&& + \frac{e^2}{s^2} \left\{ 1 +\frac{3}{2} \left( \frac{x_1}{x^2} \right) 
\right\} (p^2 -M_W^2) F_2(M_{\rho^0},M_{\rho^\pm} ;p^2) 
\,.,
\end{eqnarray}
where $\Pi_{WW}$ is computed in Section \ref{sec:oneloopcharged},
and the  function $F_2$ is defined in Appendix~\ref{formula}. 
Evaluating and simplifying, we find
\begin{eqnarray} 
\overline{\Pi}_{WW}(p^2) &=& \frac{e^2}{(4\pi)^2 s^2} \Bigg[ 
- \frac{3}{4} (M_W^2 + M_Z^2) \log\frac{M_{\rho^\pm}^2}{M_W^2} 
\nonumber \\ 
&& \hspace{50pt} 
+ \left\{ \frac{12c^2-9}{8} M_W^2 - \frac{3}{8} M_Z^2 
            - \frac{3}{8} M^2_{\rho^\pm} 
           - \frac{3}{2} \left( \frac{x_1}{x^2} \right) M_W^2  \right\} 
     \log\frac{\Lambda^2}{M_{\rho^\pm}^2} 
\nonumber \\ 
&& \hspace{50pt} 
+ p^2 \Bigg( \frac{29}{4} \log \frac{M_{\rho^\pm}^2}{M_W^2} 
+ \left \{ \frac{215}{24} + \frac{3}{2} \left( \frac{x_1}{x^2} \right) \right\} \log \frac{\Lambda^2}{M_{\rho^\pm}^2} 
\Bigg)
\Bigg] 
\,.  \label{inv-WW}         
\end{eqnarray}
Note that, in the custodial isospin symmetric limit $\sin^2\theta \to 0$ 
({\it i.e.} $c^2 \to 1$ and $M^2_Z \to M^2_W$), Eqn. (\ref{inv-ZZ}) reduces to Eqn. (\ref{inv-WW}).


\section{Precision Electroweak Corrections}

\subsection{The $S$ Parameter and Counterterms} 

The neutral gauge boson self-energies contribute to the $S$ parameter as \cite{Peskin:1992sw}
\begin{equation} 
\frac{\alpha S_{\rm 1-loop}}{4s^2c^2} 
            = \overline{\Pi}_{ZZ}'(0) - \overline{\Pi}_{AA}'(0)
                   - \frac{c^2- s^2}{sc} \overline{\Pi}_{ZA}'(0)
                                      \,, \label{s}
\end{equation} 
By using Eqs.(\ref{inv-AA})-(\ref{inv-ZZ}), 
the leading correction to the $S$ parameter is evaluated in the limit $M_{\rho^\pm,0}^2 \gg M_W^2$ as 
\begin{eqnarray} 
\alpha S_{\rm 1-loop} = \frac{\alpha}{12 \pi} \log\frac{M_\rho^2}{M_W^2} 
- \frac{41\alpha }{24\pi} \log\frac{\Lambda^2}{M_\rho^2}
+ \frac{3\alpha}{4\pi} \left( \frac{x_1}{x^2} \right) 
\log\frac{\Lambda^2}{M_\rho^2}
    \,. 
    \label{sone}
\end{eqnarray}
Note that the first term arises from ``scaling" between $M_W$ and $M_\rho$ -- 
and has a coefficient precisely equal to the leading-log contribution from a heavy Higgs boson\cite{Peskin:1992sw}
\begin{equation}
\alpha S_{Higgs} = \frac{\alpha}{12 \pi} \log\left(\frac{M^2_{H}}{M^2_W}\right)~.
\label{shref}
\end{equation}
This allows us to match our calculation to the phenomenological
extractions of $S$ which depend on a reference standard model Higgs-boson mass.

The dependence on the renormalization scale (here taken to be the cutoff
$\Lambda$ of the effective theory) is cancelled by the scale-dependence
of the appropriate counterms \cite{Perelstein:2004sc}.
The $ \mathcal{O}(p^4)$ counterterms relevant to  $S_{\rm 1-loop}$ are given by  
\begin{equation} 
    \mathcal{L}_{(4)} = c_1 g_1g_2 {\rm Tr}[ V_{\mu\nu} \Sigma_{(2)} B^{\mu\nu} \frac{\sigma_3}{2} \Sigma_{(2)}^\dagger ]
 + c_2 g_1g_0 {\rm Tr}[ L_{\mu\nu} \Sigma_{(1)} V^{\mu\nu} \Sigma_{(1)}^\dagger ]
 \,. \label{counter} 
\end{equation}
By using Eqn. (\ref{l3})-(\ref{b}), these may be written in terms of the mass eigenstate
fields as
\begin{eqnarray} 
\mathcal{L}_{(4)}|_{Z,A}^{\rm quad} &=&  \frac{i}{2} \delta_{ZZ} (Z_{\mu} D^{\mu\nu} Z_{\nu})  
                                 + i \delta_{ZA} (Z_{\mu} D^{\mu\nu} A_{\nu} )
                                 + \frac{i}{2} \delta_{AA} (A_{\mu} D^{\mu\nu} A_\nu)   
\,, 
\end{eqnarray}
where $D^{\mu\nu}=-\partial^2 g^{\mu\nu} + \partial^\mu \partial^\nu$ and 
\begin{eqnarray} 
       \delta_{ZZ} &=& \frac{e^2(c^2 - s^2)}{ c^2 s^2} \Bigg[ -  s^2 c_1 + c^2 c_2 \Bigg]  
       \,, \\ 
       \delta_{ZA} &=& \frac{e^2}{2s c} \Bigg[ (c^2 - 3s^2 )c_1 + (3c^2 - s^2)c_2 \Bigg] 
       \,, \\
       \delta_{AA} &=& 2 e^2 \Bigg[ c_1 + c_2 \Bigg] 
       \,. 
\end{eqnarray}
From this, applying Eqn. (\ref{s}), we find the contribution to $S$
\begin{equation}  
   \delta S_{\rm 1-loop} = - 8 \pi (c_1 + c_2) 
\,. 
\label{scounter}
\end{equation}

Adjusting for the reference Higgs mass, using Eqn. (\ref{shref}) and adding the
contribution from the counterterms in Eqn. (\ref{scounter}), we arrive at our
final result (Eqn. (\ref{result-sparam}))
\begin{eqnarray}
\alpha S_{3-site} & = & \left[\frac{4 s^2 M^2_W}{M^2_\rho}
\left(1- \frac{x_1 M^2_\rho}{2 M^2_W}\right)\right]_{\mu=\Lambda} + \frac{\alpha}{12 \pi} \log{\frac{M^2_\rho}{M^2_{Href}}}\nonumber \\
& - & \frac{41 \alpha}{24 \pi} \log{\frac{\Lambda^2}{M^2_{\rho}}}
+ \frac{3\alpha}{8\pi} \left( \frac{x_1M_\rho^2}{2M_W^2} \right) 
\log\frac{\Lambda^2}{M_{\rho}^2} 
\nonumber \\ 
&& -8 \pi \alpha (c_1(\Lambda) + c_2(\Lambda))~, \nonumber
\end{eqnarray}
where the tree-level expression
and the counterterms are now understood to be evaluated at scale $\Lambda$.

\subsection{ $\alpha T$ and a Counterterm} 

 The $T$ parameter~\cite{Peskin:1992sw} is expressed in terms of the $W$ and $Z$ boson self-energies as 
\begin{equation} 
    \alpha T_{1-loop} = \frac{\overline{\Pi}_{WW}(0)}{M_W^2} 
- \frac{\overline{\Pi}_{ZZ}(0)}{M_Z^2}
    \,. 
\end{equation} 
Noting 
\begin{equation}
   \frac{M_{\rho^\pm}^2}{M_W^2} - \frac{M_{\rho^\pm}^2}{c^2M_Z^2} 
= \frac{s^2(4c^2-1)}{c^2} + \mathcal{O}(x^2) 
\,, 
\end{equation}
 from Eqns.(\ref{inv-WW}) and (\ref{inv-ZZ}), we have 
 \begin{equation}  
   \alpha T_{1-loop} = - \frac{3 \alpha}{16 \pi c^2} \log \frac{M_\rho^2}{M_W^2} 
              - \frac{3 \alpha}{32 \pi c^2} \log \frac{\Lambda^2}{M_\rho^2} 
              \,.\label{T} 
\end{equation} 
Note that, as in the case of the $S$-parameter, the first term arises from ``scaling"
between $M_W$ and $M_\rho$ -- and has precisely the same
form as  the leading-log contribution from a heavy Higgs
boson \cite{Peskin:1992sw}
\begin{equation}
\alpha T_{Higgs} = - \frac{3 \alpha}{16 \pi c^2} \log \frac{M_H^2}{M_W^2}~.
\label{t-sm}
\end{equation}
This allows us to match our calculation to the phemenological extractions
of $T$ which depend on a reference standard model Higgs-boson mass.

The dependence on the renormalization scale (here taken to be the cutoff,
$\Lambda$, of the effective theory) is cancelled by the scale-dependence
of the appropriate counterterm. The ${\cal O}(p^4)$ counterterm\footnote{Note that
the tree-level value of $\alpha T$ is  $\mathcal{O}(M_W^4/M_\rho^4)$ 
in Higgsless models \protect \cite{SekharChivukula:2004mu},  and therefore this
counterterm is formally of ${\cal O}(p^4)$. This is manifest in Eqn. (\protect\ref{count0}) by the fact that
the counterterm is proportional to $g^2_2$. } relevant
to $T_{1-loop}$ is
\begin{equation} 
   \delta \mathcal{L} = c_0 g^2_2 f^2 \Bigg( {\rm tr}[D_\mu \Sigma_{(2)} \frac{\tau_3}{2} \Sigma_{(2)}^\dagger] \Bigg)^2 = \frac{4\pi \alpha\, c_0}{c^2}  f^2 \Bigg( {\rm tr}[D_\mu \Sigma_{(2)} \frac{\tau_3}{2} \Sigma_{(2)}^\dagger] \Bigg)^2 
   \,. \label{count0}
\end{equation}
Using Eqns. (\ref{mzee}), (\ref{v3}) and (\ref{b}), in unitary gauge
we read off a correction to the $Z$ boson mass (but {\it not} the $W$ boson mass)
from Eqn. (\ref{count0}): 
\begin{equation} 
   \delta M^2_Z = -  \frac{4\pi \alpha\,c_0}{c^2}\, M^2_Z
\,, 
\end{equation}
which leads to a contribution to $\alpha T$: 
\begin{equation} 
  [\delta \left( \alpha T \right)]_{\rm 1-loop} =\frac{4\pi\alpha\, c_0(\Lambda)}{c^2}
   \,. 
\end{equation}

Adjusting for the reference Higgs mass, $M_{Href}$, using Eqn. (\ref{t-sm}) and adding the
contribution from the counterterm in Eqn. (\ref{count0}), we then arrive at the final
result quoted in Eqn. (\ref{result-t}):
 \begin{equation} 
      \alpha T_{3-site} = - \frac{3 \alpha}{16 \pi c^2} \log \frac{M_\rho^2}{M_{Href}^2} 
              - \frac{3 \alpha}{32 \pi c^2} \log \frac{\Lambda^2}{M_\rho^2}
              + \frac{4\pi \alpha\,c_0(\Lambda)}{c^2}\nonumber
\,.
 \end{equation}
In addition to this contribution, there will typically be additional contributions
to the $T$-parameter$^2$ arising from isospin-violation in the fermion sector \cite{SekharChivukula:2006cg}.

\section{Reduction to the Two-Site Model}

\begin{figure}
\begin{center}
\vspace{10pt}
\unitlength 0.1in
\begin{picture}( 26.4000, 11.4500)(  9.9000,-14.4000)
%
\special{pn 20}%
\special{ar 1370 1070 380 360  0.0000000 6.2831853}%
%
\special{pn 20}%
\special{pa 1740 1080}%
\special{pa 2880 1080}%
\special{fp}%
%
\special{pn 20}%
\special{ar 3250 1080 380 360  0.0000000 6.2831853}%
\put(13.5000,-10.8000){\makebox(0,0){$L$}}%
\put(32.3000,-10.8000){\makebox(0,0){$R$}}%
\put(22.5000,-7.9000){\makebox(0,0){$U$}}%
\put(22.6000,-13.4000){\makebox(0,0){$v$}}%
\put(13.5000,-3.8000){\makebox(0,0){0}}%
\put(32.2000,-3.8000){\makebox(0,0){2}}%
\put(13.7000,-5.8000){\makebox(0,0){$g_0$}}%
\put(32.3000,-5.6000){\makebox(0,0){$g_2$}}%
\end{picture}%
\vspace{10pt}
\caption{
           Two-site nonlinear model which is, formally, the limit of
           the standard model as $M_{Higgs} \to \infty$ \protect\cite{Appelquist:1980ae,Appelquist:1980vg}.} 
\label{two-site}
\end{center}
\end{figure}
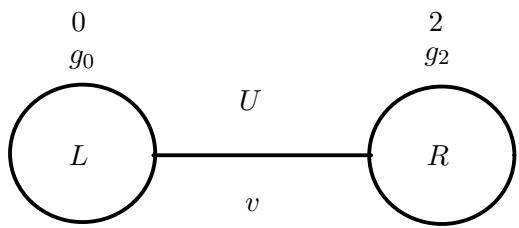

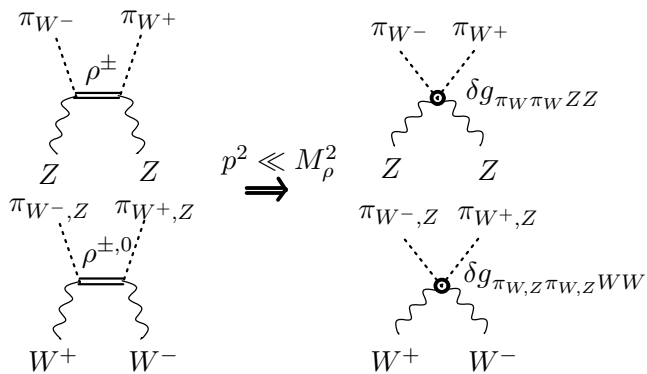
\begin{figure}
\begin{center}
\vspace{-15pt}  
\hspace{-65pt}
\unitlength 0.1in
\begin{picture}( 28.7900, 17.7700)( 11.5000,-25.0500)
%
\special{pn 13}%
\special{pa 1770 910}%
\special{pa 1876 1218}%
\special{dt 0.045}%
%
\special{pn 8}%
\special{pa 1872 1212}%
\special{pa 1828 1224}%
\special{pa 1796 1240}%
\special{pa 1796 1258}%
\special{pa 1824 1282}%
\special{pa 1842 1304}%
\special{pa 1818 1322}%
\special{pa 1780 1336}%
\special{pa 1762 1358}%
\special{pa 1784 1386}%
\special{pa 1804 1412}%
\special{pa 1786 1428}%
\special{pa 1746 1440}%
\special{pa 1730 1462}%
\special{pa 1758 1498}%
\special{pa 1776 1514}%
\special{pa 1770 1512}%
\special{sp}%
%
\special{pn 13}%
\special{pa 2212 894}%
\special{pa 2106 1202}%
\special{dt 0.045}%
%
\special{pn 13}%
\special{pa 1872 1194}%
\special{pa 2106 1194}%
\special{fp}%
%
\special{pn 13}%
\special{pa 1862 1220}%
\special{pa 2098 1220}%
\special{fp}%
%
\special{pn 8}%
\special{pa 2094 1208}%
\special{pa 2140 1220}%
\special{pa 2170 1234}%
\special{pa 2172 1254}%
\special{pa 2144 1278}%
\special{pa 2126 1300}%
\special{pa 2146 1316}%
\special{pa 2184 1332}%
\special{pa 2204 1352}%
\special{pa 2186 1380}%
\special{pa 2162 1406}%
\special{pa 2176 1424}%
\special{pa 2216 1434}%
\special{pa 2238 1456}%
\special{pa 2214 1492}%
\special{pa 2190 1512}%
\special{pa 2198 1508}%
\special{sp}%
\put(17.3100,-16.2000){\makebox(0,0){$Z$}}%
\put(22.5700,-16.0400){\makebox(0,0){$Z$}}%
\put(20.0700,-10.7700){\makebox(0,0){$\rho^\pm$}}%
\put(17.2300,-8.2100){\makebox(0,0){$\pi_{W^-}$}}%
\put(22.5700,-8.1300){\makebox(0,0){$\pi_{W^+}$}}%
%
\special{pn 20}%
\special{pa 2766 1692}%
\special{pa 2952 1692}%
\special{fp}%
%
\special{pn 20}%
\special{pa 2766 1718}%
\special{pa 2952 1718}%
\special{fp}%
%
\special{pn 20}%
\special{pa 2932 1642}%
\special{pa 2972 1704}%
\special{fp}%
%
\special{pn 20}%
\special{pa 2972 1698}%
\special{pa 2930 1754}%
\special{fp}%
\put(29.4000,-15.6000){\makebox(0,0){$p^2 \ll M_\rho^2$}}%
%
\special{pn 13}%
\special{pa 3956 974}%
\special{pa 3760 1224}%
\special{dt 0.045}%
%
\special{pn 8}%
\special{pa 3776 1244}%
\special{pa 3822 1234}%
\special{pa 3856 1234}%
\special{pa 3864 1252}%
\special{pa 3848 1286}%
\special{pa 3842 1314}%
\special{pa 3870 1318}%
\special{pa 3910 1314}%
\special{pa 3934 1326}%
\special{pa 3926 1360}%
\special{pa 3918 1394}%
\special{pa 3940 1400}%
\special{pa 3980 1390}%
\special{pa 4006 1402}%
\special{pa 4000 1448}%
\special{pa 3988 1472}%
\special{pa 3988 1466}%
\special{sp}%
\put(35.2300,-15.9000){\makebox(0,0){$Z$}}%
\put(40.3200,-16.0000){\makebox(0,0){$Z$}}%
\put(35.7000,-8.7500){\makebox(0,0){$\pi_{W^-}$}}%
\put(39.9300,-8.7500){\makebox(0,0){$\pi_{W^+}$}}%
%
\special{pn 13}%
\special{pa 3572 978}%
\special{pa 3768 1228}%
\special{dt 0.045}%
%
\special{pn 8}%
\special{pa 3750 1248}%
\special{pa 3704 1238}%
\special{pa 3670 1238}%
\special{pa 3662 1256}%
\special{pa 3680 1288}%
\special{pa 3686 1316}%
\special{pa 3660 1320}%
\special{pa 3618 1318}%
\special{pa 3592 1328}%
\special{pa 3600 1362}%
\special{pa 3610 1396}%
\special{pa 3590 1402}%
\special{pa 3548 1394}%
\special{pa 3522 1404}%
\special{pa 3530 1448}%
\special{pa 3542 1476}%
\special{pa 3538 1468}%
\special{sp}%
%
\special{pn 20}%
\special{ar 3766 1220 32 32  1.7928782 6.2831853}%
\special{ar 3766 1220 32 32  0.0000000 1.7126934}%
%
\special{pn 13}%
\special{pa 1788 1880}%
\special{pa 1894 2188}%
\special{dt 0.045}%
%
\special{pn 8}%
\special{pa 1892 2182}%
\special{pa 1846 2194}%
\special{pa 1816 2208}%
\special{pa 1814 2228}%
\special{pa 1842 2252}%
\special{pa 1860 2274}%
\special{pa 1838 2290}%
\special{pa 1800 2306}%
\special{pa 1780 2326}%
\special{pa 1800 2354}%
\special{pa 1824 2382}%
\special{pa 1808 2398}%
\special{pa 1768 2410}%
\special{pa 1748 2430}%
\special{pa 1774 2464}%
\special{pa 1796 2484}%
\special{pa 1788 2482}%
\special{sp}%
%
\special{pn 13}%
\special{pa 2230 1864}%
\special{pa 2124 2172}%
\special{dt 0.045}%
%
\special{pn 13}%
\special{pa 1892 2166}%
\special{pa 2124 2166}%
\special{fp}%
%
\special{pn 13}%
\special{pa 1886 2196}%
\special{pa 2120 2196}%
\special{fp}%
%
\special{pn 8}%
\special{pa 2114 2178}%
\special{pa 2158 2190}%
\special{pa 2190 2206}%
\special{pa 2190 2224}%
\special{pa 2162 2248}%
\special{pa 2144 2270}%
\special{pa 2166 2288}%
\special{pa 2204 2304}%
\special{pa 2222 2324}%
\special{pa 2202 2354}%
\special{pa 2180 2380}%
\special{pa 2198 2394}%
\special{pa 2238 2406}%
\special{pa 2256 2430}%
\special{pa 2230 2470}%
\special{pa 2212 2484}%
\special{pa 2216 2476}%
\special{sp}%
\put(17.5300,-25.9000){\makebox(0,0){$W^+$}}%
\put(22.7900,-25.9000){\makebox(0,0){$W^-$}}%
\put(20.3700,-20.3000){\makebox(0,0){$\rho^{\pm,0}$}}%
\put(17.3500,-18.0800){\makebox(0,0){$\pi_{W^-,Z}$}}%
\put(22.8700,-18.1600){\makebox(0,0){$\pi_{W^+,Z}$}}%
%
\special{pn 13}%
\special{pa 3978 1958}%
\special{pa 3782 2208}%
\special{dt 0.045}%
%
\special{pn 8}%
\special{pa 3798 2228}%
\special{pa 3844 2218}%
\special{pa 3878 2218}%
\special{pa 3886 2236}%
\special{pa 3872 2268}%
\special{pa 3864 2296}%
\special{pa 3892 2302}%
\special{pa 3932 2298}%
\special{pa 3956 2308}%
\special{pa 3950 2344}%
\special{pa 3940 2378}%
\special{pa 3960 2384}%
\special{pa 4002 2374}%
\special{pa 4028 2386}%
\special{pa 4022 2430}%
\special{pa 4010 2456}%
\special{pa 4010 2450}%
\special{sp}%
\put(35.4600,-25.9000){\makebox(0,0){$W^+$}}%
\put(40.5500,-25.9000){\makebox(0,0){$W^-$}}%
%
\special{pn 13}%
\special{pa 3594 1962}%
\special{pa 3790 2210}%
\special{dt 0.045}%
%
\special{pn 8}%
\special{pa 3774 2232}%
\special{pa 3728 2222}%
\special{pa 3694 2222}%
\special{pa 3686 2240}%
\special{pa 3702 2272}%
\special{pa 3708 2300}%
\special{pa 3682 2306}%
\special{pa 3640 2302}%
\special{pa 3616 2310}%
\special{pa 3622 2344}%
\special{pa 3632 2378}%
\special{pa 3612 2386}%
\special{pa 3572 2378}%
\special{pa 3544 2388}%
\special{pa 3550 2434}%
\special{pa 3562 2460}%
\special{pa 3562 2452}%
\special{sp}%
%
\special{pn 20}%
\special{ar 3788 2204 34 32  1.7619808 6.2831853}%
\special{ar 3788 2204 34 32  0.0000000 1.6775320}%
\put(42.6300,-12.0000){\makebox(0,0){${\delta{g}}_{\pi_W\pi_W ZZ}$}}%
\put(43.8100,-21.7000){\makebox(0,0){${\delta{g}}_{\pi_{W,Z}\pi_{W,Z} WW}$}}%
\put(35.6400,-18.4100){\makebox(0,0){$\pi_{W^-,Z}$}}%
\put(40.8100,-18.4900){\makebox(0,0){$\pi_{W^+,Z}$}}%
\end{picture}%
\vspace{15pt}
\caption{ $\rho$-exchange diagrams contributing to the $\pi_{W^+}\pi_{W^-}ZZ$, 
          $\pi_{W^+}\pi_{W^-}W^+W^-$ and 
          $\pi_Z\pi_Z W^+W^-$ interaction terms at energy scales 
$p^2 \ll M_\rho^2$.   
          }
\label{rhoex}
\end{center}

\end{figure}

In the limit $M_{Higgs} \to \infty$, which corresponds to taking the self-coupling of
the Higgs to infinity, the standard model formally reduces to the
electroweak chiral lagrangian \cite{Appelquist:1980ae,Appelquist:1980vg} which may be
viewed as the ``two-site" model illustrated in Fig. \ref{two-site}. Consider the limit $M_\rho
\to \infty$ in the three-site model, obtained by taking the coupling
$g_1$ in Fig. \ref{threesite} to infinity. In either case one is taking a {\it dimensionless} coupling
to infinity, and the ordinary decoupling theorem \cite{Appelquist:1974tg}  does
not apply. 

Indeed, as noted above, there are interaction vertices proportional to 
$M_{\rho}$ (see Eqn. (A.1)) in the three-site model.
These yield corrections to $\overline{\Pi}_{ZZ}$ and $\overline{\Pi}_{WW}$ that are proportional to $\ln{M_\rho^2/M_W^2}$ without any $1/M^2_\rho$ suppression factor.  Ordinarily, one would expect such contributions to the low-energy behavior of the theory to arise only from diagrams without propagating $\rho$ bosons.  In this case, however, the non-decoupling is manifested by the presence of terms
proportional to $\ln{M^2_\rho/M^2_W}$ in the amplitudes for diagram $(N)_{ZZ}$ of Fig. 2 and diagrams $(M)_{WW}$ and $(O)_{WW}$ of Fig. 4, all of which include propagating $\rho$ fields.  Examining the contributions of these diagrams in detail (see Eqns. (\ref{eq:nzz}, \ref{eq:mww}, \ref{eq:oww})) reveals that these diagrams contribute only to  $\overline{\Pi}_{ZZ,WW}(0),$ and therefore affect $\alpha T$ but not $\alpha S$.

Nonetheless, we have seen above that the one-loop leading-log
contribution to $T$ arising from scaling between $M_W$ and $M_\rho$ has
precisely the same form as the leading-log contribution from a heavy
Higgs boson. In retrospect, this is an expected result.  The chiral-logarithmic
contributions of this kind depend only on the low-energy theory valid
at energy scales between $M_W$ and $M_\rho$. The leading order
-- ${\cal O}(p^2)$ --  interactions in this energy regime are determined entirely
by gauge-invariance and chiral low-energy theorems. Since the gauge- and chiral-symmetries
of the three-site model at energies below $M_\rho$ are precisely the same as those
in the standard model, the ${\cal O}(p^2)$ interactions must be the same in both
theories --- and therefore the chiral-logarithmic corrections arising from this
energy regime must also be the same in both theories \cite{Bando:1985ej,Bando:1985rf,Bando:1988ym,Bando:1988br}.

\begin{table}

\begin{center} 
\begin{tabular}{|c|c|c|c|c|}
\hline 
       & $\pi_{W^+}\pi_{W^-} ZZ$ & $\pi_{W^+}\pi_{W^-} W^+ W^-$ & $\pi_Z \pi_Z W^+ W^-$ & $\pi_Z \pi_Z Z Z$ \\ 
\hline \hline  
2-site & $-e^2$ 
       & $0$ 
       & $0$ 
       & $0$ \\ 
\hline 
3-site & $\frac{e^2(c^2-s^2)^2}{4s^2c^2}$
       & $\frac{e^2}{4s^2}$ 
       & $\frac{e^2}{4s^2}$
       & $0$ \\
\hline
\end{tabular}
\end{center} 
\caption{ This table lists the coupling constants of the tree-level two-pion/two-gauge-boson interactions
arising from the ${\cal O}(p^2)$ interactions in the 2-site and 3-site models. Adding
the non-decoupling contributions arising from $\rho$-exchange illustrated in Fig. \protect\ref{rhoex}, 
Eqn. (\protect\ref{nondecoupling}),
we see that the 3-site interactions reduce to those of the 2-site model at energies
less than $M_\rho$.
         }
\label{differences}
\end{table}

Examining the pion and Goldstone boson interactions in the ${\cal O}(p^2)$ lagrangian, 
we find that the only differences between the three-site and two-site model relevant to 
the calculation of the gauge-boson self-energies occur 
in the two-pion/two-gauge-boson interactions summarized in table 1.
We may  see how the three-site to two-site reduction occurs 
explicitly.\footnote{See Figs. 2 and 3 of ref. 
\protect\cite{Harada:2003jx}.}
Starting in the three-site model, we have the ${\cal L}_{\pi\pi AA}$ vertices
listed in table \ref{differences}. We also have $\pi \pi AA$ interactions
mediated by $\rho$-exchange, as illustrated in the left panel of Fig.~\ref{rhoex}. 
At energies $E < M_\rho$, we may integrate out the $\rho$-mesons and,
at tree-level, we obtain additional $\pi\pi AA$ interactions
as illustrated in the right panel of Fig.~\ref{rhoex}.
The correspondingly induced
couplings  are evaluated to be
           \begin{eqnarray} 
                 {\delta{g}}_{\pi_W \pi_W WW} &=& - \frac{e^2}{4s^2} \,, \qquad 
                   \delta{g}_{\pi_Z \pi_Z WW} = - \frac{e^2}{4s^2} 
                \,, \nonumber \\ 
                  {\delta{g}}_{\pi_W \pi_W ZZ} &=& - \frac{e^2}{4s^2c^2} ~.
                  \label{nondecoupling}
           \end{eqnarray} 
           Combining these contributions with the three-site couplings 
           given in table \ref{differences}, we find that the three-site model interactions reduce
           to those of the two-site model at energies less than $M_\rho$.


\section{Discussion}

We have computed the one-loop corrections to the $S$ and $T$ parameters
in a highly-deconstructed three site Higgless model. Higgsless models may be
considered as dual \cite{Maldacena:1998re,Gubser:1998bc,Witten:1998qj,Aharony:1999ti} to models of dynamical symmetry breaking \cite{Weinberg:1979bn,Susskind:1979ms} akin to ``walking technicolor" \cite{Holdom:1981rm,Holdom:1985sk,Yamawaki:1986zg,Appelquist:1986an,Appelquist:1987tr,Appelquist:1987fc}, and in these terms our calculation is the first to compute the 
subleading  $1/N$ corrections to the $S$ and $T$ parameters.   We find that the chiral-logarithmic corrections
naturally separate into  a model-independent part arising from scaling below the $\rho$ mass, which has the same form as the large Higgs-mass dependence of the $S$ or $T$ parameter in the standard model, and a second model-dependent contribution arising from scaling between the $\rho$ mass and the cutoff of the model. The former allows us to correctly interpret the phenomenologically derived limits on the $S$ and $T$ parameters (in terms of a ``reference" Higgs-boson mass) in this three-site Higgsless model. 
We also discussed the reduction of the model
to the ``two-site" model, which is the usual electroweak chiral lagrangian, noting
the ``non-decoupling" contributions present in the limit $M_\rho \to \infty$.

Our analysis has focused on contributions to the $S$ and $T$ parameters from the extended electroweak gauge sector.  In principle, there would also be contributions from the extended fermion sector of the model.  We calculated these contributions$^2$ to $\alpha T$ in the three-site model \cite{SekharChivukula:2006cg} and demonstrated that they are sizable enough to place strong lower bounds on the masses of the KK fermions.  Specifically, the enlarged fermion sector results from adding fermions with Dirac masses ($M$) and the bound from $\alpha T$ is $M \geq 1.8$ TeV.  The contributions of these Dirac fermions to the $S$ parameter decouple in the large-$M$ limit;  the lower bound on $M$ is large enough to  render their  ${\cal O}(M^2_W /16\pi^2M^2)$ contributions to $\alpha S$ negligible.

In the limit in which the vector fields at sites 0 and 2 are treated as 
external gauge fields ({\it i.e.}, not as dynamical fields)  the three-site model is equivalent to the
``vector limit" \cite{Georgi:1989xy} (with $a$=1) of  ``Hidden
Local Symmetry" \cite{Bando:1985ej,Bando:1985rf,Bando:1988ym,Bando:1988br,Harada:2003jx}
models of chiral dynamics in QCD. The calculation of one-loop corrections to the $S$-parameter
presented here -- in the limit of ``brane-localized" fermions, $x_1=0$ --
 is equivalent to those presented in \cite{Tanabashi:1993sr,Harada:2003jx}.
Note, however, the contributions to the $T$ parameter arise from one-loop diagrams
involving the vector-boson at site 2, and cannot be reproduced in the limit that one treats
this gauge-boson as external.

In a forthcoming publication
\cite{newpaper} we will report the results of a full renormalization-group analysis of the ${\cal O}(p^4)$
terms in the three-site Higgsless model effective theory, allowing us to independently
confirm the results in Eqns.
(\protect\ref{result-sparam}) and (\protect\ref{result-t}), and to express the values of $\alpha S$
and $\alpha T$ in terms of low-energy parameters.

\bigskip

\centerline{\bf Acknowledgments}

We thank Chris Jackson for correspondence about the calculation of $\alpha S$
in Higgsless models, and Sally Dawson, Nick Evans, and Koichi Yamawaki
for discussions. The visit of S.M. to Michigan State University which made
this collaboration possible was fully supported by
the fund of The Mitsubishi Foundation through Koichi Yamawaki. 
R.S.C. and E.H.S. are supported in part by the US National Science Foundation under
grant  PHY-0354226; they thank the Aspen Center for Physics for its hospitality while
this work was being completed. M.T.'s work is supported in part by the JSPS Grant-in-Aid for
Scientific Research No.16540226

\newpage

\appendix 
\renewcommand\theequation{\Alph{section}.\arabic{equation}}

\section{Interactions of Mass Eigenstate Fields}
\label{interactions}

In this appendix, we rewrite the gauge-sector interactions from Section
\ref{gaugelagrangian} in terms
of mass eigenstate fields using  Eqns. (\ref{lpm}) -- (\ref{vpm}) and 
(\ref{l3}) -- (\ref{c1}), expanding those in powers of $x$ and 
keeping the terms which give rise to a non-trivial contribution 
to the gauge boson self-energy functions in the leading log approximation, up to 
corrections of order ${\cal O}(\alpha x^2 M^2_W)$.

\subsection{Three-Point Vertices Proportional to Gauge Boson Masses: $\mathcal{L}_{\pi AA}$ } 

Rewriting eqn. (\ref{re pi aa}) in terms of mass eigenstate fields yields
\begin{eqnarray} 
\mathcal{L}_{\pi AA} &=& i \Bigg[ 
\sum_{n=A,Z,\rho^0} \Bigg\{
\left( g_{n W^+}^{\pi_{W^-}} \, n_\mu W^{\mu +}  
+ g_{n \rho^+}^{\pi_{W^-}} \, n_\mu \rho^{\mu +} \right) \pi_{W^-} 
                    \nonumber\\ 
&& \hspace{55pt} 
+ \left( g_{n W^+}^{\pi_{\rho^-}} \, n_\mu W^{\mu +} 
+ g_{n \rho^+}^{\pi_{\rho^-}} \, n_\mu \rho^{\mu +} \right) \pi_{\rho^-}  
\Bigg\} 
                     \nonumber \\ 
&& 
+ \sum_{\alpha,\beta=W^\pm, \rho^\pm}^{\alpha \neq \beta} \Bigg\{ 
g_{\alpha \beta}^{\pi_Z} \, \alpha_\mu \beta^\mu  \pi_Z  
+ g_{\alpha \beta}^{\pi_{\rho^-}} \, \alpha_\mu \beta^\mu  \pi_{\rho^-} 
\Bigg) 
\Bigg] + {\rm h.c.} \,.
\label{eq:a1}
\end{eqnarray} 
The couplings are expressed in terms of the gauge and Goldstone 
boson wavefunctions given in Eqns.(\ref{wavefunc:Wrhopm}), 
(\ref{wavefunc:AZrho0}) and (\ref{wavefunc:Pi}) as 
\begin{eqnarray} 
g_{n \alpha}^{\pi_{W^-}} &=& 
\frac{e M_{\rho^\pm}}{\sqrt{2} s} \left( 1 + \frac{4s^2-1}{8}x^2 \right) 
\left( \left[ v_n^L v_{\alpha}^V - v_n^V v_{\alpha}^L \right] v_{\pi_{W^\pm}}^{(1)} 
- t \cdot v_n^B v_{\alpha}^V v_{\pi_{W^\pm}}^{(2)}  
\right)
\,, \\
g_{n \alpha}^{\pi_{\rho^-}} &=& 
\frac{e M_{\rho^\pm}}{\sqrt{2} s} 
\left( 1 + \frac{4s^2-1}{8}x^2 \right) 
\left( \left[ v_n^L v_{\alpha}^V - v_n^V v_{\alpha}^L \right] v_{\pi_{\rho^\pm}}^{(1)} 
- t \cdot v_n^B v_{\alpha}^V v_{\pi_{\rho^\pm}}^{(2)}  \right)
\,, \\
g_{\alpha \beta}^{\pi_Z} &=& 
\frac{e M_{\rho^\pm}}{\sqrt{2} s} 
\left( 1 + \frac{4s^2-1}{8}x^2 \right) 
\left( v_\alpha^L v_{\beta}^V - v_\alpha^V v_\beta^L \right) v_{\pi_Z}^{(1)}  
\qquad (\alpha \neq \beta) \,,\\ 
g_{\alpha \beta}^{\pi_{\rho^0}} &=& 
\frac{e M_{\rho^\pm}}{\sqrt{2} s} 
\left( 1 + \frac{4s^2-1}{8}x^2 \right) 
\left( v_\alpha^L v_{\beta}^V - v_\alpha^V v_\beta^L \right) v_{\pi_{\rho^0}}^{(1)}  \qquad (\alpha \neq \beta)
\,,   
\end{eqnarray}
for $\alpha,\beta=W^\pm, \rho^\pm$ and $n=A,Z,\rho^0$. 
In obtaining the expressions for these couplings, we have used 
Eqn. (\ref{e-g0}) and 
\begin{equation} 
g_1 f = \sqrt{2} M_{\rho^\pm} 
\left( 1 - \frac{x^2}{8} + \cdots \right) 
  \,, 
\end{equation} 
which follows from Eqn. (\ref{mrhopm}). 
The explicit expressions for each of these couplings is shown 
in table~\ref{mass-couplings}. From table~\ref{mass-couplings}, 
in the isospin symmetric limit $s \to 0$ (or $c \to 1$) with $(e/s)$ fixed, 
we find 
\begin{eqnarray} 
g_{Z W^+}^{\pi_{\rho^-}}|_{c \to 1} 
&=& 
g_{W^+W^-}^{\pi_{\rho^0}}=0 \,, \qquad 
g_{\rho^0 W^+}^{\pi_{W^-}}|_{c \to 1} 
= g_{Z \rho^+}^{\pi_{W^-}}|_{c \to 1} 
= g_{W^- \rho^+}^{\pi_Z}|_{c \to 1}
\,, \nonumber \\ 
g_{\rho^0 W^+}^{\pi_{\rho^-}} |_{c \to 1} 
&=& g_{Z \rho^-}^{\pi_{\rho^+}}|_{c \to 1} 
= g_{W^- \rho^+}^{\pi_{\rho^0}}|_{c \to 1}
\, , 
\end{eqnarray} 
as expected.

 The interaction vertices in eqn. (\ref{eq:a1}) include terms that explicitly mix the standard model and new-physics sectors of the model.  The second term in line one of Eqn. (\ref{eq:a1}) includes an interaction ($Z \rho \pi_W$) contributing to diagram $(N)_{ZZ}$ of Fig. \ref{selfenergy}. The first term of line one includes an interaction ($W \rho \pi_W$) contributing to diagram $(M)_{WW}$ of Fig. \ref{selfenergyW}; 
the first term of line three includes a $W \rho \pi_Z$ interaction contributing to diagram $(O)_{WW}$ of the same figure.  All three of these terms contribute to diagrams whose amplitudes are explicitly found to have non-decoupling contributions proportional to $\ln (M_\rho^2/M_W^2)$, unsuppressed by factors of $1/M_\rho^2$. 

\begin{table}

\begin{center} 
\begin{tabular}{|l||c|c|} 
\hline 
 $n$ & $g_{n W^+}^{ \pi_{W^-}}$ & $g_{n W^+}^{ \pi_{\rho^-}}$ \\ 
 \hline \hline  
 $A$ & $-eM_W$ & $0$  \\ 
\hline 
 $Z$ & $\frac{esM_W}{c}$ & $ \frac{eM_W x^2}{8sc} \left( 1- \frac{1}{c^2} \right) $ \\ 
\hline 
 $\rho^0$ & $ - \frac{e M_{\rho^\pm}}{2s} \left( 1 - \frac{x^2(  4c^2 -4 + \frac{3}{c^2})}{8} \right)$ 
        & $\frac{eM_{\rho^\pm}}{2s} \left( 1 + \frac{x^2( - 4 c^2 + 4 + \frac{1}{c^2})}{8} \right)$ \\ 
\hline 
\end{tabular} 
\end{center} 

\begin{center} 
\begin{tabular}{|l||c|c|} 
\hline 
 $n$ & $g_{n \rho^+}^{ \pi_{W^-}}$ & $g_{n \rho^+}^{ \pi_{\rho^-}}$ \\ 
\hline \hline  
 $A$ & $0$ & $-e M_{\rho^\pm}$ \\ 
\hline 
 $Z$ & $\frac{e M_{\rho^\pm}}{2sc} \left( 1 - \frac{x^2( 8 c^2  - 6 + \frac{1}{c^2})}{8} \right) $  &
         $ - \frac{e (c^2-s^2) M_{\rho^\pm}}{2sc} \left( 1 + \frac{x^2}{8(c^2-s^2)c^2} \right) $ \\ 
\hline 
 $\rho^0$ & $\frac{es M_W}{2c^2} $ & $\frac{es M_W}{2c^2} $ \\ 
\hline   
\end{tabular}
\end{center} 

\begin{center} 
\begin{tabular}{|l||c|c|} 
\hline 
 $g_{\alpha \beta}^{\pi_Z}$ & $\beta=$ $W^+$ & $\beta=$ $\rho^+$ \\
\hline \hline
 $\alpha=$ $W^-$ & $0$ & $\frac{eM_{\rho^\pm}}{2s} \left( 1 - \frac{x^2( 4c^2  + 1 - \frac{2}{c^2})}{8} \right)$ \\ 
\hline 
 $\alpha=$ $\rho^-$ & $-\frac{eM_{\rho^\pm}}{2s} \left( 1 - \frac{x^2( 4c^2  + 1 - \frac{2}{c^2})}{8} \right)$  & $0$ \\  
\hline 
\end{tabular} 
\end{center}

\begin{center} 
\begin{tabular}{|l||c|c|} 
\hline 
 $g_{\alpha \beta}^{\pi_{\rho^0}}$ & $\beta=W^+$ & $\beta=\rho^+$ \\
\hline \hline 
 $\alpha=W^-$ & $0$ & $- \frac{eM_{\rho^\pm}}{2s}\left( 1 + \frac{x^2 (-4c^2 + 7 - \frac{2}{c^2})}{8} \right) $ \\ 
\hline 
 $\alpha=\rho^-$ &  $ \frac{eM_{\rho^\pm}}{2s}\left( 1 + \frac{x^2 (-4c^2 + 7 - \frac{2}{c^2})}{8} \right) $ & $0$ \\  
\hline 
\end{tabular} 
\end{center}

\caption{ The Goldstone boson couplings proportional to gauge boson masses. } 
\label{mass-couplings}
\end{table}

\subsection{Three-Point Vertices Dependent on Derivatives: $\mathcal{L}_{\pi \pi A}$ } 

Rewriting Eqn. (\ref{re pipi a}) in terms of mass eigenstate fields
yields
\begin{eqnarray} 
   \mathcal{L}_{\pi\pi A} &=& 
i \Bigg\{ \sum_{n=A,Z,\rho^0} n_\mu \Bigg[   
g_{\pi_{W^+} \pi_{W^-}}^n \partial^\mu \pi_{W^+} \pi_{W^-}
+ g_{\pi_{\rho^+} \pi_{\rho^-}}^n \partial^\mu \pi_{\rho^+} \pi_{\rho^-} 
 \nonumber \\ 
&& + 
g_{\pi_{W^+} \pi_{\rho^-}}^n \left( \partial^\mu \pi_{W^+} \pi_{\rho^-} 
+ \partial^\mu \pi_{\rho^+} \pi_{W^-} \right)   
\Bigg] 
\nonumber \\ 
&&  + 
\sum_{\alpha=W^\pm,\rho^\pm} \alpha^\mu \Bigg[ 
 g_{\pi_Z\pi_{W^-}}^\alpha \left(\pi_Z \stackrel{\leftrightarrow}{\partial}_\mu \pi_{W^-} \right) + g_{\pi_{\rho^0}\pi_{\rho^-}}^\alpha \left(\pi_{\rho^0} \stackrel{\leftrightarrow}{\partial}_\mu \pi_{\rho^-} \right) 
\nonumber \\ 
&& + 
g_{\pi_Z\pi_{\rho^-}}^\alpha 
\left(\pi_Z \stackrel{\leftrightarrow}{\partial}_\mu \pi_{\rho^-} \right)
+ g_{\pi_{\rho^0}\pi_{W^-}}^\alpha 
\left(\pi_{\rho^0} \stackrel{\leftrightarrow}{\partial}_\mu \pi_{W^-} \right)
\Bigg] \Bigg\} 
\nonumber \\ 
&& 
+ {\rm h.c.} \,.
\label{eq:a8}
\end{eqnarray} 
The couplings are expressed in terms of the gauge and Goldstone 
boson wavefunctions given in Eqns.(\ref{wavefunc:Wrhopm}), 
(\ref{wavefunc:AZrho0}) and (\ref{wavefunc:Pi}) as 
\begin{eqnarray} 
   g_{\pi_\alpha \pi_\beta}^n &=& \frac{e}{2s} \left(  1 + \frac{s^2}{2}x^2 \right)\left[ v_{\pi_\alpha}^{(1)}v_{\pi_\beta}^{(1)} 
                    (v_n^L + \frac{1}{x} v_n^V) + v_{\pi_\alpha}^{(2)}v_{\pi_\beta}^{(2)} 
                    ( t \cdot  v_n^B + \frac{1}{x} v_n^V)\right] 
                    \,, \\ 
   g_{\pi_Z \pi_{\beta}}^\alpha &=& - \frac{e}{2s} \left(  1 + \frac{s^2}{2}x^2 \right) \left[ v_{\pi_Z}^{(1)}v_{\pi_{\beta}}^{(1)} 
                    (v_\alpha^L + \frac{1}{x} v_\alpha^V) + \frac{1}{x} v_{\pi_Z}^{(2)}v_{\pi_{\beta}}^{(2)} 
                    v_\alpha^V \right] 
                    \,, \\ 
    g_{\pi_{\rho^0} \pi_{\beta}}^\alpha &=& - \frac{e}{2s} \left(  1 + \frac{s^2}{2}x^2 \right) \left[ v_{\pi_{\rho^0}}^{(1)}v_{\pi_{\beta}}^{(1)} 
                    (v_\alpha^L + \frac{1}{x} v_\alpha^V) + \frac{1}{x} v_{\pi_{\rho^0}}^{(2)}v_{\pi_{\beta}}^{(2)} 
                     v_\alpha^V \right] 
                    \,,              
\end{eqnarray}
for $\alpha,\beta=W^\pm, \rho^\pm$ and $n=A,Z,\rho^0$. 
In obtaining the expressions for these couplings, we have used 
Eqn. (\ref{e-g0}). 
The explicit expression for each of these couplings is shown in table~\ref{deriv-couplings}. From table~\ref{deriv-couplings}, 
in the isospin symmetric limit $s \to 0$ (or $c \to 1$) with $(e/s)$ fixed, 
we find 
\begin{eqnarray} 
g_{\pi_{W^+} \pi_{W^-}}^Z |_{c \to 1} 
&=& 
g_{\pi_Z\pi_{W^-}}^{W^+}|_{c \to 1} \,, \qquad 
g_{\pi_Z \pi_{\rho^-}}^{W^+} |_{c \to 1} 
= g^Z_{\pi_{W^+} \pi_{\rho^-}} |_{c \to 1} 
= g^{W^+}_{\pi_{\rho^0} \pi_{W^-}} |_{c \to 1} 
\,, \nonumber \\ 
g_{\pi_{\rho^+} \pi_{\rho^-}}^Z |_{c \to 1} 
&=& g_{\pi_{\rho^0} \pi_{\rho^-}}^{W^+} |_{c \to 1} \,, \qquad  
g_{\pi_{W^+} \pi_{\rho^-}}^{\rho^0} |_{c \to 1} 
= g_{\pi_Z \pi_{\rho^-}}^{\rho^+} 
= g_{\pi_{\rho^0}\pi_{W^-}}^{\rho^+} |_{c \to 1}
\,, 
\end{eqnarray} 
up to an overall sign, 
as expected.

  The interaction vertices in Eqn. (\ref{eq:a8}) include terms that explicitly mix the standard model and new-physics sectors of the model.  The terms on line two of (\ref{eq:a8}) include interactions ($Z \pi_W \pi_\rho$) contributing to diagram $(M)_{ZZ}$ of Fig. \ref{selfenergy}; the terms of line four include $W \pi_Z \pi_\rho$ and $W\pi_\rho \pi_W$ interactions contributing, respectively, to diagrams $(R)_{WW}$ and $(S)_{WW}$ of Fig. \ref{selfenergyW}. 

\begin{table}

\begin{center} 
\begin{tabular}{|l||c|c|c|} 
\hline 
 $n$ & $g_{\pi_{W^+}\pi_{W^-}}^n$ & $g^n_{\pi_{W^+} \pi_{\rho^-}}$ & $g^n_{\pi_{\rho^+} \pi_{\rho^-}}$ \\ 
\hline \hline  
 $A$ & $e$ & $0$& $e$ \\ 
\hline 
 $Z$ & $\frac{e(c^2-s^2)}{2sc}$  & 
         $ - \frac{e}{4sc} \left( 1 + \frac{x^2( - 8c^2 + 8 - \frac{1}{c^2})}{8} \right) $ & 
         $ \frac{e(c^2-s^2)}{2sc} \left( 1 + \frac{x^2}{4(c^2-s^2)}  \right) $ \\ 
\hline 
 $\rho^0$ & $\frac{e}{2sx}$  & $ \frac{e(c^2-s^2) x}{8sc^2} $ & $\frac{e}{2sx} $ \\ 
\hline   
\end{tabular}
\end{center} 

\begin{center} 

\begin{tabular}{|l||c|c|} 
\hline 
$g_{\pi_Z \pi_{\beta}}^\alpha$ & $\beta=W^-$  & $\beta=\rho^-$ \\ 
\hline \hline 
$\alpha=W^+$ & $- \frac{e}{2s}$ & $\frac{e}{4s} \left( 1 + \frac{x^2( - 4c^2 -1 + \frac{4}{c^2})}{8} \right)$ \\ 
\hline 
$\alpha=\rho^+$ & irrelevant  & $- \frac{e (c^2-s^2) x}{8sc^2} $  \\ 
\hline 
\end{tabular}

\end{center}

\begin{center} 

\begin{tabular}{|l||c|c|} 
\hline 
$g_{\pi_{\rho^0} \pi_{\beta}}^\alpha$ & $\beta=W^-$  & $\beta=\rho^-$ \\ 
\hline \hline 
$\alpha=W^+$ & $ \frac{e}{4s} \left( 1 + \frac{x^2( -4c^2 +7  - \frac{4}{c^2})}{8} \right) $ & $ - \frac{e}{2s} \left( 1 + \frac{x^2( - 4c^2 + 7 - \frac{1}{c^2})}{8} \right)$ \\ 
\hline 
$\alpha=\rho^+$ & $ - \frac{e x}{8sc^2} $  & $- \frac{e}{2sx}$ \\ 
\hline 
\end{tabular}

\end{center}

\caption{ The Goldstone boson couplings dependent on derivatives. 
The expression of the $g_{\pi_Z\pi_W^-}^{\rho^+}$ coupling 
is not shown (denoted as ``irrelevant") 
since the vertex constructed from this coupling 
does not contribute to the self-energy functions of the 
standard model gauge bosons $W,Z$ and photon. }
\label{deriv-couplings}
\end{table}

\subsection{Four-Point Vertices: $\mathcal{L}_{\pi \pi AA}$ } 

Rewriting Eqn. (\ref{re pipi aa}) in terms of mass eigenstate fields
yields

 \begin{eqnarray} 
   \mathcal{L}_{\pi\pi AA} &=& 
              \pi_{W^+} \pi_{W^-} \Bigg( 
                 \sum_{n = A,Z ,\rho^0} g^{nn}_{\pi_{W^+}\pi_{W^-}} n_\mu n^\mu + 
                  \sum_{n = A,Z ,\rho^0}^{ n \neq m} g^{nm}_{\pi_{W^+}\pi_{W^-}} n_\mu m^\mu \Bigg) 
                  \nonumber \\ 
              && + \pi_{\rho^+} \pi_{\rho^-} \Bigg( 
                 \sum_{n = A,Z ,\rho^0} g^{nn}_{\pi_{\rho^+}\pi_{\rho^-}} n_\mu n^\mu + 
                  \sum_{n, m= A,Z ,\rho^0}^{ n \neq m} g^{nm}_{\pi_{\rho^+}\pi_{\rho^-}} n_\mu m^\mu \Bigg) 
                  \nonumber \\ 
              && + \pi_{W^+} \pi_{W^-} \Bigg( 
                     \sum_{\alpha=W^\pm , \rho^\pm} g^{\alpha \alpha}_{\pi_{W^+}\pi_{W^-}}  
                      \, \alpha_\mu \alpha^\mu    
                   + \sum_{\alpha, \beta=W^\pm ,\rho^\pm}^{\alpha \neq \beta} g^{\alpha \beta}_{\pi_{W^+}\pi_{W^-}} 
                      \, \alpha_\mu \beta^\mu  \Bigg)
                  \nonumber \\ 
              &&  + \pi_Z \pi_Z \Bigg( 
                     \sum_{\alpha=W^\pm , \rho^\pm} g^{\alpha \alpha}_{\pi_Z\pi_Z}  
                      \, \alpha_\mu \alpha^\mu    
                   + \sum_{\alpha, \beta=W^\pm ,\rho^\pm}^{\alpha \neq \beta} g^{\alpha \beta}_{\pi_Z \pi_Z} 
                      \, \alpha_\mu \beta^\mu  \Bigg)
                  \nonumber \\ 
              && + \pi_{\rho^0} \pi_{\rho^0} \Bigg( 
                     \sum_{\alpha=W^\pm , \rho^\pm} g^{\alpha \alpha}_{\pi_{\rho^0}\pi_{\rho^0}}  
                      \, \alpha_\mu \alpha^\mu    
                   + \sum_{\alpha, \beta=W^\pm ,\rho^\pm}^{\alpha \neq \beta} g^{\alpha \beta}_{\pi_{\rho^0}\pi_{\rho^0}} 
                      \, \alpha_\mu \beta^\mu  \Bigg)
                  \,. 
 \end{eqnarray} 
These couplings are expressed in terms of the gauge and Goldstone 
boson wavefunctions given in Eqns. (\ref{wavefunc:Wrhopm}), 
(\ref{wavefunc:AZrho0}) and (\ref{wavefunc:Pi}) as 
\begin{eqnarray} 
     g^{nn}_{\pi_{\alpha}\pi_{\alpha}} &=& \frac{e^2}{s^2x} \left(  1 + s^2x^2 \right) \left[ 
                                v_{\pi_{\alpha}}^{(1)} v_{\pi_{\alpha}}^{(1)} v_n^L v_n^V 
                            + t \cdot v_{\pi_{\alpha}}^{(2)} v_{\pi_{\alpha}}^{(2)} v_n^B v_n^V  \right]
       \,, \\ 
     g^{nm}_{\pi_{\alpha}\pi_{\alpha}} &=& \frac{e^2}{s^2x} \left(  1 + s^2x^2 \right) \left[ 
                                v_{\pi_{\alpha}}^{(1)} v_{\pi_{\alpha}}^{(1)} ( v_n^L v_m^V + v_n^V v_m^L )  
                            + t \cdot v_{\pi_{\rho^+}}^{(2)} v_{\pi_{\alpha}}^{(2)} 
                           ( v_n^B v_m^V + v_n^V v_m^B) \right]
                           \,, \nonumber \\
\\ 
     g^{\beta \beta}_{\pi_{\alpha}\pi_{\alpha}} &=& \frac{e^2}{s^2x} \left(  1 + s^2x^2 \right)
                                v_{\pi_{\alpha}}^{(1)} v_{\pi_{\alpha}}^{(1)} v_\beta^L v_\beta^V 
                           \,, \\ 
     g^{\beta \gamma}_{\pi_{\alpha}\pi_{\alpha}} &=& \frac{e^2}{2s^2x} \left(  1 + s^2x^2 \right)
                                v_{\pi_{\alpha}}^{(1)} v_{\pi_{\alpha}}^{(1)} 
                            \left( v_\beta^L v_\gamma^V + v_\gamma^L v_\beta^V  \right) 
                            \,, \qquad \textrm{for $\beta \neq \gamma$} \\ 
     g^{\beta \beta}_{\pi_Z \pi_Z} &=& \frac{e^2}{s^2x} \left(  1 + s^2x^2 \right)
                                v_{\pi_Z}^{(1)} v_{\pi_Z}^{(1)} v_\beta^L v_\beta^V 
                           \,, \\ 
     g^{\beta \gamma}_{\pi_Z \pi_Z} &=& \frac{e^2}{2s^2x} \left(  1 + s^2x^2 \right)
                                v_{\pi_Z}^{(1)} v_{\pi_Z}^{(1)} 
                            \left( v_\beta^L v_\gamma^V + v_\gamma^L v_\beta^V  \right) 
                            \,, \qquad \textrm{for $\beta \neq \gamma$} \\  
     g^{\beta \beta}_{\pi_{\rho^0} \pi_{\rho^0}} &=& \frac{e^2}{s^2x} \left(  1 + s^2x^2 \right)
                                v_{\pi_{\rho^0}}^{(1)} v_{\pi_{\rho^0}}^{(1)} v_\beta^L v_\beta^V 
                           \,, \\ 
     g^{\beta \gamma}_{\pi_{\rho^0} \pi_{\rho^0}} &=& \frac{e^2}{2s^2x} \left(  1 + s^2x^2 \right) 
                                v_{\pi_{\rho^0}}^{(1)} v_{\pi_{\rho^0}}^{(1)} 
                            \left( v_\beta^L v_\gamma^V + v_\gamma^L v_\beta^V  \right) 
                            \,, \qquad \textrm{for $\beta \neq \gamma$} \,,                          
\end{eqnarray} 
where subscripts $\alpha,\beta$ and $\gamma$ 
denote $W^\pm$ and $\rho^\pm$ and $n$ does $A,Z$ and $\rho^0$. 
In obtaining the expressions for these couplings, we have used 
Eqn. (\ref{e-g0}). 
 The explicit expression for each of these couplings is shown 
in table~\ref{contact-couplings}. From table~\ref{contact-couplings}, 
in the isospin symmetric limit $s \to 0$ (or $c \to 1$) with $(e/s)$ fixed, 
we find 
\begin{eqnarray} 
g_{\pi_{\rho^+} \pi_{\rho^-}}^{ZZ} |_{c \to 1} 
&=& g_{\pi_{\rho^+} \pi_{\rho^-}}^{W^+W^-} |_{c \to 1} 
= g_{\pi_{\rho^0} \pi_{\rho^0}}^{W^+W^-} |_{c \to 1}
\,, \nonumber \\ 
g_{\pi_{\rho^+} \pi_{\rho^-}}^{\rho^0Z} |_{c \to 1} 
&=& 2g_{\pi_{\rho^+} \pi_{\rho^-}}^{W^+\rho^-} |_{c \to 1} 
= 2g_{\pi_{\rho^0} \pi_{\rho^0}}^{W^+\rho^-} |_{c \to 1} 
\,,  
\end{eqnarray}
up to an overall sign, 
as expected. 

\begin{table}

\begin{center} 

\begin{tabular}{|l||c|c|c|}
\hline  
$g_{\pi_{W^+} \pi_{W^-}}^{mn}$ & $n=A$ & $n=Z$ & $n=\rho^0$  \\ 
\hline \hline  
$m=A$ & $e^2$ & $\frac{e^2(c^2-s^2)}{sc}$ & $\frac{e^2(c^2-s^2)^2}{4s^2c^2}$ \\
\hline 
$m=Z$ & $\frac{e^2(c^2-s^2)}{sc}$ & $\frac{e^2(c^2-s^2)^2}{4s^2c^2}$ & irrelevant \\ 
\hline 
$m=\rho^0$ & irrelevant & irrelevant & irrelevant \\  
\hline 
\end{tabular}
\end{center}

\begin{center} 
\begin{tabular}{|l||c|c|c|}
\hline 
 $g^{mn}_{\pi_{\rho^+} \pi_{\rho^-}}$ & $n=A$ & $n=Z$ & $n=\rho^0$ \\ 
\hline \hline  
 $m=A$ & $e^2$& $\frac{e^2(c^2-s^2)}{sc} \left( 1 + \frac{x^2}{4(c^2-s^2)} \right)$ & $\frac{e^2(c^2-s^2)^2}{4s^2c^2} \left( 1 + \frac{x^2}{2(c^2-s^2)}  \right) $ \\ 
\hline 
 $m=Z$ & $ \frac{e^2(c^2-s^2)}{sc} \left( 1 + \frac{x^2}{4(c^2-s^2)} \right) $ 
& $ \frac{e^2(c^2-s^2)^2}{4s^2c^2} \left( 1 + \frac{x^2}{2(c^2-s^2)}  \right) $ & $\frac{e^2(c^2-s^2)}{2s^2cx}$ \\ 
\hline 
 $m=\rho^0$ & $\frac{e^2}{sx}$ & $\frac{e^2(c^2-s^2)}{2s^2cx} $ & irrelevant \\
\hline   
\end{tabular}
\end{center}

\begin{minipage}[t]{200pt}

\begin{center} 

\begin{tabular}{|l||c|c|} 
\hline 
$g_{\pi_{W^+}\pi_{W^-}}^{\alpha \beta}$ & $\beta=W^-$ & $\beta=\rho^-$ \\ 
 \hline \hline 
$\alpha=W^+$  & $\frac{e^2}{4s^2}$ & irrelevant \\ 
\hline 
$\alpha=\rho^+$ &irrelevant & irrelevant \\ 
 \hline 
\end{tabular}

\end{center}

\end{minipage} 

\begin{minipage}[t]{420pt}

\begin{flushright} 
\vspace{-50pt}
\begin{tabular}{|l||c|c|} 
\hline 
$g_{\pi_{\rho^+}\pi_{\rho^-}}^{\alpha \beta}$ & $\beta=W^-$ & $\beta=\rho^-$ \\
 \hline \hline 
$\alpha=W^+$ & $\frac{e^2}{4s^2} \left( 1 + \frac{x^2 ( -2c^2 + 3)}{2} \right)$ & $\frac{e^2}{4s^2x}$ \\ 
\hline 
$\alpha=\rho^+$ & $\frac{e^2}{4s^2x}$ & irrelevant \\ 
 \hline 
\end{tabular}

\end{flushright}

\end{minipage}

\begin{minipage}[t]{200pt}
\vspace{10pt}
\begin{center} 

\begin{tabular}{|l||c|c|} 
\hline 
$g_{\pi_{Z}\pi_{Z}}^{\alpha \beta}$ & $\beta=W^-$ & $\beta=\rho^-$ \\ 
 \hline \hline 
$\alpha=W^+$  & $\frac{e^2}{4s^2}$ & irrelevant \\ 
\hline 
$\alpha=\rho^+$ & irrelevant & irrelevant \\ 
 \hline 
\end{tabular}

\end{center}

\end{minipage}

\begin{minipage}[t]{430pt}

\begin{flushright} 
\vspace{-50pt}
\begin{tabular}{|l||c|c|} 
\hline 
$g_{\pi_{\rho^0}\pi_{\rho^0}}^{\alpha \beta}$ & $\beta=W^-$ & $\beta=\rho^-$ \\
 \hline \hline 
$\alpha=W^+$  & $\frac{e^2}{4s^2} \left( 1 + \frac{x^2( - 2c^2 + 4 -  \frac{1}{c^2})}{2} \right)$ & $\frac{e^2}{4s^2x}$ \\ 
\hline 
$\alpha=\rho^+$ & $\frac{e^2}{4s^2x} $ & irrelevant \\ 
 \hline 
\end{tabular}

\end{flushright}

\end{minipage} 

\caption{ The Goldstone boson couplings between two gauge bosons and two Goldstone bosons. 
The vertices denoted as ``irrelevant" do not generate leading log corrections to the gauge boson self-energy functions we are concerned with. } 
\label{contact-couplings}
\end{table}

\subsection{Three-Point Vertices among the Gauge Bosons: $\mathcal{L}_{AAA}$} 
\label{sec:3pt}

Rewriting Eqn. (\ref{re aaa}) in terms of mass eigenstate fields yields
\begin{eqnarray} 
  \mathcal{L}_{AAA} & =  & i \Bigg\{ \sum_{n = A,Z,\rho^0} 
                             g_{W^+W^-}^n \left( W^+_{\mu\nu} W^{\mu -} n^\nu  
                                    + \frac{1}{2} n_{\mu\nu} W^{+ \mu} W^{- \nu}  \right) 
                            \nonumber \\ 
                                 && +  
                                 g_{W^+ \rho^-}^n \left( 
                                   \left( W_{\mu\nu}^+ \rho^{\mu -} +  \rho_{\mu\nu}^+ W^{\mu -} \right) n^\nu  
                                          + \frac{1}{2} n_{\mu\nu} \left( W^{\mu +} \rho^{\nu -} + \rho^{\mu +} W^{\nu -} \right) 
                                                       \right) 
                            \nonumber \\ 
                                 && + g_{\rho^+\rho^-}^n \left( \rho^+_{\mu\nu} \rho^{\mu -} n^\nu  
                                    + \frac{1}{2} n_{\mu\nu} \rho^{+ \mu} \rho^{- \nu}  \right) 
                             \Bigg\}+ {\rm h.c.} \,,  
\end{eqnarray}
where $\mathcal{A}_{\mu\nu}=\partial_\mu \mathcal{A}_\mu - \partial_\nu \mathcal{A}_\mu$ ($\mathcal{A}=A,Z,\rho^{\pm,0},W^\pm$). 
These couplings are expressed by using the wavefunctions 
given in Eqns.(\ref{wavefunc:Wrhopm}) and (\ref{wavefunc:AZrho0}) as 
\begin{eqnarray} 
  g_{W^+W^-}^n &=& \frac{e}{s} \left( 1 + \frac{1}{2}s^2x^2 \right) \left[ (v_W^L)^2 v_n^L + \frac{1}{x} (v_W^V)^2 v_n^V \right] 
  \,, \\
  g_{W^+\rho^-}^n &=& \frac{e}{s} \left( 1 + \frac{1}{2}s^2x^2 \right) \left[ v_W^L v_{\rho^\pm}^L v_n^L + \frac{1}{x} v_W^V v_{\rho^\pm}^V v_n^V \right] 
  \,, \\
  g_{\rho^+ \rho^-}^n &=& \frac{e}{s} \left( 1 + \frac{1}{2}s^2x^2 \right)  \left[ (v_{\rho^\pm}^L)^2 v_n^L + \frac{1}{x} (v_{\rho^\pm}^V)^2 v_n^V \right] 
  \,, 
\end{eqnarray}
for $n=A,Z,\rho$. 
In obtaining the expressions for these couplings, we have used 
Eqn. (\ref{e-g0}). 
The explicit expression for each  of these couplings is shown in table~\ref{3-point-gauge-couplings}. 
From table~\ref{3-point-gauge-couplings}, 
in the isospin symmetric limit $s \to 0$ (or $c \to 1$) with $(e/s)$ fixed, 
we find 
\begin{equation} 
g_{\rho^+ \rho^-}^Z|_{c\to 1} 
= g_{W^+ \rho^-}^{\rho^0}|_{c \to 1}
\,, \qquad  
g_{W^+W^-}^{\rho^0}|_{c\to 1} = g_{W^+\rho^-}^Z|_{c\to 1}
\, ,
\end{equation}
as expected. 

\begin{table}

\begin{center} 
\begin{tabular}{|c||c|c|c|} 
\hline 
$n$  & $g_{W^+W^-}^n $ & $g_{W^+ \rho^-}^n$ & $g_{\rho^+ \rho^-}^n$  \\ 
  \hline\hline  
$A$  & $e$ & 0  & $e$ \\ 
\hline 
$Z$  & $\frac{ec}{s}$  & $- \frac{ex}{4sc}$ & $\frac{e(c^2-s^2)}{2sc} \left(1 + \frac{x^2(4- \frac{1}{c^2} ) }{8(c^2-s^2)} \right)$ \\ 
\hline 
$\rho^0$ & - $\frac{ex}{4s}$ & $\frac{e}{2s} \left( 1 + \frac{x^2( - 4c^2 + 8 - \frac{1}{c^2})}{8} \right)$ & $\frac{e}{sx}$  \\
\hline 
\end{tabular}
\end{center} 
\caption{ The three-point couplings among the gauge fields. }
\label{3-point-gauge-couplings}
\end{table}

\subsection{Four-Point Vertices among the Gauge Bosons: $\mathcal{L}_{AAAA}$}

Rewriting Eqn. (\ref{re aaaa}) in terms of the mass eigenstate fields yields
\begin{eqnarray} 
 \mathcal{L}_{AAAA} & =  & \sum_{n=A,Z,\rho^0} \Bigg\{ 
                                  g_{W^+ W^-}^{nn} (W_\mu^+ W_\nu^- n^\mu n^\nu  
                                    - W_\mu^+ W_\mu^- n^\nu m^\nu ) 
                                    \nonumber \\ 
                                    && 
                                  \hspace{55pt}  + g_{\rho^+ \rho^-}^{nn} (\rho_\mu^+ \rho_\nu^- n^\mu n^\nu  
                                    - \rho_\mu^+ \rho_\mu^- n^\nu m^\nu) \Bigg\}
                                    \nonumber \\ 
                                  && + \sum_{n,m=A,Z,\rho^0}^{ n \neq m} \Bigg\{ 
                                  g_{W^+ W^-}^{nm} 
                                    \left( W_\mu^+ W_\nu^- (n^\mu m^\nu + m^\mu n^\nu) 
                                       - 2 W_\mu^+ W_\mu^- n^\nu m^\nu \right) 
                                  \nonumber \\ 
                                  && \hspace{55pt}   
                                  + g_{\rho^+ \rho^-}^{nm} 
                                    \left( \rho_\mu^+ \rho_\nu^- (n^\mu m^\nu + m^\mu n^\nu) 
                                       - 2 \rho_\mu^+ \rho_\mu^- n^\nu m^\nu \right) \Bigg\} 
                                  \nonumber \\ 
                                  && + g_{W^+ \rho^-}^{\rho^0 \rho^0} \left( \rho_\mu^0 \rho^0_\nu W_{\mu +} \rho^{\nu -} 
                                                      - \rho_\mu^0 \rho^{\mu 0} W_\nu^+ \rho^{\nu -} + {\rm h.c.}  
                                                      \right) 
                                  \nonumber \\ 
                                  && + g_{W^+ \rho^-}^{\rho^+ \rho^-} \left( \rho_\mu^+ \rho_\nu^- W^{\mu +} \rho^{\nu -} 
                                                      - \rho_\mu^+ \rho^{\mu -} W_\nu^+ \rho^{\nu -} + {\rm h.c.}  \right)
                                  \nonumber \\ 
                                  && + g_{W^+W^-}^{\rho^+\rho^-} \left( 2 W_\mu^+ W_\nu^- \rho^{\mu +} \rho^{\nu -} 
                                      - W_\mu^+ W_\nu^- \rho^{\mu -}\rho^{\nu +} 
                                      - W_\mu^+ W^{\nu -} \rho_\nu^+ \rho^{\nu -}  \right)  
                                     \nonumber \\ 
                                  && + g_{W^+W^-}^{W^+W^-} W_\mu^+ W_\nu^- \left( W^{\mu +} W^{\nu -} 
                                      - W^{\mu -} W^{\nu +} \right) 
                                     \,,    
\end{eqnarray}
where we have neglected terms which do not affect
the gauge boson self-energy functions at the one-loop level. 
These relevant couplings are expressed by using the wavefunctions 
given in Eqns.(\ref{wavefunc:Wrhopm}) and (\ref{wavefunc:AZrho0}) as 
\begin{eqnarray} 
   g_{W^+ W^-}^{nm} &=& \frac{e^2}{s^2} \left( 1 + s^2x^2 \right) \left[ (v_{W^\pm}^L)^2 v_n^L v_m^L 
                            + \frac{1}{x^2} (v_{W^\pm}^V)^2 v_n^V v_m^V \right]
\,, \\ 
   g_{\rho^+ \rho^-}^{nm} &=& \frac{e^2}{s^2} \left( 1 + s^2x^2 \right)  \left[ (v_{\rho^\pm}^L)^2 v_n^L v_m^L 
                            + \frac{1}{x^2} (v_{\rho^\pm}^V)^2 v_n^V v_m^V \right]
\,, \\ 
   g_{W^+ \rho^-}^{\rho^0\rho^0} &=& \frac{e^2}{s^2} \left( 1 + s^2x^2 \right)  \left[ v_{W^\pm}^L v_{\rho^\pm}^L (v_{\rho^0}^L)^2  
                            + \frac{1}{x^2} v_{W^\pm}^V v_{\rho^\pm}^V (v_{\rho^0}^V)^2  \right] 
\,, \\ 
   g_{W^+ \rho^-}^{\rho^+ \rho^-} &=& \frac{e^2}{s^2} \left( 1 + s^2x^2 \right)  \left[  v_{W^\pm}^L (v_{\rho^\pm}^L)^3   
                            + \frac{1}{x^2} v_{W^\pm}^V (v_{\rho^\pm}^V)^3   \right] 
\,, \\ 
    g_{W^+ W^-}^{\rho^+ \rho^-} &=& \frac{e^2}{s^2} \left( 1 + s^2x^2 \right)  \left[ (v_{W^\pm}^L)^2 (v_{\rho^\pm}^L)^2    
                            + \frac{1}{x^2} (v_{W^\pm}^V)^2 (v_{\rho^\pm}^V)^2   \right]                           
\,, \\
    g_{W^+ W^-}^{W^+ W^-} &=& \frac{e^2}{2s^2} \left( 1 + s^2x^2 \right)  \left[ (v_{W^\pm}^L)^4    
                            + \frac{1}{x^2} (v_{W^\pm}^V)^4 \right]   
\,,  
\end{eqnarray} 
for $m, n=A,Z,\rho$. 
In obtaining the expressions for these couplings, we have used 
Eqn. (\ref{e-g0}). 
The explicit expression for each  of these couplings is shown 
in table~\ref{4-point-gauge-couplings}.  
From table~\ref{4-point-gauge-couplings}, 
in the isospin symmetric limit $s \to 0$ (or $c \to 1$) with $(e/s)$ fixed, 
we find 
\begin{eqnarray} 
g_{W^+W^-}^{\rho^0\rho^0}|_{c \to 1} 
&=& g_{W^+W^-}^{\rho^+ \rho^-}|_{c \to 1} 
= g_{ZZ}^{\rho^+\rho^-}|_{c \to 1}
\,, \nonumber \\ 
g_{\rho^+\rho^-}^{\rho^0 Z}|_{c \to 1} 
&=& g_{W^+ \rho^-}^{\rho^0 \rho^0} = g_{W^+ \rho^-}^{\rho^+ \rho^-}
\,, 
\end{eqnarray} 
as expected. 
\begin{table}

\begin{center} 
\begin{tabular}{|l||c|c|c|}
\hline  
$g_{W^+ W^-}^{mn}$ & $n=A$ & $n=Z$ & $n=\rho^0$ \\  
  \hline\hline  
$m=A$ & $e^2$  & $\frac{e^2c}{s}$ & irrelevant \\  
\hline 
$m=Z$ & $\frac{e^2c}{s}$ & $\frac{e^2c^2}{s^2}$ & irrelavant \\ 
\hline 
$m=\rho^0$ & irrelevant & irrelevant & $\frac{e^2}{4s^2} \left( 1 + \frac{x^2( - 4c^2 + 9 - \frac{1}{c^2})}{4} \right)$ \\ 
\hline 
\end{tabular}
\end{center}

\begin{center} 

\begin{tabular}{|l||c|c|} 
\hline 
$g_{\rho^+ \rho^-}^{mn}$ & $n=A$ & $n=Z$ \\  
  \hline\hline  
$m=A$ & $e^2$  &  $\frac{e^2(c^2-s^2)}{2sc}\left( 1 + \frac{x^2(4-\frac{1}{c^2})}{8(c^2-s^2)} \right)$ \\  
\hline 
$m=Z$ & $\frac{e^2(c^2-s^2)}{2sc}\left( 1 + \frac{x^2(4-\frac{1}{c^2})}{8(c^2-s^2)} \right)$ 
    & $\frac{e^2(c^2-s^2)^2}{4s^2c^2} \left(1 + \frac{x^2(8c^2 - 5 + \frac{1}{c^2} ) }{4(c^2-s^2)^2} \right)$ 
    \\ 
\hline 
$m=\rho^0$ & $\frac{e^2}{sx}$ & $\frac{e^2(c^2-s^2)}{2s^2cx}$ \\ 
\hline 
\end{tabular}
\end{center} 
\begin{center} 
 \begin{tabular}{cccc}  
   $g_{W^+ \rho^-}^{\rho^0\rho^0} = \frac{e^2}{2s^2x}\,, $ & $g_{W^+ \rho^-}^{\rho^+\rho^-} = \frac{e^2}{2s^2 x}\,,$ 
   & $g_{W^+ W^-}^{\rho^+ \rho^-} = \frac{e^2}{4s^2} \left( 1 + (2-c^2)x^2 \right)\,, $ & $g_{W^+W^-}^{W^+W^-}=\frac{e^2}{2s^2}~.$  
 \end{tabular}
\end{center}
\caption{ The four-point couplings among the gauge fields. 
The vertices denoted as ``irrelevant" do not generate leading log corrections to the gauge boson self-energy functions. }
\label{4-point-gauge-couplings}
\end{table}

\subsection{FP Ghost Terms in the Mass Eigenstate Basis} 

Rewriting Eqn. (\ref{re fpint}) in terms of mass eigenstate fields yields
\begin{eqnarray} 
   \mathcal{L}_{FP}^{\rm int} & =  & 
                                 i \sum_{n=A,Z,\rho^0} \Bigg[ 
                 g_{W^+ W^-}^n n_\mu \partial^\mu \bar{C}_{W^+} C_{W^-} 
                 + g_{\rho^+\rho^-}^n n_\mu \partial^\mu \bar{C}_{\rho^+} C_{\rho^-} 
                 \nonumber \\ 
                              && \hspace{60pt} + g_{W^+ \rho^-}^n n_\mu (\partial^\mu \bar{C}_{W^+} C_{\rho^-} + \partial^\mu \bar{C}_{\rho^+} C_{W^-})  
                                                       \Bigg] 
                  + {\rm h.c.} 
                  \,. 
\end{eqnarray}
Note that these couplings are equal to 
the three-point couplings among the gauge bosons 
listed in table~\ref{3-point-gauge-couplings} 
due to gauge invariance.

\section{Feynman Integral Formulae}
\label{formula}

We define the following Feynman integrals: 
\begin{eqnarray}
i F_1^{\mu\nu} (M_A, M_B; p^2) 
&\equiv& 
\int \frac{d^4 k}{(2\pi)^4} \frac{(2k+p)^{\mu}(2k+p)^{\nu}}{\left[ k^2 - M_A^2 \right]
\left[ (k+p)^2 - M^2_B \right]} 
\,, \label{l1} \\
i F_2(M_A, M_B; p^2) 
&\equiv& 
\int \frac{d^4 k}{(2\pi)^4} \frac{1}{\left[ k^2 - M_A^2 \right]
\left[ (k+p)^2 - M^2_B \right]}
\,, \label{l2} \\
i F_3(M) 
&\equiv& 
\int \frac{d^4 k}{(2\pi)^4} \frac{1}{\left( k^2 - M^2 \right)}
\,. 
\end{eqnarray}
By introducing Feynman parameters, and performing dimensional regularization, 
these integrals are evaluated as   
\begin{eqnarray}
(4\pi)^2 F_1^{\mu\nu} (M_A, M_B; p^2)
&=& g^{\mu\nu}
\Bigg[ 
\left( M_A^2 + M_B^2 - \frac{1}{3} p^2 \right) \cdot \left( \frac{1}{\bar{\epsilon}} + 1 \right)  
\nonumber \\ 
&& - 2 \int_0^1 dx \Delta_A^B (p^2) \log \Delta_{A}^{B}(p^2) \Bigg]  
\nonumber \\ 
&& 
+ \left(  p^\mu p^\nu \, {\rm term} \right) 
\,,  
\label{l1 ev} \\
(4\pi)^2 F_2(M_A, M_B; p^2)
&=& 
\frac{1}{\bar{\epsilon}}  
- \int_0^1 dx \log \Delta_{A}^{B}(p^2) 
\,, \label{l2 ev} \\
(4\pi)^2 F_3(M) 
&=& 
M^2 \cdot \frac{1}{\bar{\epsilon}}  
- \left( M^2 \log M^2 + M^2 \right) 
\,, \label{l3 ev} 
\end{eqnarray}
where 
\begin{eqnarray} 
\Delta_A^B(p^2) 
&=& 
x M_A^2 + (1-x) M^2_B - x(1-x) p^2 
\,, 
\\
\frac{1}{\bar{\epsilon}} 
&=& \frac{2}{\epsilon} - \gamma + \log 4\pi 
\,. 
\end{eqnarray}
 Interpreting the results in terms of a dimensional cutoff representing the cutoff
of the effective theory, we make the replacement
\begin{equation}
\frac{1}{\bar{\epsilon}} \rightarrow \log \Lambda^2\, .
 \end{equation}
Equivalently, the replacement above may be viewed as evaluating the counterterms, which cancel divergences, renormalized at the scale of the cutoff.

In applying these formulae to extract the leading-log corrections to
the $S$ and $T$ parameters arising from loops of $\rho$-mesons,  we encounter 
two separate cases. In one case, the loop-corrections involve only propagating $\rho$-mesons .
We refer to this as the degenerate heavy mass case, 
in which $M^2_A \simeq M^2_B \gg |p^2|= {\cal O}(M^2_W)$.
In the other case -- which we refer to as the hierarchical mass case --
loop-corrections involve one $\rho$-meson and one light gauge-boson, and $M^2_A \gg M^2_B
\simeq |p^2| = {\cal O}(M^2_W)$.

We first examine the degenerate heavy mass case in which,
from Eqns. (\ref{l1 ev})-(\ref{l3 ev}), 
we find the approximate formulae, 
\begin{eqnarray} 
(4\pi)^2 F_1^{\mu\nu} (M_A, M_B; p^2) 
& \simeq &  
g^{\mu\nu} \Bigg[ \left(M_A^2 +M_B^2 - \frac{1}{3} p^2 \right) 
\log \frac{\Lambda^2}{M_A^2} 
\nonumber \\ 
&& 
+ M_A^2 + M_B^2 + \frac{1}{3}p^2 
+ \mathcal{O}(p^4/M_A^2) \Bigg]
\nonumber \\ 
&& 
+ \left(  p^\mu p^\nu \, {\rm term} \right) 
\,, \label{f1} \\ 
(4\pi)^2 F_2(M_A, M_B; p^2) 
& \simeq & 
\log \frac{\Lambda^2}{M_A^2} 
+ \frac{1}{6}\frac{p^2}{M_A^2} + \mathcal{O}(p^4/M_A^4)
\,, \\ 
(4\pi)^2 F_3(M_A) 
& = & 
M_A^2 \log \frac{\Lambda^2}{M_A^2} - M_A^2 
\,. \label{f3} 
\end{eqnarray}

Consider next the hierarchical mass case.
In this case, 
the expression for the function $F_3$ 
remains the same as in Eqn. (\ref{f3}). 
In order to derive the approximate formulae for 
the functions $F_1^{\mu\nu}$ and $F_2$, 
we may rewrite 
the expressions (\ref{l1 ev}) and (\ref{l2 ev}) 
as follows: 
\begin{eqnarray} 
(4\pi)^2 F_1^{\mu\nu}(M_A, M_B; p^2) 
&=& g^{\mu\nu} \Bigg[ \left( M_A^2 + M_B^2 - \frac{1}{3} p^2 \right) 
\log\frac{\Lambda^2}{M_A^2} + M_A^2 + M_B^2 - \frac{1}{3}p^2
\nonumber \\ 
&& 
- 2 \int_0^1 dx \left( x M_A^2 + (1-x)M_B^2 - x(1-x)p^2 \right) 
\nonumber \\ 
&&
\hspace{35pt} \times
\log \left( x + (1-x) \frac{M_B^2}{M_A^2} -x(1-x) \frac{p^2}{M_A^2} \right) 
\Bigg]
\nonumber \\ 
&& 
+ \left(  p^\mu p^\nu \, {\rm term} \right) 
\,, \\ 
(4\pi)^2 F_2(M_A, M_B; p^2) 
&=& 
\log\frac{\Lambda^2}{M_A^2} 
- \int_0^1 dx 
\log \left( x + (1-x) \frac{M_B^2}{M_A^2} -x(1-x) \frac{p^2}{M_A^2} \right) 
\,.  \nonumber \\ 
\end{eqnarray}
Expanding the right hand sides in terms of $1/M_A^2$, 
we find 
\begin{eqnarray} 
(4\pi)^2 F_1^{\mu\nu}(M_A, M_B; p^2) 
&=& g^{\mu\nu} \Bigg[ \left( M_A^2 + M_B^2 - \frac{1}{3} p^2 \right) 
\log\frac{\Lambda^2}{M_A^2} 
+ \frac{3}{2} (M_A^2+M_B^2-\frac{5}{18}p^2)
\Bigg] 
\nonumber \\ 
&& 
+ \mathcal{O}(\frac{M_B^4}{M_A^2})
+ \left(  p^\mu p^\nu \, {\rm term} \right) 
\,, \label{f1:h-approx} \\ 
(4\pi)^2 F_2(M_A, M_B; p^2) 
&=& 
\log\frac{\Lambda^2}{M_A^2} + 1 
+ \frac{M_B^2}{M_A^2} \log\frac{M_B^2}{M_A^2} 
+ \frac{1}{2} \frac{p^2}{M_A^2} 
\nonumber \\ 
&& 
+ \mathcal{O}(\frac{M_B^4}{M_A^4} \log\frac{M_B^2}{M_A^2})
\,, \label{f2:h-approx} 
\end{eqnarray}
where we have used \cite{Aoki:1982ed}
\begin{eqnarray}
\int_0^1 dx \Delta_1^{\epsilon_1} (\epsilon_2) 
&=& - 1 - \epsilon_1 \log \epsilon_1 - \frac{1}{2} \epsilon_2 + \cdots, 
\\ 
\int_0^1 dx x \Delta_1^{\epsilon_1} (\epsilon_2) 
&=& - \frac{1}{4} + \frac{1}{2} \epsilon_1 
- \frac{1}{6} \epsilon_2 + \cdots, 
\\ 
\int_0^1 dx (1-x) \Delta_1^{\epsilon_1} (\epsilon_2) 
&=& - \frac{3}{4} - \epsilon_1 \log \epsilon_1 
- \frac{1}{2} \epsilon_1 
- \frac{1}{3} \epsilon_2 + \cdots, 
\\ 
\int_0^1 dx x(1-x) \Delta_1^{\epsilon_1} (\epsilon_2) 
&=& - \frac{5}{36} 
+ \frac{1}{3} \epsilon_1 
- \frac{1}{12} \epsilon_2 + \cdots \, , 
\end{eqnarray} 
with $\Delta_1^{\epsilon_1}(\epsilon_2)=x \cdot 1 
+ (1-x) \epsilon_1 -x(1-x)\epsilon_2$. 
Note the large logarithm, 
$\log(M_B^2/M_A^2)$, in the expression for $F_2$.

As a sample application of Eqns.(\ref{f1:h-approx}) and (\ref{f2:h-approx}), 
consider evaluating the amplitudes $(M)_{ZZ}$ and $(N)_{ZZ}$, 
from diagrams (M) and (N) 
in Fig. \ref{selfenergy} , 
in the limit $M_{\rho^\pm}^2 \gg M_W^2$: 
\begin{eqnarray} 
 (M)_{ZZ} &\simeq& 
 \frac{ie^2}{8s^2c^2}
 \Bigg\{ 1 + \frac{x^2\left( -8c^2 + 8 -\frac{1}{c^2}\right)}{4} \Bigg\} 
 F_1^{\mu\nu}(M_{\rho^\pm}, M_W; p^2) + \mathcal{O}(\alpha x^2 M_W^2)
 \nonumber \\
  &\simeq &
\frac{ie^2}{8s^2c^2} 
\Bigg[  F_1^{\mu\nu}(M_{\rho^\pm}, M_W; p^2) + \frac{x^2\left( -8c^2 + 8 -\frac{1}{c^2}\right)}{4} 
 g^{\mu\nu} F_3(M_{\rho^\pm}) \Bigg] 
\nonumber \\ 
&& 
 + \mathcal{O}(\alpha x^2 M_W^2) 
\,,  \\ 
 (N)_{ZZ} & \simeq & 
 \frac{-i e^2 g^{\mu\nu}}{2s^2c^2} M_{\rho^\pm}^2 
\Bigg\{ 1 -  \frac{x^2\left( 8c^2 - 6 + \frac{1}{c^2} \right)}{4} \Bigg\}  
F_2(M_{\rho^\pm}, M_W; p^2) + \mathcal{O}(\alpha x^2 M_W^2) 
\nonumber \\ 
 &\simeq &
  \frac{-i e^2 g^{\mu\nu}}{2s^2c^2} 
\Bigg[ F_3(M_{\rho^\pm}) - F_3(M_W) - \frac{x^2\left( 8c^2 - 7 + \frac{1}{c^2} \right)}{4} 
F_3(M_{\rho^\pm})  \Bigg] 
\nonumber \\ 
&& 
+ \mathcal{O}(\alpha x^2 M_W^2) 
\,,  
\end{eqnarray}
where $x^2 \approx 4M_W^2/M_{\rho^\pm}^2$. 
It is easy to see that, in the leading log approximation, 
these expressions precisely equal
Eqns. (\ref{eq:cmzz}) and (\ref{eq:nzz}) in Appendix \ref{feynman-nc}.

\section{Feynman Graph Results: Neutral Gauge Bosons}
\label{feynman-nc}

In this appendix, we present the results of each contribution to the neutral gauge-boson self-energy functions $\Pi_{AA,ZA,ZZ}$, as shown in Fig. \ref{selfenergy}.

\subsection{ Photon Self-Energy Amplitude $ \Pi_{AA}$}

\begin{eqnarray} 
(A)_{\gamma\gamma} 
&=& \frac{ie^2g^{\mu\nu}}{(4\pi)^2} 
\Bigg[ 9 M_{\rho^\pm}^2 + \frac{19}{6} p^2 \Bigg] 
\log \frac{\Lambda^2}{M_{\rho^\pm}^2} 
         \,, \\ 
(B)_{\gamma\gamma} 
&=& \frac{ie^2g^{\mu\nu}}{(4\pi)^2} 
\Bigg[ 9 M_W^2 + \frac{19}{6} p^2 \Bigg] 
\log \frac{\Lambda^2}{M_W^2} 
         \,, \\ 
(C)_{\gamma\gamma} 
&=& \frac{ie^2g^{\mu\nu}}{(4\pi)^2} 
\Bigg[ - 6 M_{\rho^\pm}^2 \Bigg] 
\log \frac{\Lambda^2}{M_{\rho^\pm}^2} 
         \,, \\ 
(D)_{\gamma\gamma} 
&=&  \frac{ie^2g^{\mu\nu}}{(4\pi)^2} 
\Bigg[- 6 M_W^2 \Bigg] 
\log \frac{\Lambda^2}{M_W^2} 
         \,, \\ 
(E)_{\gamma\gamma} 
&=& \frac{ie^2g^{\mu\nu}}{(4\pi)^2} 
\Bigg[ 2 M_{\rho^\pm}^2 - \frac{1}{3} p^2 \Bigg] 
\log \frac{\Lambda^2}{M_{\rho^\pm}^2}
         \,, \\ 
(F)_{\gamma\gamma} 
&=&  \frac{ie^2g^{\mu\nu}}{(4\pi)^2} 
\Bigg[2 M_W^2 - \frac{1}{3} p^2 \Bigg] 
\log \frac{\Lambda^2}{M_W^2}
         \,, \\ 
(G)_{\gamma\gamma} 
&=& \frac{ie^2g^{\mu\nu}}{(4\pi)^2} 
\Bigg[- 2 M_{\rho^\pm}^2 \Bigg] 
\log \frac{\Lambda^2}{M_{\rho^\pm}^2} 
         \,, \\ 
(H)_{\gamma\gamma} 
&=&  \frac{ie^2g^{\mu\nu}}{(4\pi)^2} 
\Bigg[- 2 M_W^2 \Bigg] 
\log \frac{\Lambda^2}{M_W^2}
         \,,\\ 
(I)_{\gamma\gamma} 
&=&  \frac{ie^2g^{\mu\nu}}{(4\pi)^2} 
\Bigg[- 2 M_{\rho^\pm}^2 \Bigg] 
\log \frac{\Lambda^2}{M_{\rho^\pm}^2}
         \,,\\  
(J)_{\gamma\gamma} 
&=&  \frac{ie^2g^{\mu\nu}}{(4\pi)^2} 
\Bigg[ - 2 M_W^2 \Bigg] 
\log \frac{\Lambda^2}{M_W^2}
         \,,\\  
(K)_{\gamma\gamma} 
&=& \frac{ie^2g^{\mu\nu}}{(4\pi)^2} 
\Bigg[- M_{\rho^\pm}^2 + \frac{1}{6} p^2 \Bigg] 
\log \frac{\Lambda^2}{M_{\rho^\pm}^2}
         \,, \\  
(L)_{\gamma\gamma} 
&=& \frac{ie^2g^{\mu\nu}}{(4\pi)^2} 
\Bigg[- M_W^2 + \frac{1}{6} p^2 \Bigg] 
\log \frac{\Lambda^2}{M_W^2}
         \,,  
\end{eqnarray} 
where we have neglected  terms of $\mathcal{O}(\alpha x^2 M^2_W )$. 

\subsection{ Photon/Z boson Mixing Amplitude $ \Pi_{ZA}$}

\begin{eqnarray} 
(A)_{Z\gamma} 
&=& \frac{ie^2g^{\mu\nu}}{(4\pi)^2sc} \Bigg[  \frac{c^2-s^2}{4} 
\left( 18 M_{\rho^\pm}^2 + \frac{19}{3} p^2 \right) 
+ \frac{9}{4} \left( 4c^2 - 1 \right) M_Z^2 \Bigg] 
\log \frac{\Lambda^2}{M_{\rho^\pm}^2} 
         \,, \\ 
(B)_{Z\gamma} 
&=& \frac{ie^2g^{\mu\nu}}{(4\pi)^2sc} \Bigg[ 
c^2 \left( 9 M_W^2 + \frac{19}{6} p^2 \right) \Bigg] 
\log \frac{\Lambda^2}{M_W^2} 
         \,, \\ 
(C)_{Z\gamma} 
&=& \frac{ie^2g^{\mu\nu}}{(4\pi)^2sc} \Bigg[  
\frac{c^2-s^2}{4} \left( -12 M_{\rho^\pm}^2 \right) 
- \frac{3}{2} \left( 4c^2 - 1 \right) M_Z^2 \Bigg] 
\log \frac{\Lambda^2}{M_{\rho^\pm}^2} 
         \,, \\ 
(D)_{Z\gamma} 
&=&  \frac{ie^2g^{\mu\nu}}{(4\pi)^2sc} \Bigg[ - 6 c^2 M_W^2 \Bigg] 
\log \frac{\Lambda^2}{M_W^2} 
         \,, \\ 
(E)_{Z\gamma} 
&=& \frac{ie^2g^{\mu\nu}}{(4\pi)^2sc} \Bigg[ 
\frac{c^2-s^2}{4} \left( 4 M_{\rho^\pm}^2 - \frac{2}{3} p^2 \right) 
+ c^2 M_Z^2 \Bigg] 
\log \frac{\Lambda^2}{M_{\rho^\pm}^2}
         \,, \\ 
(F)_{Z\gamma} 
&=&  \frac{ie^2g^{\mu\nu}}{(4\pi)^2sc} 
\Bigg[ \frac{c^2-s^2}{4} \left( 4 M_W^2 - \frac{2}{3} p^2 \right) \Bigg] \log \frac{\Lambda^2}{M_W^2} 
         \,, \\ 
(G)_{Z\gamma} 
&=& \frac{ie^2g^{\mu\nu}}{(4\pi)^2sc} \Bigg[ 
\frac{c^2-s^2}{4} \left( -4 M_{\rho^\pm}^2 \right) 
- c^2 M_Z^2 \Bigg] 
\log \frac{\Lambda^2}{M_{\rho^\pm}^2}
         \,, \\ 
(H)_{Z\gamma} 
&=&  \frac{ie^2g^{\mu\nu}}{(4\pi)^2 sc} 
\Bigg[ \frac{c^2-s^2}{4} \left(- 4 M_W^2 \right) \Bigg] 
\log \frac{\Lambda^2}{M_W^2}
         \,,\\ 
(I)_{Z\gamma} 
&=& \frac{ie^2g^{\mu\nu}}{(4\pi)^2sc} \Bigg[ 
\frac{c^2-s^2}{4} \left( - 4 M_{\rho^\pm}^2 \right) 
 - \frac{1}{2} M_Z^2 \Bigg] 
\log \frac{\Lambda^2}{M_{\rho^\pm}^2}
         \,, \\  
(J)_{Z\gamma} 
&=&  \frac{ie^2g^{\mu\nu}}{(4\pi)^2sc} \Bigg[ 2 (1- c^2) M_W^2 \Bigg] 
\log \frac{\Lambda^2}{M_W^2}
         \,,\\  
(K)_{Z\gamma} 
&=& \frac{ie^2g^{\mu\nu}}{(4\pi)^2sc} \Bigg[ 
\frac{c^2-s^2}{4} \left( -2 M_{\rho^\pm}^2 + \frac{1}{3} p^2 \right) 
- \frac{1}{4} (4c^2-1) M_Z^2 \Bigg] 
\log \frac{\Lambda^2}{M_{\rho^\pm}^2}
         \,, \\  
(L)_{Z\gamma} 
&=& \frac{ie^2g^{\mu\nu}}{(4\pi)^2sc} \Bigg[ 
c^2 \left( - M_W^2 + \frac{1}{6} p^2 \right) \Bigg] 
\log \frac{\Lambda^2}{M_W^2}
         \,,  
\end{eqnarray} 
where we have used $x^2 M_{\rho^\pm}^2 \approx 4M_W^2$ and 
neglected terms of $\mathcal{O}(\alpha x^2 M^2_W )$.

\subsection{ Z Boson Self-Energy Amplitude $\Pi_{ZZ}$}

\begin{eqnarray} 
(A)_{ZZ} 
&=& \frac{ie^2g^{\mu\nu}}{(4\pi)^2 s^2c^2} \Bigg[ 
\frac{(c^2-s^2)^2}{4} \left( 9 M_{\rho^\pm}^2 + \frac{19}{6} p^2 \right) 
+ \left( 18c^2-\frac{27}{2} \right) M_W^2 
+\frac{9}{4}M_Z^2
\Bigg] \log \frac{\Lambda^2}{M_{\rho^\pm}^2} 
         \,, \nonumber \\ 
\\ 
(B)_{ZZ} 
&=& \frac{ie^2g^{\mu\nu}}{(4\pi)^2 s^2 c^2} \Bigg[c^4 \left( 
 9 M_W^2 + \frac{19}{6} p^2\right) 
\Bigg] \log \frac{\Lambda^2}{M_W^2} 
         \,, \\ 
(C)_{ZZ} 
&=& \frac{ie^2g^{\mu\nu}}{(4\pi)^2s^2c^2} \Bigg[ 
- \frac{3(c^2-s^2)^2}{2} M_{\rho^\pm}^2 
- \left( 12c^2 -\frac{15}{2} \right) M_W^2
- \frac{3}{2}M_Z^2
\Bigg] \log \frac{\Lambda^2}{M_{\rho^\pm}^2} 
         \,, \\ 
(D)_{ZZ} 
 &=& \frac{ie^2g^{\mu\nu}}{(4\pi)s^2c^2} 
\Bigg[ - 6 c^4 M_W^2 \Bigg] \log \frac{\Lambda^2}{M_W^2} 
         \,, \\ 
(E)_{ZZ} 
&=& \frac{ie^2g^{\mu\nu}}{(4\pi)^2s^2c^2} \Bigg[ 
\frac{(c^2-s^2)^2}{4} \left( 2 M_{\rho^\pm}^2 - \frac{1}{3} p^2 \right) 
 + (2c^2-1) M_W^2 
\Bigg] \log \frac{\Lambda^2}{M_{\rho^\pm}^2}
         \,, \\ 
(F)_{ZZ} 
&=&  \frac{ie^2g^{\mu\nu}}{(4\pi)^2s^2c^2} \Bigg[ 
\frac{(c^2-s^2)^2}{4} 
\left( 2 M_W^2 - \frac{1}{3} p^2 \right) 
\Bigg] \log \frac{\Lambda^2}{M_W^2}
         \,, \\ 
(G)_{ZZ} 
&=& \frac{ie^2}{(4\pi)^2s^2c^2} \Bigg[ 
- \frac{(c^2-s^2)^2}{2} M_{\rho^\pm}^2  
-  (2c^2-1) M_W^2 \Bigg] 
\log \frac{\Lambda^2}{M_{\rho^\pm}^2} 
         \,, \\ 
(H)_{ZZ} 
&=&  \frac{ie^2g^{\mu\nu}}{(4\pi)^2s^2c^2} \Bigg[ 
- \frac{(c^2-s^2)^2}{2} M_W^2 
\Bigg] \log \frac{\Lambda^2}{M_W^2}
         \,,\\ 
(I)_{ZZ} 
&=& \frac{ie^2g^{\mu\nu}}{(4\pi)^2s^2c^2} \Bigg[ 
- \frac{(c^2-s^2)^2}{2} M_{\rho^\pm}^2 
- M_W^2 + \frac{1}{2}M_Z^2  
\Bigg] 
\log \frac{\Lambda^2}{M_{\rho^\pm}^2}
         \,,\\  
(J)_{ZZ} 
&=&  \frac{ie^2g^{\mu\nu}}{(4\pi)^2s^2c^2} \Bigg[ 
- 2 (c^4-2c^2+1) M_W^2  
\Bigg] \log \frac{\Lambda^2}{M_W^2}
         \,,\\  
(K)_{ZZ} 
&=& \frac{ie^2g^{\mu\nu}}{(4\pi)^2s^2c^2} \Bigg[ 
\frac{(c^2-s^2)^2}{4} \left( - M_{\rho^\pm}^2 + \frac{1}{6} p^2 \right) 
- \left( 2c^2 - \frac{3}{2}  \right) M_W^2 
- \frac{1}{4}M_Z^2
\Bigg] \log \frac{\Lambda^2}{M_{\rho^\pm}^2}
         \,, \nonumber \\  
\\ 
(L)_{ZZ} 
&=& \frac{ie^2g^{\mu\nu}}{(4\pi)^2s^2c^2} \Bigg[ 
c^4 \left( - M_W^2 + \frac{1}{6} p^2 \right) 
\Bigg] \log \frac{\Lambda^2}{M_W^2}
         \,, 
\end{eqnarray}

\begin{eqnarray}
(M)_{ZZ} 
&=& \frac{ie^2g^{\mu\nu}}{(4\pi)^2s^2c^2} \Bigg[ 
\frac{1}{8} \left( M_{\rho^\pm}^2 - \frac{1}{3} p^2 \right) 
- \left( c^2 - \frac{9}{8}  \right) M_W^2 
-  \frac{1}{8}M_Z^2
\Bigg] \log \frac{\Lambda^2}{M_{\rho^\pm}^2}  
\,, \label{eq:cmzz} \\ 
(N)_{ZZ} 
&=& \frac{ie^2g^{\mu\nu}}{(4\pi)^2s^2c^2} \Bigg\{ 
\frac{1}{2} M_W^2 \log \frac{\Lambda^2}{M_W^2} 
+ \Bigg[ - \frac{1}{2} M_{\rho^\pm}^2 
+ \left( 4c^2 - \frac{7}{2}  \right)M_W^2 
+ \frac{1}{2} M_Z^2
\Bigg] \log \frac{\Lambda^2}{M_{\rho^\pm}^2} 
\Bigg\} \,,         
\nonumber \\ 
\label{eq:nzz} \\
(O)_{ZZ} 
&=& \frac{ie^2g^{\mu\nu}}{(4\pi)^2s^2c^2} \Bigg[ 
\frac{9}{4} M_W^2 \Bigg] \log \frac{\Lambda^2}{M_{\rho^\pm}^2} 
\,, \\ 
(P)_{ZZ} 
&=& \frac{ie^2g^{\mu\nu}}{(4\pi)^2s^2c^2} \Bigg[ 
 - \frac{1}{4} M_W^2 \Bigg] \log \frac{\Lambda^2}{M_{\rho^\pm}^2} 
\,,
\end{eqnarray} 
where we have neglected  terms of $\mathcal{O}(\alpha x^2 M^2_W )$ and 
used $M_W^2 \approx c^2 M_Z^2$ which is  valid to leading order in $x$.


\section{Feynman Graph Results: Charged Gauge Bosons}
\label{feynman-cc}

In this appendix, we present the results of each diagram contributing to the
charged gauge-boson self-energy function $\Pi_{WW}$, as shown in Fig. \ref{selfenergyW}.  

\begin{eqnarray} 
(A)_{WW} &=& 
\frac{ie^2g^{\mu\nu}}{(4 \pi)^2s^2} \Bigg[ 
(1-c^2)\Bigg(\frac{9}{2} M_W^2 + \frac{19}{6} p^2 \Bigg) \Bigg] 
\log \frac{\Lambda^2}{M_W^2} 
\,, \\
(B)_{WW} &=& 
\frac{ie^2g^{\mu\nu}}{(4 \pi)^2s^2} \Bigg[ 
c^2 \Bigg( \frac{9}{2} M_W^2 + \frac{9}{2} M_Z^2 
+ \frac{19}{6} p^2 \Bigg) \Bigg]  
\log \frac{\Lambda^2}{M_W^2} 
\,, \\
(C)_{WW} &=& 
\frac{ie^2g^{\mu\nu}}{(4 \pi)^2s^2} \Bigg[ 
\frac{1}{4} \Bigg( \frac{9}{2} M_{\rho^\pm}^2  + \frac{9}{2} M_{\rho^0}^2 + \frac{19}{6} p^2 \Bigg) 
+ \left( -9c^2 + 18  \right) M_W^2
-\frac{9}{4} M_Z^2
\Bigg] \log \frac{\Lambda^2}{M_{\rho^\pm}^2} 
\nonumber \\ 
&=& 
\frac{ie^2g^{\mu\nu}}{(4 \pi)^2s^2} \Bigg[ 
\frac{1}{4} \Bigg( 9 M_{\rho^\pm}^2 + \frac{19}{6} p^2 \Bigg) 
+ \left( -9c^2 + \frac{135}{8}  \right) M_W^2
-\frac{9}{8} M_Z^2
\Bigg] \log \frac{\Lambda^2}{M_{\rho^\pm}^2} 
\,, \nonumber \\ 
\label{eq:cww} \\
(D)_{WW} &=& 
\frac{ie^2g^{\mu\nu}}{(4 \pi)^2s^2} \Bigg[ 
- 3 M_W^2 \Bigg] 
\log \frac{\Lambda^2}{M_W^2} 
   \,, 
\end{eqnarray}

\begin{eqnarray}
(E)_{WW} &=& 
\frac{ie^2g^{\mu\nu}}{(4 \pi)^2s^2} \Bigg[ 
- 3 c^2 M_Z^2 \Bigg] 
\log \frac{\Lambda^2}{M_W^2} 
\,, \\
(F)_{WW} &=& 
\frac{ie^2g^{\mu\nu}}{(4 \pi)^2s^2} \Bigg[ 
- \frac{3}{4}  M_{\rho^\pm}^2 + 3 (c^2-2)M_W^2 
\Bigg] \log \frac{\Lambda^2}{M_{\rho^\pm}^2} 
\,, \\                              
(G)_{WW} &=& 
\frac{ie^2g^{\mu\nu}}{(4 \pi)^2s^2} \Bigg[ 
- \frac{3}{4} M_{\rho^0}^2 + \left( 3c^2 - \frac{27}{4} \right)M_W^2 
+ \frac{3}{4}M_Z^2
\Bigg]
\log \frac{\Lambda^2}{M_{\rho^\pm}^2}
\nonumber \\ 
&=&
\frac{ie^2g^{\mu\nu}}{(4 \pi)^2s^2} \Bigg[ 
- \frac{3}{4} M_{\rho^\pm}^2 + 3 (c^2-2)M_W^2 
\Bigg]
\log \frac{\Lambda^2}{M_{\rho^\pm}^2} 
\,, \label{eq:gww}\\  
(H)_{WW} &=& 
\frac{ie^2g^{\mu\nu}}{(4 \pi)^2s^2} \Bigg[ 
- \frac{(1-c^2)}{2} \Bigg( M_W^2 - \frac{1}{3} p^2 \Bigg) 
\Bigg]  \log \frac{\Lambda^2}{M_W^2} 
\,, \\
(I)_{WW} &=& 
\frac{ie^2g^{\mu\nu}}{(4 \pi)^2s^2} \Bigg[ 
- \frac{c^2}{2} \Bigg( M_W^2 + M_Z^2 - \frac{1}{3} p^2 \Bigg) 
\Bigg]  \log \frac{\Lambda^2}{M_W^2} 
\,, \\
(J)_{WW} &=& 
\frac{ie^2g^{\mu\nu}}{(4 \pi)^2s^2} \Bigg[ 
- \frac{1}{8} \left( M_{\rho^\pm}^2 + M_{\rho^0}^2 - \frac{1}{3} p^2 \right) 
+ \left( c^2 - 2  \right) M_W^2 
+ \frac{1}{4} M_Z^2 
\Bigg] \log \frac{\Lambda^2}{M_{\rho^\pm}^2} 
\nonumber \\ 
&=& 
\frac{ie^2g^{\mu\nu}}{(4 \pi)^2s^2} \Bigg[ 
- \frac{1}{4} \left( M_{\rho^\pm}^2 - \frac{1}{6} p^2 \right) 
+ \left( c^2 - \frac{15}{8}  \right) M_W^2 
+ \frac{1}{8} M_Z^2 
\Bigg] \log \frac{\Lambda^2}{M_{\rho^\pm}^2} 
\,, \label{eq:jww} \\
(K)_{WW} &=& 
\frac{ie^2g^{\mu\nu}}{(4 \pi)^2s^2} \Bigg[ 
\left( -c^2 + 2  \right) M_W^2 
- M_Z^2
\Bigg] \log \frac{\Lambda^2}{M_W^2} 
\,, \\
(L)_{WW} &=& 
\frac{ie^2g^{\mu\nu}}{(4 \pi)^2s^2}  \Bigg[ 
- (1-c^2) M_W^2 
\Bigg] \log \frac{\Lambda^2}{M_W^2}
\,, \\
(M)_{WW} &=& 
\frac{ie^2g^{\mu\nu}}{(4 \pi)^2s^2} \Bigg\{ 
\frac{1}{4}M_W^2 \log \frac{\Lambda^2}{M_W^2} 
+ \Bigg[ -\frac{1}{4} M_{\rho^\pm}^2 
+ \left( c^2 - \frac{5}{4}  \right) M_W^2 
+ \frac{3}{4} M_Z^2 
\Bigg] \log \frac{\Lambda^2}{M_{\rho^\pm}^2} 
\Bigg\} 
 \,, \nonumber \\ 
\label{eq:mww}\\
(N)_{WW} &=& 
\frac{ie^2g^{\mu\nu}}{(4 \pi)^2s^2} \Bigg[ 
- \frac{1}{4}M_{\rho^\pm}^2 + \left( c^2 -1 \right) M_W^2 
- \frac{1}{4} M_Z^2
\Bigg] \log \frac{\Lambda^2}{M_{\rho^\pm}^2} 
\,, \\
(O)_{WW} &=& 
\frac{ie^2g^{\mu\nu}}{(4 \pi)^2s^2} \Bigg\{ 
\frac{1}{4} M_Z^2 \log \frac{\Lambda^2}{M_W^2}  
+ \Bigg[ - \frac{1}{4}M_{\rho^\pm}^2 
+ \left( c^2 + \frac{1}{4}  \right)M_W^2 
- \frac{3}{4} M_Z^2 
\Bigg ] \log\frac{\Lambda^2}{M_{\rho^\pm}^2} 
\Bigg\} 
\,, \nonumber\\ 
\label{eq:oww} \\
(P)_{WW} &=& 
\frac{ie^2g^{\mu\nu}}{(4 \pi)^2s^2} \Bigg[ 
- \frac{1}{4} M_{\rho^\pm}^2 
+ \left( c^2 - \frac{7}{4}  \right) M_W^2 
+ \frac{1}{2} M_Z^2  
\Bigg] \log \frac{\Lambda^2}{M_{\rho^\pm}^2} 
\,, \\
(Q)_{WW} &=& 
\frac{ie^2g^{\mu\nu}}{(4 \pi)^2s^2} \Bigg[ 
\frac{1}{4} \Bigg( M_W^2  + M_Z^2 - \frac{1}{3} p^2 \Bigg) 
\Bigg]  \log \frac{\Lambda^2}{M_W^2} 
\,, \\
(R)_{WW} &=& 
\frac{ie^2g^{\mu\nu}}{(4 \pi)^2s^2} \Bigg[ 
\frac{1}{16} \Bigg( M_{\rho^\pm}^2 - \frac{1}{3} p^2 \Bigg) 
- \left( \frac{1}{4}c^2 + \frac{1}{16} \right) M_W^2
+ \frac{5}{16} M_Z^2 
\Bigg]  \log \frac{\Lambda^2}{M_{\rho^\pm}^2} 
 \,, 
\end{eqnarray}

\begin{eqnarray}
(S)_{WW} &=& 
\frac{ie^2g^{\mu\nu}}{(4 \pi)^2s^2} \Bigg[ 
\frac{1}{16} \Bigg( M_{\rho^0}^2 + M_W^2 - \frac{1}{3} p^2 \Bigg)  
- \left( \frac{1}{4}c^2 - \frac{7}{16} \right) M_W^2 
- \frac{1}{4} M_Z^2 
\Bigg]  \log \frac{\Lambda^2}{M_{\rho^\pm}^2}
\nonumber \\ 
&=& 
\frac{ie^2g^{\mu\nu}}{(4 \pi)^2s^2} \Bigg[ 
\frac{1}{16} \Bigg( M_{\rho^\pm}^2 - \frac{1}{3} p^2 \Bigg)  
- \left( \frac{1}{4}c^2 - \frac{7}{16} \right) M_W^2 
- \frac{3}{16} M_Z^2 
\Bigg]  \log \frac{\Lambda^2}{M_{\rho^\pm}^2} 
\,, \label{eq:sww}\\
(T)_{WW} &=& 
\frac{ie^2g^{\mu\nu}}{(4 \pi)^2s^2} \Bigg[ 
\frac{1}{4}\Bigg( M_{\rho^\pm}^2 + M_{\rho^0}^2 - \frac{1}{3} p^2 \Bigg) 
- \left( 2c^2 - \frac{7}{2}  \right)M_W^2 
- \frac{1}{2} M_Z^2 
\Bigg]  \log \frac{\Lambda^2}{M_{\rho^\pm}^2} 
\nonumber \\ 
&=&
\frac{ie^2g^{\mu\nu}}{(4 \pi)^2s^2} \Bigg[ 
\frac{1}{2}\Bigg( M_{\rho^\pm}^2 - \frac{1}{6} p^2 \Bigg) 
- \left( 2c^2 - \frac{13}{4}  \right)M_W^2 
- \frac{1}{4} M_Z^2 
\Bigg]  \log \frac{\Lambda^2}{M_{\rho^\pm}^2} 
\,, \label{eq:tww} \\
 (U)_{WW} &=& 
\frac{ie^2g^{\mu\nu}}{(4 \pi)^2s^2} \Bigg[ 
- \frac{1}{4}M_W^2 
\Bigg] \log \frac{\Lambda^2}{M_W^2}
\,, \\
(V)_{WW} &=& 
\frac{ie^2g^{\mu\nu}}{(4 \pi)^2s^2} \Bigg[ 
- \frac{1}{4}M_Z^2 
\Bigg] \log \frac{\Lambda^2}{M_W^2}
\,, \\
(W)_{WW} &=& 
\frac{ie^2g^{\mu\nu}}{(4 \pi)^2s^2} \Bigg[ 
- \frac{1}{4} M_{\rho^0}^2 
+ \left( c^2 - 2 \right)M_W^2 
+ \frac{1}{2} M_Z^2 
\Bigg] \log \frac{\Lambda^2}{M_{\rho^\pm}^2}
\nonumber \\ 
&=& 
\frac{ie^2g^{\mu\nu}}{(4 \pi)^2s^2} \Bigg[ 
- \frac{1}{4} M_{\rho^\pm}^2 
+ \left( c^2 - \frac{7}{4} \right)M_W^2 
+ \frac{1}{4} M_Z^2 
\Bigg] \log \frac{\Lambda^2}{M_{\rho^\pm}^2}
\,, \label{eq:www}\\
(X)_{WW} &=& 
\frac{ie^2g^{\mu\nu}}{(4 \pi)^2s^2} \Bigg[ 
- \frac{1}{4} M_{\rho^\pm}^2 
+ \left( c^2 - \frac{3}{2} \right) M_W^2 
\Bigg] \log \frac{\Lambda^2}{M_{\rho^\pm}^2}
\,, \\ 
(Y)_{WW} &=& 
\frac{ie^2g^{\mu\nu}}{(4 \pi)^2s^2} \Bigg[ 
\frac{9}{8} M_Z^2 
\Bigg] \log \frac{\Lambda^2}{M_{\rho^\pm}^2}
\,, \\  
(Z)_{WW} &=& 
\frac{ie^2g^{\mu\nu}}{(4 \pi)^2s^2} \Bigg[ 
\frac{9}{8} M_W^2 
\Bigg] \log \frac{\Lambda^2}{M_{\rho^\pm}^2}
\,, \\  
(\alpha)_{WW} &=& 
\frac{ie^2g^{\mu\nu}}{(4 \pi)^2s^2} \Bigg[ 
- \frac{1}{8} M_Z^2 
\Bigg] \log \frac{\Lambda^2}{M_{\rho^\pm}^2}
\,, \\  
(\beta
)_{WW} &=& 
\frac{ie^2g^{\mu\nu}}{(4 \pi)^2s^2} \Bigg[ 
- \frac{1}{8} M_W^2 
\Bigg] \log \frac{\Lambda^2}{M_{\rho^\pm}^2}
\,, 
\end{eqnarray} 
where we have neglected  terms of $\mathcal{O}(\alpha x^2 M^2_W )$ and 
used $M_W^2 \approx c^2 M_Z^2$ which is valid to leading order in $x$. 
Note that, in Eqn. (\ref{eq:piwwnaive}), we have used
the relation $M_{\rho^0}^2 \approx M_{\rho^\pm}^2 + \frac{s^2}{c^2} M_W^2$ 
which follows from Eqns.(\ref{mrhopm}) and 
(\ref{mrho0}) to leading order in $x=g_0/g_1$. 

\section{Pinch Contributions and 
$\gamma$-$\rho$, $Z$-$\rho$, $W$-$\rho$ Mixing Amplitudes}
\label{appendix:mixing}

As discussed in Section \ref{subsec:pinch}, the 
$\rho$-pinch contributions of Eqns. (\ref{eq:pinchi}) -- (\ref{eq:pinchv}) 
arise diagramatically from the $\gamma$-$\rho$, $Z$-$\rho$ and $W$-$\rho$
mixing contributions to  the 
scattering amplitudes of ordinary fermions.
From Eqn. (\ref{eq:rhocouplings}), we see that the couplings
of the neutral $\rho$-meson may be written
\begin{equation}
J^\mu_\rho = {e \over sx} {J^\mu_3}' -s x\left(1-{x_1\over x^2}\right)J^\mu_A
- x {{c^2-s^2}\over 2 c }\left(1-{2c^2 \over c^2-s^2}{x_1\over x^2}\right)J^\mu_Z~.
\end{equation}
For ordinary fermions, whose couplings to the gauge-eigenstate $V^\mu$
at site 1 are suppressed by $x_1={\cal O}(x^2)$, the $\gamma-\rho$ and
$Z-\rho$ mixing amplitudes give rise to corrections to four-fermion
scattering amplitudes analogous to Eqns. (\ref{eq:mrhoa}) and (\ref{eq:mrhoz}):
\begin{eqnarray}
{\cal M}\vert^{\gamma-\rho} & \propto &
{\cal A} \cdot {1\over p^2}\cdot \Pi_{A\rho} \cdot {1\over p^2-M^2_\rho} \cdot
\left[
-sx\left(1-{x_1\over x^2}\right){\cal A}' -x  {{c^2-s^2}\over 2 c }\left(1-{2c^2 \over c^2-s^2}{x_1\over x^2}\right)
{\cal Z}'\right] \nonumber \\
& +& ({\cal A, Z} \leftrightarrow {\cal A}', {\cal Z}')~,
\\
{\cal M}\vert^{Z-\rho} & \propto &
{\cal Z} \cdot {1\over p^2-M^2_Z} \cdot \Pi_{Z\rho} \cdot {1\over p^2-M^2_\rho}
\cdot \left[
-sx \left(1-{x_1\over x^2}\right){\cal A}' -x   {{c^2-s^2}\over 2 c }\left(1-{2c^2 \over c^2-s^2}{x_1\over x^2}\right)
{\cal Z}'\right] \nonumber \\
&+&({\cal A, Z} \leftrightarrow {\cal A}', {\cal Z}')~.
\end{eqnarray}
Comparing to Eqns. (\ref{eq:maa}) and (\ref{eq:mzz}), we see
that these corrections may be absorbed into redefinitions of the
neutral boson self-energy contributions  through
\begin{eqnarray}
  \Delta\Pi_{AA} 
    &=& 2s \frac{x}{M_\rho^2} \left(1-\frac{x_1}{x^2}\right) 
        p^2 \Pi_{A\rho}(0), 
\label{eq:rhopinchAA}
  \\
  \Delta\Pi_{ZA}^\gamma 
    &=& \frac{c^2-s^2}{2c} \frac{x}{M_\rho^2} 
        \left(1-\frac{2c^2}{c^2-s^2}\frac{x_1}{x^2}\right)
        (p^2 - M_Z^2) \Pi_{A\rho}(0), 
\label{eq:rhopinchZA1}
  \\
  \Delta\Pi_{ZA}^Z
    &=& s \frac{x}{M_\rho^2} \left(1-\frac{x_1}{x^2}\right) 
        p^2 \Pi_{Z\rho}(0), 
\label{eq:rhopinchZA2}
  \\
  \Delta\Pi_{ZZ}
    &=& \frac{c^2-s^2}{c} \frac{x}{M_\rho^2} 
        \left(1-\frac{2c^2}{c^2-s^2}\frac{x_1}{x^2}\right)
        (p^2 - M_Z^2) \Pi_{Z\rho}(0)~,
\label{eq:rhopinchZZ}
\end{eqnarray}
where we have assumed $p^2 \simeq M_{W,Z}^2 \ll M_\rho^2$.
Similar considerations  (or, alternatively, taking the limit $s\to 0$ and
$M_Z \to M_W$ in Eqn. (\ref{eq:rhopinchZZ}))
lead to contributions to the charged-boson self-energy
\begin{equation}
\Delta \Pi_{WW} = {x \over M^2_\rho} \left(1-2{x_1 \over x^2}\right)(p^2-M^2_W)
\Pi_{W\rho}(0)~.
\label{eq:rhopinchWW}
\end{equation}

In this appendix, we present the results of the mixing amplitudes
$\Pi_{A\rho}$, $\Pi_{Z\rho}$,  and $\Pi_{W\rho}$ and we confirm that the relations 
Eqns.~(\ref{eq:rhopinchAA})--(\ref{eq:rhopinchWW}) reproduce the 
results of Eqns. (\ref{eq:pinchi}) -- (\ref{eq:pinchv}).

\subsection{Photon-$\rho$ and $Z$-$\rho$ Mixing Amplitudes 
$\Pi_{A\rho}$ and $\Pi_{Z\rho}$} 

\begin{figure} 
 \begin{center} 
  \input{zrho_mixing.tex}
  \vspace{15pt} 
  \caption{One-loop diagrams contributing to $Z-\rho$ mixing amplitude}
  \label{mixing:zrho} 
 \end{center} 
\end{figure} 

The photon-$\rho$ mixing amplitude $\Pi_{A\rho}$ 
and the $Z$-$\rho$ mixing amplitude $\Pi_{Z\rho}$ 
arise from the  
diagrams illustrated in Fig. \ref{mixing:zrho}.
We find 
\begin{eqnarray} 
  (A)_{\gamma\rho} &=& \frac{ie^2 g^{\mu\nu}}{(4\pi)^2 s} \left[ \frac{9M_\rho^2}{x} \right] 
                      \log\frac{\Lambda^2}{M_\rho^2} 
                      \,, \\ 
  (C)_{\gamma\rho} &=& \frac{ie^2 g^{\mu\nu}}{(4\pi)^2 s} \left[ \frac{-6M_\rho^2}{x} \right] 
                      \log\frac{\Lambda^2}{M_\rho^2} 
                      \,, \\ 
  (E)_{\gamma\rho} &=& \frac{ie^2 g^{\mu\nu}}{(4\pi)^2 s} \left[ \frac{M_\rho^2}{x} \right] 
                      \log\frac{\Lambda^2}{M_\rho^2} 
                      \,, \\ 
  (G)_{\gamma\rho} &=& \frac{ie^2 g^{\mu\nu}}{(4\pi)^2 s} \left[ -\frac{M_\rho^2}{x} \right] 
                      \log\frac{\Lambda^2}{M_\rho^2} 
                      \,, \\ 
  (K)_{\gamma\rho} &=& \frac{ie^2 g^{\mu\nu}}{(4\pi)^2 s} \left[ -\frac{M_\rho^2}{x} \right] 
                      \log\frac{\Lambda^2}{M_\rho^2} 
                      \,,
\end{eqnarray}
for $\Pi_{A\rho}$ and
\begin{eqnarray} 
  (A)_{Z \rho} &=& \frac{ie^2 (c^2-s^2) g^{\mu\nu}}{ 2(4\pi)^2 s^2 c} \left[ \frac{9M_\rho^2}{x} \right] 
                      \log\frac{\Lambda^2}{M_\rho^2} 
                      \,, \\ 
  (C)_{Z \rho} &=& \frac{ie^2 (c^2-s^2) g^{\mu\nu}}{2 (4\pi)^2 s^2 c} \left[ \frac{- 6M_\rho^2}{x} \right] 
                      \log\frac{\Lambda^2}{M_\rho^2} 
                      \,, \\ 
  (E)_{Z \rho} &=& \frac{ie^2 (c^2-s^2) g^{\mu\nu}}{2 (4\pi)^2 s^2 c} \left[ \frac{ M_\rho^2}{x} \right] 
                      \log\frac{\Lambda^2}{M_\rho^2} 
                      \,, \\ 
  (G)_{Z \rho} &=& \frac{ie^2 (c^2-s^2) g^{\mu\nu}}{ 2 (4\pi)^2 s^2 c} \left[ - \frac{ M_\rho^2}{x} \right] 
                      \log\frac{\Lambda^2}{M_\rho^2} 
                      \,, \\ 
  (K)_{Z \rho} &=& \frac{ie^2 (c^2-s^2) g^{\mu\nu}}{ 2(4\pi)^2 s^2 c} \left[ - \frac{ M_\rho^2}{x} \right] 
                      \log\frac{\Lambda^2}{M_\rho^2} 
                      \,,
\end{eqnarray}
for $\Pi_{Z\rho}$.
Here we keep only  the leading $M_\rho^2/x$ terms, and
neglect other subleading contributions.
Putting these contributions together, 
we find
\begin{eqnarray}
  \Pi_{A\rho} (0) &=& \frac{2e^2}{(4 \pi)^2 s} \frac{M_\rho^2}{x}      
                               \log \frac{\Lambda^2}{M_\rho^2} 
                               \, , \label{eq:Arho}
  \\
  \Pi_{Z \rho} (0) &=& \frac{ e^2 (c^2 -s^2) }{(4 \pi)^2 s^2 c } \frac{M_\rho^2}{x}      
                               \log \frac{\Lambda^2}{M_\rho^2} 
                               \, . \label{eq:Zrho}
\end{eqnarray}
Combining Eqns.~(\ref{eq:Arho})--(\ref{eq:Zrho})
with Eqns.~(\ref{eq:rhopinchAA})--(\ref{eq:rhopinchZZ})
yield the results presented in Section \ref{subsec:pinch}.

\subsection{$W$-$\rho$ Mixing Amplitude $\Pi_{W\rho}$} 

\begin{figure} 
 \begin{center} 
  \input{wrho_mixing.tex}
  \vspace{15pt} 
  \caption{One-loop diagrams contributing to $W-\rho$ mixing amplitude}
  \label{mixing:wrho} 
 \end{center} 
\end{figure} 

The $W$-$\rho$ mixing amplitude $\Pi_{W\rho}$ 
arises from the diagrams illustrated in Fig. \ref{mixing:wrho}.
We find 
\begin{eqnarray} 
  (C)_{W\rho} &=& \frac{ie^2 g^{\mu\nu}}{2 (4\pi)^2 s^2} \left[ \frac{9M_\rho^2}{x} \right] 
                      \log\frac{\Lambda^2}{M_\rho^2} 
                      \,, \\ 
  (F)_{W\rho} &=& \frac{ie^2 g^{\mu\nu}}{2 (4\pi)^2 s^2} \left[ - \frac{3M_\rho^2}{x} \right] 
                      \log\frac{\Lambda^2}{M_\rho^2} 
                      \,, \\ 
  (G)_{W\rho} &=& \frac{ie^2 g^{\mu\nu}}{2 (4\pi)^2 s^2} \left[ - \frac{3M_\rho^2}{x} \right] 
                      \log\frac{\Lambda^2}{M_\rho^2} 
                      \,, \\ 
  (J)_{W\rho} &=& \frac{ie^2 g^{\mu\nu}}{2 (4\pi)^2 s^2} \left[ - \frac{M_\rho^2}{x} \right] 
                      \log\frac{\Lambda^2}{M_\rho^2} 
                      \,, \\   
  (T)_{W\rho} &=& \frac{ ie^2 g^{\mu\nu}}{2 (4\pi)^2 s^2} \left[ \frac{M_\rho^2}{x} \right] 
                      \log\frac{\Lambda^2}{M_\rho^2} 
                      \,, \\ 
  (W)_{W\rho} &=& \frac{ ie^2 g^{\mu\nu}}{2 (4\pi)^2 s^2} \left[ - \frac{M_\rho^2}{2x} \right] 
                      \log\frac{\Lambda^2}{M_\rho^2} 
                      \,, \\ 
  (X)_{W\rho} &=& \frac{ ie^2 g^{\mu\nu}}{2 (4\pi)^2 s^2} \left[ - \frac{M_\rho^2}{2x} \right] 
                      \log\frac{\Lambda^2}{M_\rho^2} 
                      \,. 
\end{eqnarray}
Here we keep only  the leading $M_\rho^2/x$ terms, and
neglect other subleading contributions.
Putting these contributions together, 
we find
\begin{equation} 
  \Pi_{W\rho}(0) = \frac{ e^2 }{ (4 \pi)^2 s^2 } \frac{M_\rho^2}{x}      
                               \log \frac{\Lambda^2}{M_\rho^2} 
                               \, . \label{eq:Wrho}
\end{equation}
The amplitude Eq.~(\ref{eq:Wrho}), used
in Eq.~(\ref{eq:rhopinchWW}), yields the results presented in 
Section \ref{subsec:pinch}.

\section{Vertex Corrections from Fermion Delocalization Operator}
\label{appendix:delocal}

 From the forms of the $x_1$-dependent interactions of Eqn. (\ref{delocal-op:me}), 
we see it is straightforward to compute the vertex correction amplitudes 
$J^3_L-A$ and $J^3_L-Z$. 
Note that the vertex correction arising from an interaction 
of the first term of line two in Eqn. (\ref{delocal-op:me}) is 
suppressed by a factor of $(M_W^2/M^2_{\rho^\pm})$ due to 
the $\pi_{W^\pm}$ loop;  therefore, it may be neglected since  we are 
concerned with the leading log contributions of the form 
$\log(\Lambda^2/M_{W,\rho}^2)$. 
The relevant Feynman graphs are shown in fig.~\ref{x1-reno-graphs}.

By using the formulae of Feynman integral in Appendix \ref{formula}, 
the vertex correction amplitudes corresponding to 
diagrams (A)-(E) of figure~\ref{x1-reno-graphs} are evaluated to be, 
for an external photon
\begin{eqnarray} 
(B)_{J^3_LA} &=& \frac{2ie^3g^{\mu\nu}}{(4\pi)^2 s^2} 
\left( \frac{x_1}{x^2} \right) \log\frac{\Lambda^2}{M_{\rho^\pm}^2} 
\,, \label{J3-A:B} \\   
(C)_{J^3_LA} &=&  - \frac{ie^3g^{\mu\nu}}{(4\pi)^2 s^2} 
\left( \frac{x_1}{x^2} \right) \log\frac{\Lambda^2}{M_{\rho^\pm}^2} 
\,, \label{J3-A:C} \\ 
(E)_{J^3_LA} &=&  \frac{ie^3g^{\mu\nu}}{(4\pi)^2 s^2} 
\left( \frac{x_1}{x^2} \right) \log\frac{\Lambda^2}{M_{\rho^\pm}^2} 
\,,  \label{J3-A:E}
\end{eqnarray}
and, for an external $Z$-boson
\begin{eqnarray} 
(A)_{J^3_LZ} &=& \frac{ie^3g^{\mu\nu}}{(4\pi)^2 s^3c} 
\left( \frac{x_1}{x^2} \right) \log\frac{\Lambda^2}{M_{\rho^\pm}^2} 
\,, \label{J3-Z:A} \\   
(B)_{J^3_LZ} &=& \frac{ie^3(c^2-s^2) g^{\mu\nu}}{(4\pi)^2 s^3c} 
\left(\frac{x_1}{x^2} \right) \log\frac{\Lambda^2}{M_{\rho^\pm}^2} 
\,, \label{J3-Z:B} \\ 
(C)_{J^3_LZ} &=&  - \frac{ie^3g^{\mu\nu}}{4(4\pi)^2 s^3c} 
\left(\frac{x_1}{x^2} \right) \log\frac{\Lambda^2}{M_{\rho^\pm}^2} 
\,, \label{J3-Z:C} \\ 
(D)_{J^3_LZ} &=&  - \frac{ie^3(c^2-s^2) g^{\mu\nu}}{2(4\pi)^2 s^3c} 
\left( \frac{x_1}{x^2} \right) \log\frac{\Lambda^2}{M_{\rho^\pm}^2} 
\,, \label{J3-Z:D} \\ 
(E)_{J^3_LZ} &=&   \frac{ie^3(c^2-s^2) g^{\mu\nu}}{2(4\pi)^2 s^3c} 
\left( \frac{x_1}{x^2} \right) \log\frac{\Lambda^2}{M_{\rho^\pm}^2} 
\,. \label{J3-Z:E}
\end{eqnarray} 
It should be noticed that $(C)_{J^3_LA} + (E)_{J^3_LA}=0$;
the corresponding one-loop 
generated operator is written in the form, 
\begin{equation} 
   \mathcal{L}_{(C)+(E)} = - \bar{\psi}_L \gamma^\mu 
   (g_0 L_\mu - g_1 V_\mu) \psi_L \cdot 
   \Bigg[ \frac{e^2}{(4\pi)^2 s^2} 
\left( \frac{x_1}{x^2} \right) \log\frac{\Lambda^2}{M_{\rho^\pm}^2} 
\Bigg]
\,, 
\end{equation}
in which the operator $(g_0 L_\mu - g_1 V_\mu)$ is orthogonal to the photon.

Combining Eqns.(\ref{J3-Z:A})-(\ref{J3-Z:E}) and (\ref{J3-A:B}), 
we find that the vertex corrections are incorporated into 
the following operators: 
\begin{equation} 
  \mathcal{L}_{\rm eff} = 
\frac{e}{sc} \cdot G_1(M_{\rho^\pm}^2;x_1) \cdot J_L^{\mu 3} Z_\mu 
+ e \cdot G_2(M_{\rho^\pm}^2; x_1) \cdot J_L^{\mu 3} A_\mu 
\,, \label{eff-op}
\end{equation}
where 
\begin{eqnarray}
G_1(M_{\rho^\pm}^2; x_1) 
&=&  \frac{e^2}{(4\pi)^2s^2} \left( 2c^2 - \frac{1}{4} \right) 
\left( \frac{x_1}{x^2} \right) \log\frac{\Lambda^2}{M_{\rho^\pm}^2} 
\,, \label{G1} \\ 
G_2(M_{\rho^\pm}^2; x_1) 
&=&  \frac{2e^2}{(4\pi)^2s^2}  
\left( \frac{x_1}{x^2} \right) \log\frac{\Lambda^2}{M_{\rho^\pm}^2} 
\,. \label{G2}
\end{eqnarray}


\end{document}